\documentclass[english]{article}
\usepackage[T1]{fontenc}
\usepackage[latin9]{inputenc}
\usepackage{longtable}
\usepackage{textcomp}
\usepackage{bm}
\usepackage{amsmath}
\usepackage{graphicx}
\usepackage{setspace}
\usepackage{amssymb}
\usepackage{esint}

\makeatletter

\newcommand{\lyxline}[1][1pt]{%
  \par\noindent%
  \rule[.5ex]{\linewidth}{#1}\par}
\newcommand{\noun}[1]{\textsc{#1}}
\providecommand{\tabularnewline}{\\}

\usepackage[figuresright]{rotating} 
\usepackage{lscape}
\usepackage{longtable}
\usepackage{graphicx,subfigure}
\usepackage{epsfig}
\usepackage{amssymb}
\usepackage{amsmath}
\usepackage{bm}

\usepackage{fancyhdr}
\usepackage{upgreek}
\usepackage{ccaption}

\newfont{\cour}{cmtt12}

\let\newcommand=\providecommand

\newcommand{\captionfonts}{\small}
\makeatletter  
\long\def\@makecaption#1#2{%
  \vskip\abovecaptionskip
  \sbox\@tempboxa{{\captionfonts #1: #2}}%
  \ifdim \wd\@tempboxa >\hsize
    {\captionfonts #1: #2\par}
  \else
    \hbox to\hsize{\hfil\box\@tempboxa\hfil}%
  \fi
  \vskip\belowcaptionskip}
\makeatother   

\let\newcommand=\providecommand

\setlongtables
\makeatother

\usepackage{babel}

\begin{document}

\title{Long-wavelength gravitational waves and cosmic acceleration}

\author{\noindent Edmund R Schluessel%
\thanks{schluesseler@cardiff.ac.uk%
}\\
School of Physics and Astronomy, Cardiff University, 5, the Parade,\\
Cardiff, Wales, United Kingdom CF24 3AA}
\maketitle
\begin{abstract}
Strong long-scale gravitational waves can explain cosmic acceleration
within the context of general relativity without resorting to the
assumption of exotic forms of matter such as quintessence. The existence
of these gravitational waves in sufficient strength to cause observed
acceleration can be compatible with the cosmic microwave background
under reasonable physical circumstances. An instance of the Bianchi
IX cosmology is demonstrated which also explains the alignment of
low-order multipoles observed in the CMB. The model requires a closed
cosmology but is otherwise not strongly constrained. Recommendations
are made for further observations to verify and better constrain the
model.

\noindent \noun{Keywords: }Dark energy, acceleration, gravitational
waves, Bianchi model, CMB, Axis of Evil, CMB cold spot.
\end{abstract}
\tableofcontents{}

\part{Introduction}

\section{Background}

The observational confirmation that the universe has been expanding
from a condition of extreme density and minute size since some point
in the finite past represents a major triumph of Einstein's theory
of gravitation in providing an elegant explanation for cosmology,
without the addition of exotic, heretofore-unobserved substances or
fundamental forces. This notion has however faced a serious challenge
since Riess's 1998 discovery\cite{Riess 1998} of cosmic acceleration.
The purpose of this research is to evaluate the question: can the
back-reaction of long-wavelength gravitational waves in a closed universe
contribute to cosmic acceleration while remaining compatible with
observational constraints?

\subsection{Standard cosmology predicts an expanding universe}

The full Einstein equations read%
\footnote{Throughout this document, indices written with Greek letters $\mu,\nu$
etc. run over 0,1,2,3 and indices written in Roman letters \emph{i,
j} etc. run over 1,2,3. The sign of the metric tensor reads $+,-,-,-$.%
}

\begin{equation}
R_{\mu\nu}-\frac{1}{2}Rg_{\mu\nu}=kT_{\mu\nu}+\Lambda g_{\mu\nu}\label{eq:Generic complete Einstein equations}\end{equation}

\noindent where $g_{\mu\nu}$ is the metric tensor, $R_{\mu\nu}$
is the once-contracted Riemann tensor, \emph{R} is the Ricci curvature
scalar, $T_{\mu\nu}$ is the energy-momentum tensor, $\Lambda$ is
the {}``cosmological constant'' and the constant $k\equiv8\pi G/c^{4}\approx2.08\times10^{-43}\mbox{kg}^{-1}\mbox{m}^{-1}\mbox{s}^{2}$.
The Bianchi identity guarantees $T_{11}=T_{22}=T_{33}$ so, in a Gaussian
($g_{00}=1$) and synchronous ($g_{0i}=0$) coordinate system we have\cite{Weinberg}:

\begin{align}
R_{0}^{0}-\frac{1}{2}R= & kT_{0}^{0}+\Lambda\label{Einstein equation T_00}\\
-R= & kT_{\mu}^{\mu}-2\Lambda\label{Einstein equation T^mu_mu}\end{align}

\noindent .

\subsubsection{Cosmological parameters}

In discussions of cosmology it is conventional to track the expansion
of an isotropic metric by introducing a {}``scale factor'', a positive
function of time only. In general the scale factor has no specific
geometric meaning other than to compare distances in the metric at
different points in time. Furthermore, the scale factor loses unique
meaning when the universe becomes non-isotropic. We will denote this
scale factor function as $a\left(t\right)$ in analogy with its definition
in the Robertson-Walker metric, where it appears as\cite{Bondi}

\begin{equation}
ds^{2}=dt^{2}-a^{2}\left(t\right)\frac{\left(dx^{1}\right)^{2}+\left(dx^{2}\right)^{2}+\left(dx^{3}\right)^{2}}{1+\frac{1}{4}K\left[\left(dx^{1}\right)^{2}+\left(dx^{2}\right)^{2}+\left(dx^{3}\right)^{2}\right]^{2}}\label{eq:RW metric}\end{equation}

\noindent and the symbol \emph{K} has the value 0 in a flat universe,
1 in a closed universe, and -1 in an open universe. In this case the
Einstein equations read%
\footnote{A single dot denotes a derivative with respect to \emph{t; }two dots
denote a second derivative with respect to \emph{t.}%
} \begin{align}
\frac{3}{a^{2}}\left(\dot{a}^{2}+K\right)= & k\epsilon+\Lambda\label{eq:Einstein eq energy}\\
-6\frac{\ddot{a}}{a}= & k\left(\epsilon+3p\right)-2\Lambda\label{eq:Einstein eq e+3p}\end{align}

\noindent where $\epsilon$ denotes the energy density of matter described
by the energy-momentum tensor and \emph{p} denotes the pressure of
matter described in that tensor. Let the Hubble parameter be defined
by $H\equiv\dot{a}/a$ and let $H_{0}$ be the value of \emph{H} at
the present time (this is how we will generally use the subscript
\emph{0}). Because the universe seems flat and dominated by ordinary
matter over small scales, it is common to move terms arising from
\emph{K} to the right-hand side of the equation, where they act as
elements of an {}``effective energy-momentum tensor'', and to state
contributors to cosmological expansion as dimensionless parameters
$\Omega_{i}$ in comparison to the {}``critical density'' $\epsilon_{\mbox{critical}}$,
that is, the energy-density of ordinary matter required for the universe
to be flat: $k\epsilon_{\mbox{critical}}=3H_{0}^{2}$ so 

\begin{equation}
\begin{array}{c}
3H_{0}^{2}=k\epsilon_{0}+\Lambda-3\frac{K}{a_{0}^{2}}\\
\Omega_{K}+\Omega_{M}+\Omega_{R}+\Omega_{\Lambda}=1\end{array}\label{eq:Einstein Omegas}\end{equation}

\noindent where

\begin{align}
\Omega_{M}+\Omega_{R} & \equiv & k\epsilon_{0}/3H_{0}^{2}\\
\Omega_{\Lambda} & \equiv & \Lambda/3H_{0}^{2}\\
\Omega_{K} & \equiv & -K/\dot{a}_{0}^{2}\label{eq:Omegas defined-1}\end{align}

\noindent . multiple observations, most recently by WMAP, have confirmed
that $\Omega_{R}\ll\Omega_{M}$\cite{WMAP 7-year} and so to the limit
of the precision with which these quantities can be evaluated $\Omega_{K}+\Omega_{M}+\Omega_{\Lambda}=1$.%
\footnote{Chernin, in \cite{Chernin}, elegantly derives a description of the
scale factor in an open Friedmann cosmology which can be used when
there is a significant amount of relativistic matter in a cold matter-dominated
universe. Chernin's equation is easily generalized to the closed Friedmann
universe.%
}

\subsection{Simple cosmology predicts a decelerating universe}

If the scale factor \emph{a} measures a distance, it is reasonable
to say by analogy that $\dot{a}$ can be compared to a velocity and
$\ddot{a}$ an acceleration. In isotropic cosmology we define the
{}``deceleration parameter'' \emph{Q}~by%
\footnote{\emph{Q}~has been defined with a minus sign for historical reasons.
$Q>0$ denotes a decelerating universe; $Q<0$ denotes an accelerating
universe. We have avoided the more common notation \emph{q} in favor
of \emph{Q} to avoid confusion when interpreting the source material.%
}:

\begin{equation}
Q\equiv-\frac{\ddot{a}a}{\dot{a}^{2}}=\frac{d}{dt}\frac{1}{H}-1\label{eq:Deceleration parameter definition}\end{equation}

\noindent . Dividing (\ref{eq:Einstein eq e+3p}) by (\ref{eq:Einstein Omegas})
we easily obtain

\begin{equation}
Q=\frac{1}{2}\frac{k\left(\epsilon+3p\right)-2\Lambda}{k\epsilon+\Lambda-K/a^{2}}=\frac{1}{2}\Omega_{M}-\Omega_{\Lambda}\label{eq:Einstein equation energy pressure}\end{equation}

\noindent . A flat universe with no cosmological constant must always
decelerate. While the properties of so-called {}``dark matter''
remain undetermined, the localisibility of dark matter's distribution
and its slow motion implies it can be treated as $w=0$ dust.

We can also immediately say that in a universe with no cosmological
constant, acceleration is possible under the condition \begin{equation}
\frac{2}{1+3w}\left(1-K/a^{2}k\epsilon\right)<0\end{equation}

\noindent .

\subsection{Observations say the universe is accelerating}

Acceleration in and of itself is not a newcomer to cosmology. The
de Sitter cosmology\cite{de Sitter}, discovered in 1917, is driven
solely by a cosmological constant and consequently has a constant
deceleration parameter of $Q=-1$. Bondi, Gold \& Hoyle's {}``steady
state'' universe\cite{Steady state} similarly accelerates with $Q=-1$,
this value being associated with a universe whose expansion is driven
solely by a field whose energy density is not dependent on the size
of the universe. With the proposal of {}``big bang'' nucleosynthesis\cite{Big bang}
and the subsequent discovery of the cosmic microwave background\cite{CMB},
consensus came to settle on the simplest matter-filled model, the
Friedmann universe.\cite{Friedmann}

Throughout the 1990s, astronomical observations began to indicate
that the matter energy density of the universe was far below the critical
density, leading some (for example \cite{Turner 1995}) to propose
the resurrection of the cosmological constant in order to preserve
the observed near-flatness of space.

In 1998, Riess \emph{et al.} published an analysis\cite{Riess 1998}
of the light from a small number of type Ia supernovae with $0.16\leq z\leq0.62$
and concluded from this set that the recent universe is accelerating
with $Q_{0}=-1.0\pm0.4$. Further observations and analysis (see \noun{Part}
\ref{sec:Observations-of-acceleration}) have also provided evidence
that the universe has $Q_{0}<0$.

While Riess \emph{et al.}~did not exclude the possibility of a universe
with $K\neq0$, the assumption of a flat universe remains predominant
throughout the field of cosmology as observations, both from supernova
data and WMAP\emph{,} have shown that the universe is, on observable
scales, very close to flat -- although it is impossible to distinguish
between a universe that is genuinely flat, with $\Omega_{K}=0$ and
one with $\Omega_{K}$ very close to but not equal to zero.

\section{Dark energy}

Since the discovery of acceleration, numerous explanations for the
phenomenon have been proposed, all depending on an isotropic field
creating additional, invisible energy. Turner and Huterer\cite{Turner}
introduce the term {}``dark energy'', analogous to dark matter in
the sense that dark energy does not interact electromagnetically with
ordinary matter and has the property of an energy density, to term
this additional energy, which appears to make up over 70\% of the
total energy content of the universe\cite{WMAP 7-year}.

The assumption of a flat homogeneous cosmology demands that cosmic
acceleration comes from a cosmological constant or a scalar field.
Most scalar theories for explaining cosmic acceleration fall into
two classes: an exotic form of matter with negative energy density,
and surrender of the cosmological principle. Other scalar theories
sacrifice different assumptions, such as homogeneity, or invoke more
exotic explanations unsupported by laboratory physics.

\subsection{Cosmological constant}

The simplest most familiar variation on the Robertson-Walker cosmological
model which allows an accelerating universe is the {}``$\Lambda$CDM''
model -- a universe dominated by {}``cold'' (non-relativistic, $p=0$)
matter with both baryonic and dark components, and with the existence
of a non-zero cosmological constant. In such a universe the Einstein
equations read\cite{Bondi}

\begin{align}
3H^{2}= & k\epsilon+\Lambda\\
-6\frac{\ddot{a}}{a}= & k\epsilon-2\Lambda\end{align}

\noindent so when $k\epsilon/\Lambda$ is small such that $\left(k\epsilon/\Lambda\right)^{2}$
is negligible, that is, the universe is dominated by a cosmological
constant,

\begin{equation}
Q=\frac{1}{2}\frac{k\epsilon-2\Lambda}{k\epsilon+\Lambda}\approx-1+\frac{3k\epsilon}{2\Lambda}\end{equation}

\noindent which at first glance appears to neatly explain Riesse \emph{et
al.}'s result. However, as will be shown (see \noun{section} \ref{sub:Analysis}),
the case for a cosmological constant is not definite. Furthermore,
the theoretical background explaining the strength of the cosmological
constant is not well developed, relying on an understanding of quantum
gravity which does not yet exist\cite{Carroll}. While the cosmological
constant can always be said to have a {}``right to exist'' in the
Einstein equations, current physics does not explain why it should
have any particular strength and as such the cosmological constant
should be treated as the simplest form of a scalar field of exotic
matter.

\subsection{Quintessence}

More general than the cosmological constant but similar in structure
is the proposal of {}``quintessence''\cite{Turner}, a novel form
of matter with a time-dependent equation of state that can take on
negative values. Many forms of these have been proposed; one form
of these, for example, is the {}``Chaplygin gas''\cite{Chaplygin gas},
which has equation of state $p=-A/\epsilon$ for $A>0$. Quintessence
theories are particularly motivated by the idea that acceleration
is a cosmologically recent phenomenon, noting limited data (see \noun{Part}
\ref{sec:Observations-of-acceleration}) that the equation of state
of dark energy may be evolving with time.

Any formulation of quintessence must be regarded as highly speculative.
At the most fundamental level, all theories of quintessence propose
the existence of a kind of matter which:
\begin{itemize}
\item has never been observed experimentally;
\item does not interact with ordinary matter via the electromagnetic force;
\item has a negative equation of state, that is, a positive energy density
produces a negative pressure;
\item plays a prominent role at current energy levels, as opposed to exotic
effects (e.g. unification of forces) thought to have taken place only
in the very early universe.
\end{itemize}
In the absence of any compelling experimental evidence whatsoever
for any kind of quintessence, quintessence and quintessence-like models
should be rejected as definitive explanations for dark energy.

\subsection{Local inhomogeneity}

A more mundane explanation which has been offered for acceleration
is the {}``Hubble bubble''\cite{Riess 1998,Hubble bubble}, regions
of lower density in the intergalactic medium. If the vicinity of the
Milky Way had lower matter energy density, expansion in its vicinity
would increase\cite{Einstein bubble}, causing the illusion of cosmic
acceleration.

Not only would the density deficit in such a {}``bubble'' have to
be quite large in order to cause acceleration, but the theory, which
has the advantage of requiring no new physics, supposes either the
existence of a rare or unique void that the Milky Way happens to be
in -- a violation of the cosmological principle in the sense that
it makes observers in the Milky Way privileged -- or a preponderance
of voids whose presence makes the universe inhomogeneous not just
in small patches but on average.\cite{really cosmological constant?,Swiss cheese}

\subsection{Exotic models}

\subsubsection{Modified relativity}

Some proposals to explain dark energy propose modifications to the
Einstein equations. The best-known of these is the Cardassian Expansion
model\cite{Cardassian}, which proposes time-dependent variation of
the equation of state of matter. The Cardassian model is of particular
interest in that it proposes an equation for the density perturbation 

\begin{equation}
\kappa^{\prime\prime}\left(x\right)+2\frac{s}{x}\kappa^{\prime}-\frac{3}{2}s^{2}\kappa=0\end{equation}

\noindent for unknown constant \emph{s, }which equation begins to
resemble that for weak gravitational waves in a closed universe (cf.
\noun{equation} (\ref{eq:Einstein 1st order})). Like\emph{ }Chaplygin
gas and the {}``DGP'' model\cite{DGP}, the Cardassian model justifies
itself based on theories about higher-dimensional manifolds which
remain untested.

\subsubsection{Topological defects}

The existence of cosmic strings would change the overall equation
of state of the matter in the universe by a constant\cite{SNAP paper,cosmic strings},
creating acceleration through simple deviation from the Friedmann
model. While theories of cosmic inflation predict the formation of
cosmic strings and other topological defects, such defects remain
completely undetected.

\section{Tensorial theories for acceleration in a flat universe}

If we wish to preserve the theory of relativity and cosmic homogeneity,
while at the same time relying only on effects with good experimental
basis, scalar fields appear to be excluded as an explanation for acceleration.
Ergo within the context of general relativity the next place to search
for an answer is in tensor theories, which include the possibility
of gravitational waves.

Lifshitz's theory of cosmological perturbations\cite{Lifshitz 46}
appears to exclude tensorial answers to the problem of acceleration:
gravitational waves have the same equation of state as radiation,
and local clumps gravitational waves in the theory (where {}``local''
means bounded within an region smaller than the radius of curvature
of the universe) both decay rapidly and collapse spatially. Rodrigues\cite{Rodrigues}
takes a first step in discussing anisotropic dark energy, but limits
his analysis to a flat universe and thus creates the problem of an
anisotropic {}``big rip''.

A high-frequency gravitational wave background has been proposed\cite{accelerating}
as the source of cosmic acceleration. While the authors' analysis
appears initially promising, similar to many scalar dark energy candidates
the theory relies on the existence of an inflation-induced gravitational
wave background that remains only hypothetical. Furthermore, the authors
obtain their result by selection of an averaging scheme without mathematical
rigor -- surely choosing a mathematical model based on the desired
results cannot be considered physics. At any rate, the strength of
the background inflationary theory predicts is not sufficiently great
to explain the observed large acceleration.

\part{Evidence for acceleration\label{sec:Observations-of-acceleration}}

\section{Introduction}

The theory of tests to evaluate the deceleration parameter using supernovae
as standard candles began with Wagoner\cite{Wagoner} in 1977. Starting
from assumptions of an isotropic Friedmannian cosmology which is not
necessarily flat, Wagoner notes the approximate relation

\begin{equation}
d_{E}=H_{0}^{-1}\left[z-\frac{1}{2}\left(1+Q_{0}\right)z^{2}+\mathcal{O}\left(z^{3}\right)\right]\label{eq:Wagoner redshift}\end{equation}

\noindent which, when $H_{0}$ and \emph{z} are known, relates the
deceleration parameter to the distance $d_{E}$ as determined by the
dimming of the supernova (where Wagoner was originally discussing
Type II supernova events)%
\footnote{\noun{Equation} (\ref{eq:Wagoner redshift}) is of course a generalization
of the famous distance-redshift approximation $H_{0}d_{E}\approx z$.\cite{Weinberg}%
}. This relation is valid when $z$ is small such that $z^{3}$ is
negligible, limiting its usefulness above $z\sim1$, and requires
the assumption of only small changes in the Hubble constant $H_{0}$
(that is, in a Friedmann cosmology $\dot{a}_{F}/a_{F}$ evaluated
near the observer) on the interval from $z\approx0$ to $z\approx1$.

Type Ia supernovae are thought to be a {}``standard candle'' for
the measurement of distance and redshift; that is, supernovae of that
type are thought to possess spectral and luminosity curves which are
nearly identical. Therefore, observation of extragalactic type Ia
supernovae is believed to produce reliable information on both the
distance of the event (noting that brightness diminishes as the inverse
square of distance) and the redshift of the distance associated with
the event (through the change in the peak of the supernovae's spectra),
with redshift \emph{z }related to the scale factor \emph{$a_{F}$}
by\begin{equation}
z+1=\frac{a_{F}\left(t_{\mbox{observation}}\right)}{a_{F}\left(t_{\mbox{emission}}\right)}\label{eq:Redshift definition}\end{equation}

\noindent . Analysis of a statistically unbiased dataset of $z\left(t\right)$
therefore gives empirical information on $Q\left(t\right)$.

Colgate\cite{Colgate} proposed that Type I supernovae should be used
to measure the deceleration parameter in preference to Type II supernovae.
Type I supernovae, specifically the {}``Type Ia'' whose mechanism
is thought to be the accretion of matter onto the surface of a white
dwarf star, are understood to have a well-defined typical absolute
magnitude and spectrum, and assuming this is true the distance to
and redshift of a given Type Ia supernova event (SNe) can easily be
determined by fitting its light curve to standard templates. Therefore,
with a sufficient sample of extragalactic supernovae of $z\lesssim1$,
the parameters $H$ and $Q$ can be measured directly. When an isotropic
cosmology with constant deceleration parameter $Q=Q_{0}$ is assumed,
knowledge of $H_{0}$ and $Q_{0}$ are sufficient to typify the parameters
of the universe\cite{Weinberg}.

With the advent of modern optical astronomy such as adaptive optics\cite{Adaptive optics}
and space-based optical telescopy\cite{Spitzer}, such surveys have
become possible, but have produced results contradicting the standard,
cold matter-filled Friedmann model of cosmology.

\section{Surveys of acceleration}

Cosmological studies measuring \emph{Q}~have been ongoing since 1997
and consist of analysis of redshifts\cite{Riess 1998,SDSS-II technical,SDSS-II,ESSENCE,SNLS data,SNLS3,WiggleZ,WiggleZ design,Perlmutter,Perlmutter 2,Kowalski,SNLS results,Essence 2,ESSENCE Exotic,Perlmutter 1997,Riess far supernovae,Riess Gold updated,WiggleZ first release,WiggleZ test}
of type Ia supernovae.

The High-\emph{z} Supernova Search Team's initial study of the deceleration
parameter\cite{Riess 1998} was the first large study to call attention
to the problem of acceleration. Working from a sample of sixteen supernovae
(four of which were well-observed {}``high confidence'' sources),
the most distant with $z=0.97$, Riess concluded that the universe
has $Q_{0}<0$ to high confidence, although the measurement of $Q_{0}$
itself possessed a high degree of uncertainty. Riess also noted the
high sensitivity of the result to individual data points. Oddly, the
authors dismiss the closed cosmology despite the data indicating it
as preferred\cite[Fig. 7]{Riess 1998}; however the size of their
experimental error precludes real evaluation of spatial curvature.

The Supernova Cosmology Project (SCP) made an earlier attempt to evaluate
$Q_{0}$ with the use of supernovae\cite{Perlmutter 1997}. This small
survey ($n=7$) on relatively nearby supernovae found a result inconsistent
with those that followed it, giving results consistent with a universe
with no dark energy and with too high a degree of error to meaningfully
evaluate the geometry of the universe.

In contrast, the Supernova Cosmology Project's 1998 evaluation\cite{Perlmutter,Perlmutter 2}
of 42 Type Ia SNe added further evidence that the universe was accelerating,
and also makes note of the surprising coincidence of the energy density
$\Omega_{\Lambda}$'s near-equivalence with the total energy density
in the current epoch. The SCP also failed to consider the closed cosmology
despite supernova data favoring it\cite[Fig. 7]{Perlmutter 2}.

The ESSENCE\cite{ESSENCE} survey was expressly designed to examine
cosmic acceleration and detected 102 type-Ia supernovae from $0.10\leq z\leq0.78$,
of which 60 were used for cosmological analysis. The initial analysis
of ESSENCE assumed flatness of the universe. ESSENCE's observational
fields were deliberately chosen to overlap the areas of previous surveys
and to lie within ten degrees of the celestial equator; all were also
between 23:25 and 02:33 Right Ascension. Combining data from ESSENCE,
SNLS and other sources\cite{Essence 2} led to a conclusion consistent
with other analyses. Exploration of more exotic models\cite{ESSENCE Exotic}
found that no model of those tested was a good fit for ESSENCE's data.

The Supernova Legacy Survey (SNLS)\cite{SNLS data} recorded 472 type-Ia
supernovae. While analysis of the SNLS dataset\cite{SNLS3} provides
results consistent with a universe driven by cosmological constant,
the uncertainty on analysis of a time-dependent component to the equation
of state of dark energy is very large; their analysis also does not
consider a closed universe as a possible model\cite{SNLS results}.
Furthermore, the SNLS team also note the presence of two outliers
and only 125 of 472 events were used to evaluate cosmology. SNLS observed
SNe in four fields, one of which (field 3) is far above the plane
of the celestial equator at 52 degrees declination; this and \cite{Riess Gold updated}'s
northern field are the only fields with multiple observations in a
small area more than 20 degrees from the celestial equator surveyed
to date. SNLS also notes\cite[section 5.4]{SNLS results} that the
values of $\Omega_{M}$ evaluated in the four fields are compatible
only at a 37\% confidence level -- a surprising result given that
each SNLS field contains at least 60 SNe in quite small (one square
degree) areas.

The Hubble Space Telescope or HST survey of supernovae, published
in 2004 and reviewed by the Supernova Cosmology Project\cite{HST}
observed twenty type-Ia supernovae with redshifts $0.63<z<1.42$.
While the number of SNe observed is small, the HST survey has the
advantage of covering a wider area of sky than other SNe surveys.
Analysis of the HST dataset suggests a rapidly-evolving dark energy
field, although with very high error on measurements greater than
$z=1$ due to the small ($n=10$) sample size it is impossible to
take these results as anything more than suggestive. HST slightly
favored a closed model of the universe, when considering interpretations
of data that allowed $\Omega_{K}\neq0$. 

The Supernova Cosmology Project's 2008 analysis of supernova data\cite{Kowalski}
made a analysis of combined SNLS, ESSENCE and HST data, and attempted
to analyze the data in the context of a theory of a time-dependent
equation of state for dark energy but concluded {}``present SN data
sets do not have the sensitivity to answer the questions of whether
dark energy persists to \emph{z} > 1, or whether it had negative pressure
then.'' The analysis rejected 10\% of all SNe from the combined data
sets as outliers, many based on their failure to fit with a nearby
$H_{0}$; Kowalski \emph{et al.'}s rejection of outliers also shifts
their analysis from one favoring a closed universe to one favoring
a flat one\cite[Fig. 11]{Kowalski}.

Further work by Riess \emph{et. al.}\cite{Riess Gold updated,Riess far supernovae}\emph{
}produced the so-called {}``gold'' dataset of SNe, a group of supernova
events with particularly clear light curves with 33 at $z>1$. These
supernovae were observed in two small (one square degree) fields.
\cite{Riess Gold updated} claims a great reduction in the uncertainty
of the Hubble parameter at $z>1$ but the Hubble parameter measured
in the extended {}``gold'' set gives a value for the Hubble parameter
not reconcilable with that in the \cite{Riess far supernovae} dataset.
Riess \emph{et. al.~}conclude that \emph{w }is negative (with large
experimental error) in the region $1<z<2$, then attempt to extrapolate
the behavior of dark energy back to $z=1089$.

Sollerman \emph{et al'}s analysis of the Sloan Digital Sky Survey-II
supernova data\cite{SDSS-II technical,SDSS-II} is the most recent
analysis indicating cosmic acceleration. SDSS-II observed 103 type-Ia
supernovae in a long, narrow strip along the celestial equator, including
many from lower redshifts than had been previously examined in detail;
Sollerman \emph{et. al.} also made use of data from the HST, ESSENCE
and SNLS surveys, bringing the total number of SNe examined to 288.
The primary conclusion to be drawn from SDSS is the sensitivity of
cosmological measurements to the specific analysis technique used\cite{SDSS K09};
analysis of the data with two different curve-fitting algorithms produce
two different, albeit somewhat compatible, results. 

Further obscuring the neatness of measuring \emph{Q}, Jha \emph{et.
al.} noted\cite{Jha} that the uneven local distribution of galaxies,
specifically the existence of voids, can lead to a mis-estimation
of $H_{0}$ on the order of 6.5\% for a given galaxy.

Finally, of note is the WiggleZ dark energy survey\cite{WiggleZ,WiggleZ design}.
WiggleZ is the most extensive redshift survey thus far conducted,
with some 280,000 galaxies with $0.2<z<1.0$ used as sources. WiggleZ
also covers a wider area of sky than previous surveys, examining some
1000 square degrees in multiple windows around the sky. Two of WiggleZ's
windows overlap with SDSS-II's survey area, so while WiggleZ is ongoing
preliminary results \cite{WiggleZ first release,WiggleZ test} can
be used to improve the evaluation of \emph{Q} by improving precision
on measurements of \emph{z }of SNe host galaxies. The authors of \cite{WiggleZ test}
note that {}``the redshift-space clustering pattern is not isotropic
in the true cosmological model'', attributing the variation to {}``the
coherent, bulk flows of galaxies toward clusters and superclusters''.
Analysis by the WiggleZ team of pre-existing SNe datasets, using the
new, more precise data on galaxy redshifts they obtained, reconfirms
the fact of acceleration, and generates results consistent with other
surveys, but the data lack sufficient precision to determine the history
of \emph{Q}.

\noun{Table} \ref{tab:Locations-of-observations} details the sky
locations of SNe and galaxies used in the determination of acceleration;
\noun{Figure} \ref{Flo:Fig1} presents these locations graphically.
\noun{Table} \ref{tab:Summary-of-results} in \noun{Appendix} summarizes
the results of these surveys.

\begin{table}
\begin{tabular}{|c|c|c|}
\hline 
Survey & No of SNe & \emph{z} \tabularnewline
\hline
\hline 
Supernova Cosmology Project 1997\cite{Perlmutter 1997} & 7 & $0.35<z<0.46$\tabularnewline
\hline 
High-\emph{z}~Supernova Search Team\cite{Riess 1998}  & 16 & $0.16<z<0.97$\tabularnewline
\hline 
Supernova Cosmology Project 1998\cite{Perlmutter 2} & 42 & $0.18<z<0.86$\tabularnewline
\hline 
HST\cite{HST} & 20 & $0.63<z<1.42$\tabularnewline
\hline 
ESSENCE\cite{ESSENCE} & 102 & $0.10\leq z\leq0.78$\tabularnewline
\hline 
Supernova Legacy Survey\cite{SNLS data,SNLS results} & 125 & $0.015<z<1$\tabularnewline
\hline 
ESSENCE + SNLS\cite{Essence 2} & 162 & $0.015<z<1$\tabularnewline
\hline 
Supernova Cosmology Project combined\cite{Kowalski} & 307 & $0.015<z<1$\tabularnewline
\hline 
Riess {}``gold'' sample\cite{Riess far supernovae,Riess Gold updated} & 16 & $1.25<z<2$\tabularnewline
\hline 
WiggleZ\cite{WiggleZ test} & 557 & $0.1<z<0.9$\tabularnewline
\hline
\end{tabular}

\lyxline{\normalsize}\begin{tabular}{|c|c|c|c|}
\hline 
Survey & $\Omega_{\Lambda}^{\dagger}$ & $\Omega_{M}$ & $\Omega_{K}$\tabularnewline
\hline
\hline 
Supernova Cosmology Project 1997 & $0.06_{-0.34}^{+0.28}$ & $0.94_{-0.28}^{+0.34}$ & (dne)\tabularnewline
\hline 
High-\emph{z}~Supernova Search Team & $0.72_{-0.48}^{+0.72}$ & $0.24_{-0.24}^{+0.56}$ & (dne)\tabularnewline
\hline 
Supernova Cosmology Project 1998 & $0.72_{-0.09}^{+0.08}$ & $0.28_{-0.08}^{+0.09}$ & (dne)\tabularnewline
\hline 
HST & $0.715_{-0.057}^{+0.036}$ & $0.286_{-0.023}^{+0.022}$ & $-0.001_{-0.013}^{+0.037}$\tabularnewline
\hline 
ESSENCE & $0.726_{-0.032}^{+0.020}$ & $0.274_{-0.020}^{+0.032}$ & (dne)\tabularnewline
\hline 
Supernova Legacy Survey & $0.751\pm0.080$ & $0.271\pm0.020$ & (dne)\tabularnewline
\hline 
ESSENCE + SNLS & $0.733_{-0.028}^{+0.018}$ & $0.267_{-0.018}^{+0.028}$ & (dne)\tabularnewline
\hline 
Supernova Cosmology Project combined & $0.785_{-0.045}^{+0.046}$ & $0.285_{-0.030}^{+0.030}$ & $-0.010_{-0.015}^{+0.016}$\tabularnewline
\hline 
Riess {}``gold'' sample & $0.71_{-0.05}^{+0.03}$ & $0.29_{-0.03}^{+0.05}$ & (dne)\tabularnewline
\hline 
WiggleZ & $0.71\pm0.03$ & $0.29\pm0.03$ & (dne)\tabularnewline
\hline 
Sloane Digital Sky Survey-II & $\begin{array}{c}
0.693\pm.042^{1}\\
0.744\pm.041^{2}\\
0.74\pm.02^{3}\end{array}$ & $\begin{array}{c}
0.307\pm.042^{1}\\
0.256\pm.041^{2}\\
0.25\pm.02^{3}\end{array}$ & $.04\pm.04$\tabularnewline
\hline
\end{tabular}

\lyxline{\normalsize}\begin{tabular}{|c|c|c|c|}
\hline 
Survey & $Q_{0}^{\mbox{flat}\ddagger}$ & $w_{X0}$ & $w_{Xa}$\tabularnewline
\hline
\hline 
Supernova Cosmology Project 1997 & $0.41_{-0.17}^{+0.14}$ & (dne) & (dne)\tabularnewline
\hline 
High-\emph{z}~Supernova Search Team & $-1.0\pm0.4$ & (dne) & (dne)\tabularnewline
\hline 
Supernova Cosmology Project 1998 & $-0.58_{-0.05}^{+0.04}$ & (dne) & (dne)\tabularnewline
\hline 
HST & $-0.572_{-0.025}^{+0.045}$ & $-0.997_{-0.293}^{+0.266}$ & $0.13_{-1.57}^{+1.16}$\tabularnewline
\hline 
ESSENCE & $-0.589_{-0.004}^{+0.022}$ & $-1.047_{-0.124}^{+0.125}$ & (dne)\tabularnewline
\hline 
Supernova Legacy Survey & $-0.620\pm0.060$ & $-1.023\pm0.087$ & (dne)\tabularnewline
\hline 
ESSENCE + SNLS & $-0.600_{-0.014}^{+0.009}$ & $-1.069_{-0.093}^{+0.091}$ & (dne)\tabularnewline
\hline 
Supernova Cosmology Project combined & $-0.642_{-0.031}^{+0.030}$ & $-1.001_{-0.155}^{+0.149}$ & (dne)\tabularnewline
\hline 
Riess {}``gold'' sample & $-0.56_{-0.02}^{+0.05}$ & $-1.02_{-0.19}^{+0.13}$ & {*}\tabularnewline
\hline 
WiggleZ & $-0.56\pm0.02$ & (dne) & (dne)\tabularnewline
\hline 
Sloane Digital Sky Survey-II & $\begin{array}{c}
-0.539\pm.021^{1}\\
-0.616\pm.021^{2}\\
-0.61\pm.01^{3}\end{array}$ & $\begin{array}{c}
-0.76\pm.18^{1}\\
-0.96\pm.18^{2}\end{array}$ & (dne)\tabularnewline
\hline
\end{tabular}\caption{Summary of results from surveys indicating acceleration}
\label{tab:Summary-of-results}\emph{{}``dne'' = {}``Does not evaluate''.
}{*}\emph{: \cite{Riess far supernovae} attempts to analyze $w_{a}$
with several different constraints but provides no numerical figure
for its estimate of $w_{a}$'s value. $\dagger$: Where not explicitly
stated in the source, $\Omega_{\Lambda}$ is evaluated from $\Omega_{M}+\Omega_{K}+\Omega_{\Lambda}=1$.
$\ddagger$: $Q_{0}^{\mbox{flat}}=\frac{1}{2}\Omega_{M}-\Omega_{\Lambda}$.
(1): MLCS2K2 evaluation. (2): SALT-II evaluation. (3): $\Lambda$CDM
model evaluation.}
\end{table}

\begin{figure}
\label{Flo:Fig1}

\includegraphics[width=1\columnwidth]{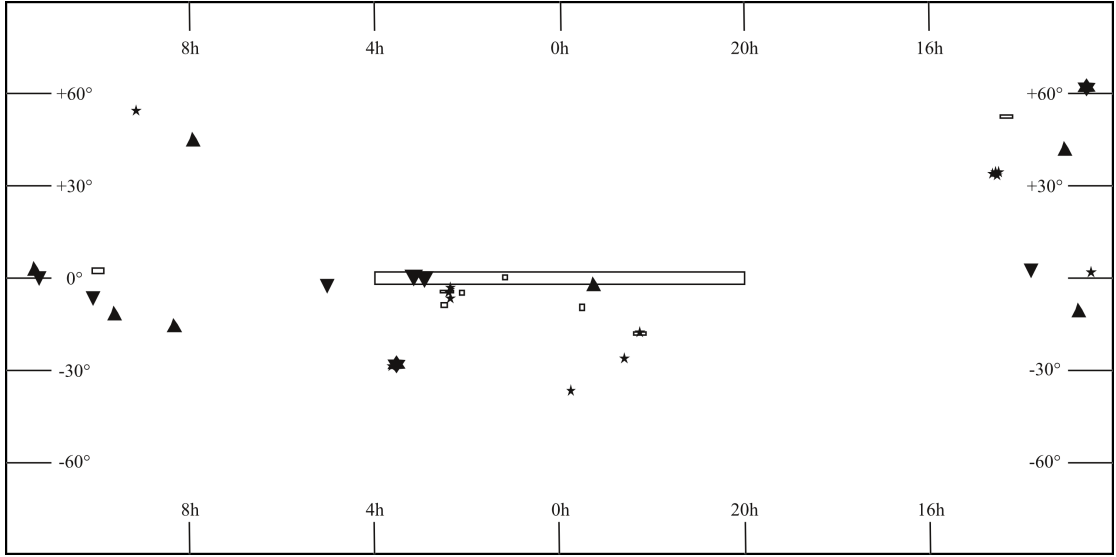}\caption{Sky positions of supernovae used as evidence for acceleration}

Surveys of cosmic acceleration cover a limited portion of the sky,
and data are divided into two contiguous, antipodal regions. Most
data has been collected in a small area of the sky near the equator.
\emph{Triangles: Riess 1998 supernovae. Five-pointed stars: HST SNe.
Six-pointed stars: Riess {}``gold'' dataset. The long, thin strip
centered on 0,0 is the SDSS-II survey area. Other boxes are the SNLS
and ESSENCE survey areas.}
\end{figure}

\section{Analysis\label{sub:Analysis}}

Analysis of supernova data is, in one sense, quite consistent: all
surveys apart from \cite{Perlmutter 1997} agree that for $z<1$ we
have a deceleration parameter $Q_{0}=-0.6$. Deeper analysis suffers
from a lack of data at high redshifts and large numbers of free parameters
in cosmological models, especially when more exotic models are considered.
Meanwhile, while most surveys indicate that the acceleration in recent
times acts as though driven by a cosmological constant, with an equation
of state compatible with $w_{X}=-1$, the results from \cite{SDSS K09}
show that this can be the result of the prior assumptions made about
the model of dark energy.

No definitive statement can be made about the evolution of $Q_{0}$
over time from the information thus far available, particularly not
statements connecting the state of cosmic acceleration now with the
state of acceleration at the epoch of last scattering.

Nor can any definitive statement be made about cosmological models,
other than to say that the most conservative, $\Lambda$CDM model
fits the data at best inconsistently. Few studies of supernova data
on acceleration examine the question of curvature in depth.

The majority of SNe data is collected from a single patch of sky:
the field bounded by RA 22:00, RA 04:00, Dec $+1^{\circ}15'$ and
Dec $-10^{\circ}00'$ (the {}``highly-observed field''). This area
comprises 1350 square degrees, or only 2.1\% of the sky. Surveys taken
in small fields outside the highly-observed field, such as the Riess
{}``gold'' dataset, have high internal consistency, while surveys
covering larger areas of sky have much lower consistency; the {}``gold''
dataset contains the same number of SNe as the \cite{Riess 1998}
but has a standard error less than a tenth the size. It is also telling
that the four SNLS fields produced results that correlated poorly
(37\% confidence) with one another\cite{SNLS results}, where two
of the SNLS survey regions are well outside the highly-observed field.
Compounding cosmographic bias, many of the remaining SNe observations
are located in a region of sky antipodal from the highly-observed
field; any vector or tensor contribution to cosmic dynamics will be
dominated by dipole and quadrupole terms, and as such be seen with
equal or opposite magnitude in the antipodal direction (that is, if
we observe a change in \emph{Q} of $\Delta Q$ along the $x^{i}$direction,
we should expect a change of $-\Delta Q$ in the event of a vector
contribution, or $\Delta Q$ in the event of a tensor contribution,
along the $-x^{i}$ direction).

There is, furthermore, no SNe data whatsoever from above Dec $+62^{\circ}$or
below Dec $-37^{\circ}$. The authors of \cite{WiggleZ test} note
a variation in the apparent Hubble parameter for galaxies in this
equatorial band (no WiggleZ region lies further north than Dec $+8^{\circ}$
or Dec $-19^{\circ}$); variation to the Hubble flow could potentially
be even greater outside this region. There is also no evaluation of
whether the Hubble flow remains isotropic beyond $z=0.3$ \cite{anisotropic hubble flow}.

Indeed, Zehavi \emph{et. al.} comment\cite{local Hubble Bubble} on
the lack of sky coverage in their analysis of local Hubble flows,
noting that {}``sparse sampling and the incomplete sky coverage (especially
at low Galactic latitudes) may introduce a bias in the peculiar monopole
due to its covariance with higher multipoles''. While the fact of
greater redshift in the range where acceleration can be measured should
overcome the peculiar velocities of galaxies, the data problem remains.

\section{Conclusions}

Many reasonable constraints prevent a full-sky survey of supernovae.
In the optical band, much of the sky is obscured by the {}``zone
of avoidance'' created by the plane of our own galaxy\cite{ZOA}.
The so-called {}``Great Attractor'', certain to be a region of particularly
high peculiar velocities and therefore great shifts in the apparent
Hubble parameter, lies in this zone\cite{Great Attractor}. Furthermore,
with only a single space-based optical observatory (the \emph{Hubble
Space Telescope}) operating, detailed observation of the sky is restricted
to those latitudes accessible by ground-based observatories, none
of which are located in Arctic latitudes. However, the directional
deficit of SNe surveys, aggregated together, cannot be ignored.

In the light of Tegmark \emph{et. al.}'s discovery\cite{Tegmark CMB map}
of a preferred axis to the CMB quadrupole, and Land \& Maguiejo's
subsequent observation\cite{Axis of Evil} of a preferred axis in
higher multipole moments aligned with the the quadrupole (the so-called
{}``Axis of Evil''), the default assumption should be that anisotropic
acceleration is not ruled out. Indeed, the prominent CMB {}``Cold
Spot''\cite{Cold Spot II} falls within the highly-observed field,
although no surveys or SNe are located exactly in its direction.

As such, Wagoner's assertion of the cosmological principle as {}``statistically
valid''\cite{Wagoner} has been misapplied by analysts of acceleration
data. A tensorial theory of cosmic acceleration would preserve homogeneity,
in the sense that every observer sees {}``the same version of cosmic
history''\cite{Sung & Coles}, at the expense of isotropy in the
form of spherical symmetry. 

More fundamentally, all studies of cosmic acceleration to date operate
on the assumption that acceleration is isotropic, that is, that the
acceleration field is equal in every direction, and therefore must
be explained either by a cosmological constant or a scalar field.
As Mörtsell and Clarkson note, {}``{[}a{]}t best this gives a small
error to all our considerations; at worst, many of our conclusions
might be wrong''\cite{Hubble bubble}. In particular, the data as
presented cannot distinguish between a scalar-field theory of acceleration,
a vector-field theory of acceleration, a cosmological constant theory
of acceleration, and a time-dependent tensor-field theory of acceleration.

Meanwhile, the simplest theory of acceleration, a cosmological constant,
is challenged on two fronts: not only is $\Omega_{\Lambda}$'s value
far out of line with that predicted by theory\cite{Carroll}, but
while its equation of state is close to $w_{X}=-1$ measurements have
tended to favor a value slightly smaller than -1.

It is interesting to note that when $\Omega_{K}$ is evaluated, supernova
data favor a closed universe (although always in a manner compatible
with a flat universe); this conclusion is consistent with the curvature
parameter evaluated by WMAP\cite{WMAP Komatsu}.

\subsection{Recommendations}

In light of these weaknesses of the current information on cosmic
acceleration, the following program is recommended:

The data already in hand should be re-evaluated to look for signs
of angular dependence in the Hubble parameter. In particular, the
SNLS fields 2 and 3 and the {}``gold'' sample fields should be examined
in and of themselves in order to build up a map of \emph{H}~as a
function of both \emph{z} and direction in the sky. The rejection
of certain SNe in \cite{Kowalski} should be re-evaluated in light
of possible inadvertent obscuring of evidence for angular dependence
in \emph{H}.

Analyses of SNe data should always consider the possibility of a closed
or open universe as well as a flat one.

Additional SNe surveys for redshifts $.15<z<2$ should be carried
out in unexamined areas of sky not obscured by the plane of the galaxy,
such as for example the celestial north and south poles. The optimal
region for these surveys is in rings located $90^{\circ}$ from the
center of the highly-observed field, which will maximize the difference
in the event of a tensor-field (that is, gravitational-wave) acceleration.

In light of this need and the lack of ground-based observatories,
as well as the Zone of Avoidance, priority should be given to the
Wide Field Infrared Survey Telescope ({}``WFIRST'') project\cite{WFIRST},
which incorporates the Super Nova/Acceleration Probe\cite{SNAP,SNAP paper}
and Joint Dark Energy Mission\cite{JDEM,JDEM-Omega}. This telescope
is currently scheduled to be launched in 2016.

As WiggleZ continues, its data on galactic redshifts should be examined
for angular dependence as well. The completion of WiggleZ will provide
invaluable information on baryon acoustic oscillations which will
make possible the charting of the history of \emph{H }and \emph{Q}
at much higher redshifts than is possible through the examination
of supernova data.

Zhao \emph{et al. }have also noted the possibility of using the Einstein
telescope as an instrument for examining dark energy through the use
of gravitational wave emissions from colliding binary objects as a
{}``standard siren'' analogous to the standard candle of type Ia
SNe\cite{Zhao}. The largest binary systems, in the period prior to
merger, could be parameterized by radio telescopy\cite{Schluessel}.

Cooray and Caldwell\cite{Cooray}, implicitly identifying the same
problem of lack of angular coverage as we note herein, propose a program
of near-redshift surveys covering a large but practical area of sky
which could also provide the relevant information with existing facilities.

Overall, the need is underscored for new theories of acceleration,
particularly ones that attempt to explain acceleration through the
action of tensor perturbations in a closed universe. Wagoner's formula
(\noun{equation} \ref{eq:Wagoner redshift}) and its generalizations
must be generalized further, to take into account the possibility
of anisotropic fields as the cause of anisotropic cosmic acceleration.

\part{The Bianchi IX cosmology}

In pursuit of a theory within the context of unmodified general relativity
which can explain cosmic acceleration while remaining compatible with
the cosmic microwave background, we wish to relax as few constraints
on our cosmological model as necessary. Therefore while having sacrificed
the requirement of isotropy in the sense of spherical symmetry in
the dark energy field, we wish to retain a stronger\cite[ss. 116]{Landau-Lifschitz}
condition of the Copernican principle on our space, that of homogeneity\cite[Chap. 13 sec. 1]{Weinberg}.
It is also desirable to have a model whose limiting case is a Friedmann
cosmology, in order to explain the almost-isotropic (that is, almost-Friedmannian)
character of the CMB. Furthermore, a model which is spatially closed,
in order to match models favored by CMB and SNe data, is desirable;
such a model would, if complying with all other conditions, have a
flat universe as a limiting case in the limit of an infinitely large
radius of curvature.

Bianchi showed\cite{Bianchi} that there exists exactly one homogeneous%
\footnote{A homogeneous space is a space such that for any two points in the
space, there exists a geodesic, not necessarily of finite length,
connecting those two points.%
} space with a closed Friedmann universe as a limiting case: the Bianchi
type IX cosmology, for short {}``Bianchi IX''.

\section{The Bianchi classification scheme}

Bianchi observed that all three-dimensional homogeneous spaces could
be classified into nine types, based on categorization of the symmetries,
that is the Killing field, in each space. Behr noted\cite{Behr} that
this categorization scheme could be simplified to filling a parameter
space of four parameters: one running over the real numbers and three
reducible to the sign function $\mbox{sgn}\left(x\right)$.

Consider some space with metric $ds^{2}=dt^{2}-g_{ij}dx^{i}dx^{j}$
(that is, a space in Gaussian coordinates) where $g_{ij}=g_{ij}\left(t,x^{i}\right)$.
If the sub-space with metric tensor $g_{ij}$ is homogeneous, then
there exists a set of vectors that solve $\xi_{i;j}+\xi_{j;i}=0$;
these are the Killing vectors of the space\cite{Weinberg}. In a homogeneous
space, these vectors will (where $\left[\,,\,\right]$ is a commutator)
obey the commutation relationship

\begin{equation}
\left[\xi_{i},\xi_{j}\right]\equiv\xi_{i}\xi_{j}-\xi_{j}\xi_{i}=C_{\, ij}^{k}\xi_{k}\end{equation}
 where in an homogeneous space, the object $C_{\, ij}^{k}$ is a constant
pseudo-tensor, the {}``structure constants'' of an homogeneous space,
with the antisymmetry property $C_{\,\left[ij\right]}^{k}=C_{\, ij}^{k}$
\cite[ss. 116]{Landau-Lifschitz}. 

We always have the freedom to perform separation of variables the
functions $g_{ij}$; let us do so by defining the matrix $\gamma_{ab}$
such that 

\begin{equation}
g_{ij}\left(t,x^{k}\right)=-\gamma_{ab}\left(t\right)e_{i}^{\left(a\right)}\left(x^{k}\right)e_{j}^{\left(b\right)}\left(x^{k}\right)\label{eq:decomposition}\end{equation}

\noindent .%
\footnote{Indices from the beginning of the Latin alphabet (\emph{a, b, c,...})
denote triad indices; indices from the middle of the alphabet (\emph{i,
j, k,...}) denote regular indices. Where the two are mixed or the
application is otherwise ambiguous, triad indices are enclosed in
parentheses; in this work, this notation never means the tensor symmetrization
operation.%
} The $3\times3$ matrix $e_{i}^{\left(a\right)}\left(x^{k}\right)$
is a triad\cite[ss. 98]{Landau-Lifschitz,LL Russian}%
\footnote{The widespread \emph{Fourth Revised English Edition}~of \cite{Landau-Lifschitz}
contains numerous serious typographical errors in the section introducing
the tetrad formalism. The Russian-language \emph{Seventh Corrected
Edition}\cite{LL Russian} contains the correct formulas.%
} of vectors ({}``frame vectors'') which solve

\begin{equation}
\xi_{k}\left(e_{i}^{\left(a\right)}e_{j}^{\left(b\right)}dx^{i}dx^{j}\right)=0\end{equation}

\noindent (in the language of linear algebra, the quantity $e_{i}^{a}dx^{i}$
is a one-form on a homogeneous space). 

Furthermore, define the matrix $e_{\left(a\right)}^{i}$ such that
$e_{\left(a\right)}^{i}=e_{j}^{\left(a\right)}=\delta_{j}^{i}$; from
this it follows that $e_{\left(a\right)}^{i}e_{i}^{\left(b\right)}=\delta_{\left(a\right)}^{\left(b\right)}$.
From these relationships we can transform between any tensor and its
decomposition into triads by saying that for some tensor $A_{j_{1}j_{2}j_{3}\ldots j_{n}}^{i_{1}i_{2}i_{3}\ldots i_{m}}$,

\begin{equation}
A_{j_{1}j_{2}j_{3}\ldots j_{n}}^{i_{1}i_{2}i_{3}\ldots i_{m}}=A_{\left(b\right)_{1}\left(b\right)_{2}\left(b\right)_{3}\ldots\left(b\right)_{n}}^{\left(a\right)_{1}\left(a\right)_{2}\left(a\right)_{3}\ldots\left(a\right)_{m}}\left(e_{\left(a\right)_{1}}^{i_{1}}e_{\left(a\right)_{2}}^{i_{2}}e_{\left(a\right)_{3}}^{i_{3}}\ldots e_{\left(a\right)_{m}}^{i_{m}}\right)\left(e_{j_{1}}^{\left(b\right)_{1}}e_{j_{2}}^{\left(b\right)_{2}}e_{j_{3}}^{\left(b\right)_{3}}\ldots e_{j_{n}}^{\left(b\right)_{n}}\right)\end{equation}

\noindent ; therefore in an homogeneous space we can perform separation
of variables on the partial differential equations of general relativity
and solve the time-dependent parts as ordinary differential equations.

The frame vectors obey the properties

\begin{equation}
e_{i,j}^{\left(a\right)}-e_{j,i}^{\left(a\right)}=C_{\, bc}^{a}e_{i}^{\left(b\right)}e_{j}^{\left(c\right)}\end{equation}

\noindent \cite{Grishchuk curvature}. The structure constants $C_{\, bc}^{a}$
typify a homogeneous space and are given by the following rule\cite{Behr}%
\footnote{The symbol $\varepsilon_{abc}$ represents the Levi-Civita symbol
defined such that $\varepsilon_{123}=1$%
}:

\begin{equation}
C_{\, bc}^{a}=\varepsilon_{bcd}n^{ad}+\delta_{c}^{d}a_{b}-\delta_{b}^{d}a_{c}\end{equation}

\noindent where the object $n^{ab}$ is a diagonal matrix $\mbox{diag}\left(n^{\left(1\right)},n^{\left(2\right)},n^{\left(3\right)}\right)$
and $a_{a}$ is the vector $\left(a,0,0\right)$, the values of this
matrix and vector typifyied by the underlying cosmology (\noun{Table}
\ref{tab:Constants-for-the}).

\begin{table}
\begin{centering}
\label{Wra:Bianchi table}\begin{tabular}{|c|c|c|c|c|}
\hline 
Bianchi type & \emph{a} & \emph{$n^{\left(1\right)}$} & \emph{$n^{\left(2\right)}$} & \emph{$n^{\left(3\right)}$}\tabularnewline
\hline
\hline 
I & 0 & 0 & 0 & 0\tabularnewline
\hline 
II & 0 & 1 & 0 & 0\tabularnewline
\hline 
III & 1 & 0 & 1 & -1\tabularnewline
\hline 
IV & 1 & 0 & 0 & 1\tabularnewline
\hline 
V & 1 & 0 & 0 & 0\tabularnewline
\hline 
$\mbox{VI}_{0}$ & 0 & 0 & 1 & -1\tabularnewline
\hline 
$\mbox{VI}_{a}$ & \emph{a} & 0 & 1 & -1\tabularnewline
\hline 
$\mbox{VII}_{0}$ & 0 & 1 & 1 & 0\tabularnewline
\hline 
$\mbox{VII}_{a}$ & \emph{a} & 1 & 1 & 0\tabularnewline
\hline 
VIII & 0 & 1 & 1 & -1\tabularnewline
\hline 
IX & 0 & 1 & 1 & 1\tabularnewline
\hline
\end{tabular}
\par\end{centering}

\caption{The Bianchi classification scheme}
\label{tab:Constants-for-the}Constants for the different homogeneous
spaces of the Bianchi classification scheme\cite{Landau-Lifschitz,Sung & Coles,Behr,Ellis MacCallum}.
The quantity \emph{a}~runs over the real numbers. This parametrization
is not unique (we could, for example, have chosen $\left(-1,-1,-1\right)$
for $\left(n^{\left(1\right)},n^{\left(2\right)},n^{\left(3\right)}\right)$
in the type IX space).
\end{table}

The cosmologies of Bianchi types I, V, $\mbox{VII}_{0}$, $\mbox{VII}_{a}$
and IX are of particular interest as they have isotropic spaces as
limiting cases; specifically, a universe with metric

\begin{equation}
ds^{2}=dt^{2}-a^{2}\eta_{ab}e_{i}^{\left(a\right)}e_{j}^{\left(b\right)}\end{equation}

\noindent is a flat $K=0$ universe for Bianchi type I, an open $K=-1$
universe for Bianchi types V, $\mbox{VII}_{0}$, $\mbox{VII}_{a}$,
and a closed $K=1$ universe for Bianchi type IX\cite{Landau-Lifschitz,GDI,Sung & Coles}.
Bianchi IX is the only homogeneous closed cosmological model in the
context of general relativity\cite{Ellis MacCallum}.

\section{The Kasner universe}

In order to illustrate the possible effects of an anisotropic but
homogeneous cosmology on cosmic dynamics, we will consider a Bianchi
type I cosmology that generalizes the Friedmann cosmology: the Kasner
universe\cite{MTW}; \cite[ss. 117]{Landau-Lifschitz}.

Let our metric read

\begin{equation}
ds^{2}=dt^{2}-t^{2p_{1}}\left(dx^{1}\right)^{2}-t^{2p_{2}}\left(dx^{2}\right)^{2}-t^{2p_{3}}\left(dx^{3}\right)^{2}\end{equation}

\noindent where $p_{1},p_{2},p_{3}$ are constants. In a co-moving
coordinate system we quickly arrive at the following set of Einstein
equations:

\begin{align}
\left[\left(p_{1}+p_{2}+p_{3}\right)-\left(p_{1}^{2}+p_{2}^{2}+p_{3}^{2}\right)\right]t^{-2}= & \frac{1}{2}k\left(\epsilon+3p\right)\\
\left(p_{1}+p_{2}+p_{3}-1\right)p_{1}t^{-2}= & \frac{1}{2}k\left(p-\epsilon\right)\\
\left(p_{1}+p_{2}+p_{3}-1\right)p_{2}t^{-2}= & \frac{1}{2}k\left(p-\epsilon\right)\\
\left(p_{1}+p_{2}+p_{3}-1\right)p_{3}t^{-2}= & \frac{1}{2}k\left(p-\epsilon\right)\end{align}

\noindent . These equations necessitate either an isotropic but unphysical
($p=\epsilon$) universe or a vacuum ($\epsilon=p=0$) universe, in
which we have either the trivial solution $p_{1}=p_{2}=p_{3}=0$ (Minkowski
space) or the more interesting solution

\begin{equation}
p_{1}+p_{2}+p_{3}=p_{1}^{2}+p_{2}^{2}+p_{3}^{2}=1\end{equation}

\noindent . This solution admits a parametrization of $p_{1},p_{2},p_{3}$
such that (if we choose $p_{1}\leq p_{2}\leq p_{3}$)

\begin{align}
p_{1}= & -u/\left(1+u+u^{2}\right)\nonumber \\
p_{2}= & \left(1+u\right)/\left(1+u+u^{2}\right)\\
p_{3}= & u\left(1+u\right)/\left(1+u+u^{2}\right)\nonumber \end{align}

\noindent where $u>0$; these relations have the nice symmetry property
that $p_{i}\left(u\right)=p_{i}\left(1/u\right)$.

\subsection{Scale factor\label{sub:Scale-factor}}

The scale factor \emph{a} does not necessarily have an intrinsic meaning,
but instead compares distances as a function of time. In an isotropic
cosmology such as the Friedmann model \emph{a}~can be given a real
geometric meaning; in an open or closed Friedmann universe, the scale
factor appears simply in the Ricci curvature of space $R_{j}^{i}=\left(2K/a^{2}\right)\delta_{j}^{i}$
and as such can be regarded as the radius of curvature of the universe.
In particular, in a closed isotropic universe \emph{a} can be considered
to have the direct physical meaning of the radius of curvature of
the spherical space, so in a closed universe one could meaningfully
say {}``the radius of the universe is \emph{a}''.

When space is no longer isotropic, the definition of scale factor
breaks down. It is, of course, possible to define any positive function
as {}``the'' scale factor. Grishchuk \emph{et al.}\cite[section 4]{GDI},
for example, use a metric

\begin{align}
\gamma_{11} & = & \frac{1}{4}a^{2}e^{2\alpha}\nonumber \\
\gamma_{22} & = & \frac{1}{4}a^{2}e^{2\beta}\label{eq:GDI metric}\\
\gamma_{33} & = & \frac{1}{4}a^{2}e^{2\gamma}\nonumber \end{align}

\noindent and propose the definition

\begin{equation}
a^{2}\equiv\frac{1}{12}\gamma_{ab}\eta^{ab}\end{equation}

\noindent in the context of a vacuum cosmology, motivated by the coincidence
of this definition of the scale factor with one the authors introduce
in separating the Bianchi IX metric into background and gravitational-wave
parts. The authors also discuss a definition of scale factor such
that

\begin{equation}
a^{2}\equiv\left(\det\gamma_{ab}\right)^{1/3}\end{equation}

\noindent . This definition has the advantage that it relates the
scale factor to a definite physical quantity, a volume element, but
it contains a deeper flaw: with such a definition in place the Einstein
equations admit no solution other than the background solution at
quadratic and higher orders. If we define the quantity

\begin{equation}
\delta\equiv\alpha+\beta+\gamma\label{eq:Delta definition}\end{equation}

\noindent then

\begin{equation}
a^{2}\equiv\left(\det\gamma_{ab}\right)^{1/3}\implies e^{\delta}=1\implies\delta=0\end{equation}

\noindent . In either case, though, discussion of definitions of \emph{a}
attempt to solve a problem that does not exist. The question of what
definition of scale factor to select is analogous to the question
of which of the orthocenter, incenter or circumcenter of a triangle
is the {}``true'' center. Consequently, attempting to extract a
single scale factor -- and thus a single Hubble parameter or a single
deceleration parameter -- from anisotropic Einstein equations is a
fool's errand.

We can, if we wish, split the metric (\ref{eq:GDI metric}) into isotropic
and anisotropic parts by noting that the quantity $a_{F}e^{\delta}$
is isotropic and that any two of the quantities $\alpha-\beta$, $\alpha-\gamma$
and $\beta-\gamma$ combined with $a_{F}e^{\delta}$ contain all the
information needed to describe the metric\cite{Barrow}; pursuing
this route would be a distraction from our main task.

Instead, let the notion of scale factor \emph{a}, Hubble parameter
\emph{H} and deceleration parameter \emph{Q} be generalized. In a
homogeneous cosmology with a diagonal metric, define the following
matrices: the generalized scale factor,

\begin{equation}
a_{ab}\equiv\left(\begin{array}{ccc}
\left(\gamma_{11}\right)^{1/2} & 0 & 0\\
0 & \left(\gamma_{22}\right)^{1/2} & 0\\
0 & 0 & \left(\gamma_{33}\right)^{1/2}\end{array}\right)\label{eq:generalized scale factor}\end{equation}

\noindent (recalling that non-integer powers of a matrix are not defined,
so we could not simply say $a_{ab}\equiv\left(\gamma_{ab}\right)^{1/2}$).
In a Bianchi I cosmology only, from this definition we can then define
the redshift matrix (in homogeneous cosmologies other than Bianchi
I the geodesic equations are non-linear; see \noun{Part} \ref{sec:Impact-on-the}):

\begin{equation}
z_{a}^{b}\equiv a_{ac}\left(\eta_{R}\right)a^{bc}\left(\eta_{E}\right)-\delta_{a}^{b}=\left(\begin{array}{ccc}
\frac{a_{11}\left(t_{R}\right)}{a_{11}\left(t_{E}\right)}-1 & 0 & 0\\
0 & \frac{a_{22}\left(t_{R}\right)}{a_{22}\left(t_{E}\right)}-1 & 0\\
0 & 0 & \frac{a_{33}\left(t_{R}\right)}{a_{33}\left(t_{E}\right)}-1\end{array}\right)\end{equation}

\noindent where the subscript \emph{R} denotes the function evaluated
at the time of observation of light, and \emph{E} denotes the function
evaluated at the time of emission; and finally the generalized Hubble
parameter and deceleration parameter:

\begin{equation}
H_{ab}\equiv\frac{1}{2}\frac{d}{dt}\ln\gamma_{ab}=\left(\begin{array}{ccc}
\dot{a}_{11}/a_{11} & 0 & 0\\
0 & \dot{a}_{22}/a_{22} & 0\\
0 & 0 & \dot{a}_{33}/a_{33}\end{array}\right)\label{eq:Generalized Hubble}\end{equation}

\begin{equation}
Q_{a}^{b}\equiv\frac{d}{dt}H^{ac}\eta_{bc}-\delta_{b}^{a}=-\left(\begin{array}{ccc}
\ddot{a}_{11}a_{11}/\left(\dot{a}_{11}\right)^{2} & 0 & 0\\
0 & \ddot{a}_{22}a_{22}/\left(\dot{a}_{22}\right)^{2} & 0\\
0 & 0 & \ddot{a}_{33}a_{33}/\left(\dot{a}_{33}\right)^{2}\end{array}\right)\end{equation}

\noindent . This approach is essentially a generalization of that
developed by Barrow in \cite{Barrow}; the object (\ref{eq:Generalized Hubble})
is closely related to the shear tensor\cite{Sung & Coles,Anile Motta S-W}.
The practical purpose of these definitions is to provide a mathematical
description of observed quantities; let $\mathbf{e}^{i}$ be a unit
vector pointing in the direction of observation. Then the redshift
observed in the $\mathbf{e}^{i}$ direction is given by

\begin{equation}
z\left(e^{i},t\right)=z_{b}^{a}e_{\left(b\right)}^{i}e_{j}^{\left(a\right)}\mathbf{e}^{j}\mathbf{e}_{i}\end{equation}

\noindent and similarly for other functions of the scale factor. Each
of these functions can be averaged over the whole sky, these averages
denoted by a bar:

\begin{equation}
\bar{a}\equiv\frac{\int a_{ab}e_{i}^{\left(b\right)}e_{j}^{\left(a\right)}\mathbf{e}^{i}\mathbf{e}^{j}dS}{\int\eta_{ij}\mathbf{e}^{i}\mathbf{e}^{j}dS}=\frac{1}{3}a_{ab}\eta^{ab}=\frac{1}{3}\left(a_{11}+a_{22}+a_{33}\right)\label{eq:averaged scale factor}\end{equation}

\noindent \emph{etc.}

\subsection{Dynamics in the Kasner universe\label{sub:Dynamics-in-the}}

An observer in a Kasner universe will see the consequences of that
universe's evolution. Examination of the observational consequences
of the Kasner universe provides an illustrative example of potential
consequences of anisotropy in other cosmologies.

\subsubsection{Expansion}

Misner, Thorne \& Wheeler argue\cite{MTW} that the Kasner universe
is expanding, as the volume element is always increasing:

\begin{equation}
\frac{dV}{dt}=\frac{d}{dt}\sqrt{\left\Vert g_{ij}\right\Vert }dx^{1}dx^{2}dx^{3}=\frac{d}{dt}\left(t^{p_{1}+p_{2}+p_{3}}\right)dx^{1}dx^{2}dx^{3}=dx^{1}dx^{2}dx^{3}\end{equation}

\noindent . However, as noted above there is no unique way to define
the scale factor. In terms of the averaged quantity defined in (\ref{eq:averaged scale factor})
we have

\begin{equation}
\bar{a}=\frac{1}{3}\left(t^{p_{1}}+t^{p_{2}}+t^{p_{3}}\right)\end{equation}

\noindent which, when we expand around \emph{$t=1$}, is approximately

\begin{equation}
\bar{a}\left(t\approx1\right)\approx\frac{1}{3}\left(2+t\right)+\mathcal{O}\left(t^{3}\right)\end{equation}

\noindent . But in the limit of \emph{t} small, we have

\begin{equation}
\bar{a}\approx\frac{1}{3}t^{p_{1}}\end{equation}

\noindent , which is clearly a decreasing function; so the Kasner
universe is not unambiguously expanding.

\subsubsection{Redshift}

Redshift in a Kasner universe is given by

\begin{equation}
z_{j}^{i}=\left(\begin{array}{ccc}
\left(t_{R}/t_{E}\right)^{p_{1}}-1 & 0 & 0\\
0 & \left(t_{R}/t_{E}\right)^{p_{2}}-1 & 0\\
0 & 0 & \left(t_{R}/t_{E}\right)^{p_{3}}-1\end{array}\right)\end{equation}

\begin{equation}
\bar{z}=\frac{1}{3}\left[\left(\frac{t_{R}}{t_{E}}-1\right)^{p_{1}}+\left(\frac{t_{R}}{t_{E}}-1\right)^{p_{2}}+\left(\frac{t_{R}}{t_{E}}-1\right)^{p_{3}}\right]\end{equation}

\noindent . In the circumstance when $t_{R}\gg t_{E}$,

\begin{equation}
\bar{z}\approx\frac{1}{3}\left(\frac{t_{R}}{t_{E}}\right)^{p_{3}}\end{equation}

\noindent . Of particular interest is the quantity $\Delta T/T_{R}$,
the variation in CMB temperature from the average (accepting for the
moment that the vacuum Kasner universe approximates a matter-filled
one at a sufficiently young age), which is given approximately by

\begin{align}
\frac{\Delta T}{T_{R}}\approx & \left[3\left(\frac{t_{R}}{t_{E}}\right)^{-p_{3}}\left(\begin{array}{ccc}
\left(t_{R}/t_{E}\right)^{p_{1}} & 0 & 0\\
0 & \left(t_{R}/t_{E}\right)^{p_{2}} & 0\\
0 & 0 & \left(t_{R}/t_{E}\right)^{p_{3}}\end{array}\right)-\eta_{ab}\right]\mathbf{e}^{i}\mathbf{e}^{j}=\nonumber \\
= & \left(\begin{array}{ccc}
3\left(t_{R}/t_{E}\right)^{p_{1}-p_{3}}-1 & 0 & 0\\
0 & 3\left(t_{R}/t_{E}\right)^{p_{2}-p_{3}}-1 & 0\\
0 & 0 & 2\end{array}\right)\mathbf{e}^{i}\mathbf{e}^{j}\approx\label{eq:Kasner temp anisotropy}\\
\approx & \left(\begin{array}{ccc}
-1 & 0 & 0\\
0 & -1 & 0\\
0 & 0 & 2\end{array}\right)\mathbf{e}^{i}\mathbf{e}^{j}\nonumber \end{align}

\noindent (except in the case when $p_{2}=p_{3}=2/3$, in which event
the (2,2) entry in (\ref{eq:Kasner temp anisotropy}) will read 2).
The CMB in a mature Kasner universe has a pronounced anisotropy, with
the observed temperature matching the average temperature only in
a circle around the axis of anisotropy. Notably, the primary axis
of the anisotropy is at a right angle to the axis along which the
Kasner universe is contracting -- not on a parallel axis!

\subsubsection{Hubble flow \& deceleration parameter}

The Kasner universe has Hubble flow

\begin{equation}
H_{ab}=\frac{1}{t}\left(\begin{array}{ccc}
p_{1} & 0 & 0\\
0 & p_{2} & 0\\
0 & 0 & p_{3}\end{array}\right)\end{equation}

\begin{equation}
\bar{H}=\frac{1}{3}t^{-1}\end{equation}

\noindent and deceleration parameter

\begin{equation}
Q_{a}^{b}=\left(\begin{array}{ccc}
\left(1-p_{1}\right)/p_{1} & 0 & 0\\
0 & \left(1-p_{2}\right)/p_{2} & 0\\
0 & 0 & \left(1-p_{3}\right)/p_{3}\end{array}\right)\end{equation}

\begin{equation}
\bar{Q}=-1\end{equation}

\noindent . In the limit that the parameter $u\rightarrow\infty$
an observer in a Kasner universe would see a universe with a positive
Hubble flow (redshift) over most of the sky, but see blueshift in
a third direction. An observer looking only at averages, though, would
not be able to distinguish between an isotropic universe and a Kasner
universe merely by examining the Hubble flow; only with a complete
picture of the sky is such a test possible. The Hubble flow in the
case of minimal anisotropy has the form

\begin{equation}
H_{ab}\left(u=1\right)=\frac{1}{t}\left(\begin{array}{ccc}
-1/3 & 0 & 0\\
0 & 2/3 & 0\\
0 & 0 & 2/3\end{array}\right)\end{equation}

\noindent -- appearing like a Friedmannian matter-dominated universe
in two directions -- and in the case of maximal isotropy

\begin{equation}
\lim_{u\rightarrow0}H_{ab}=\frac{1}{t}\left(\begin{array}{ccc}
0 & 0 & 0\\
0 & 0 & 0\\
0 & 0 & 1\end{array}\right)\end{equation}

\noindent Similarly, an observer looking only at the averaged deceleration
parameter sees a universe accelerating as though driven by a cosmological
constant; only with good enough information will the observer notice
a strong angular dependence in the acceleration field, which in the
case of minimal anisotropy has the form

\begin{equation}
Q_{a}^{b}\left(u=1\right)=\left(\begin{array}{ccc}
-4 & 0 & 0\\
0 & 1/2 & 0\\
0 & 0 & 1/2\end{array}\right)\end{equation}

\noindent -- decelerating like a Friedmann cosmology in two directions
-- and in the case of maximal anisotropy has the form

\begin{equation}
\lim_{u\rightarrow\infty}Q_{a}^{b}=\left(\begin{array}{ccc}
-\infty & 0 & 0\\
0 & \infty & 0\\
0 & 0 & 1\end{array}\right)\end{equation}

\noindent . Moreover, even though acceleration along two axes is negative
in the least-anisotropic Kasner universe, the impact of the positive-accelerating
direction is such that the isepitach%
\footnote{A neologism denoting a path of constant acceleration, similar to {}``isobar''
or {}``isochor'', from Greek\emph{ {}``epitachounse''}, acceleration.%
} of zero acceleration, the boundary an observer sees on the sky between
regions where objects accelerate and objects decelerate, is a circle
$83^{\circ}$ from the axis of acceleration; only less than 8\% of
the sky appears close to {}``normal'' to an observer expecting to
record a Friedmann universe!

While the vacuum Kasner universe is ruled out as a possible cosmology
both for reasons of the CMB, which appears isotropic to a high degree\cite{WMAP 7-year},
and due to the Hubble flow, which appears isotropic to the limit of
the peculiar motions of galaxies below $z=0.3$ \cite{anisotropic hubble flow},
the surprising difficulties in distinguishing between its dynamics
and that of a Friedmann universe serve as a reminder that sampling
of cosmological parameters must be done in an unbiased fashion and
that isotropy must be tested rather than assumed. The Kasner universe
also has an application as a limiting case of the BKL universe\cite{BKL singularity},
to which it appears identical for observers looking over a period
of time that is small compared to the radius of curvature of the universe.
Finally, the anisotropic Kasner universe serves as a limiting case
to some types of cosmology described by the Bianchi IX model.

\section{Gravitational wave nature of Bianchi IX\label{sub:Gravitational-wave-nature}}

The Bianchi IX has been considered by cosmologists repeatedly since
the establishment of general relativity to provide possible explanations
for cosmological phenomena.

Belinsky, Khalatnikov and Lifshitz discussed\cite{BKL singularity}
a Bianchi IX cosmology (the {}``BKL cosmology'') which undergoes
several {}``bounces'' as it evolves -- rather than expanding from
or converging to a point, it contracts along one axis while expanding
along two others until the smallest metric component reaches a minimum
value, at which point the axes swap roles. Misner\cite{Mixmaster}
discussed a related form of Bianchi IX universe as the {}``mixmaster
universe'', pursuing an resolution to the horizon problem through
the non-linearity of the Bianchi IX cosmology; through the mechanism
of bounces, all parts of the universe are brought into causal connection.
Bouncing vacuum cosmologies are, like the vacuum Kasner universe,
intrinsically highly anisotropic; while in the long run they tend
to act isotropically due to the back-reaction of matter\cite{BKL singularity 2,MTW}
they will still exhibit strong CMB anisotropy\cite{DLN anisotropy}.
Supernova data (\cite{Riess 1998,Perlmutter} \emph{etc.}) and CMB
data on the value of $\Omega_{M}$ (\cite{WMAP first-year results}
\emph{etc.}) coupled with the existence of high-redshift objects\cite{Carroll}
rule out bouncing cosmologies to a high degree of confidence.

The BKL cosmology undergoes anisotropic acceleration (see \noun{section}
\ref{sub:Dynamics-in-the}). Meanwhile, numerical modeling has suggested\cite{Hobill Guo}
that a matter-filled Bianchi IX universe will also undergo periods
of acceleration. Therefore, we have good reason to suppose that a
property of Bianchi IX may be to generate anisotropic acceleration,
and that consequences of the Bianchi IX cosmology may reveal a dark
energy candidate with none of the failings of scalar or exotic models.

Wheeler showed\cite{Wheeler} that an almost-isotropic Bianchi IX
universe admitted a weak tensorial perturbation that took the form
of a wave (that is, solving an equation of the form $\ddot{f}+nf\left(t\right)=g\left(t\right)$).
Grishchuk \emph{et. al.}~ were able to generalize this result\cite{GDI}:

The Bianchi IX space has frame vectors

\begin{align}
e_{i}^{1}= & \left(\cos x^{3},\right. & \sin x^{1}\sin x^{3}, & \left.0\right)\nonumber \\
e_{i}^{2}= & \left(-\sin x^{3},\right. & \sin x^{1}\cos x^{3}, & \left.0\right)\label{eq:frame vectors}\\
e_{i}^{3}= & \left(0,\right. & \cos x^{1}, & \left.1\right)\nonumber \end{align}

\noindent . Consider the metric of a Bianchi IX cosmology:

\begin{equation}
ds^{2}=dt^{2}-\gamma_{ab}e_{i}^{a}e_{j}^{b}dx^{i}dx^{j}\end{equation}

\noindent . We can split this metric up into an isotropic (Friedmannian)
part and a non-Friedmannian part:

\begin{align}
ds^{2}= & dt^{2}-a_{F}^{2}\eta_{ab}e_{i}^{a}e_{j}^{b}dx^{i}dx^{j}-\left(\gamma_{ab}-a_{F}^{2}\eta_{ab}\right)e_{i}^{a}e_{j}^{b}dx^{i}dx^{j}=\nonumber \\
= & ds_{0}^{2}-\left(\gamma_{ab}-a_{F}^{2}\eta_{ab}\right)e_{i}^{a}e_{j}^{b}dx^{i}dx^{j}\end{align}

\noindent . Grishchuk, Doroshkevich \& Iudin showed that the object
describing the space part of the anisotropic part of the metric at
some moment in time,%
\footnote{$a_{F}$ has been scaled here to equal 1%
}

\begin{equation}
G_{ij}^{ab}\equiv2\left(e_{i}^{a}e_{j}^{b}+e_{i}^{b}e_{j}^{a}\right)-\frac{4}{3}\eta^{ab}\eta_{cd}e_{i}^{c}e_{j}^{d}\end{equation}

\noindent , obeys the property

\begin{equation}
\left(G_{ij}^{ab}\right)_{;k}^{;k}=-\left(n^{2}-3K\right)G_{ij}^{ab}\end{equation}

\noindent for $n=3$ and \textbf{$K=1$}; that is, $G_{ij}^{ab}$
is a tensor eigenfunction of the Laplace operator in a Bianchi IX
space for waves with wavenumber $n=3$. A similar property for open
spaces is true of the Bianchi type VII models.\cite{DLN anisotropy}%
\footnote{We could also choose to interpret Bianchi I as the degenerate case
of a flat universe containing gravitational waves of infinite wavelength
with $n=0$. The Kasner universe, however, is \emph{not}~such a universe:
all the anisotropy is governed by a single parameter, \emph{u}, so
the system has an insufficient number of degrees of freedom. The Kasner
universe is more like the Taub universe\cite{Taub}.%
}

Lifshitz, in his development of the theory of cosmological perturbations\cite[ss. 115]{BKL singularity 2,Lifshitz 46,Landau-Lifschitz},
claims that tensorial perturbations, including gravitational waves,
can only have diminishing effect over time. Lifshitz is, however,
considering only the class of \emph{local} tensorial perturbations.

In contrast, the gravitational waves in Bianchi IX will have wavelengths
comparable to the radius of curvature of the universe. Kristian and
Sachs note\cite{Kristian & Sachs} that the wavelength of cosmic shear
(and thus, all else being equal, of cosmological gravitational waves)
must be at least $2\times10^{10}$ years -- longer than the Hubble
radius\cite{WMAP 7-year} -- and could potentially be far longer (see
\noun{section} \ref{sub:Cosmological-parameters}).

We will consider first the regime of weak gravitational waves, in
an almost-isotropic universe, and then {}``quasi-isotropic'' waves:
that is, the regime in which components of the metric evolve at equal
powers of \emph{t}.

\subsection{Einstein equations in the tetrad formalism}

For a metric $g_{\alpha\beta}$, let the space-space part of the metric
be decomposed as in (\ref{eq:decomposition}). Similarly, the tensors

\begin{alignat}{1}
R_{ij}= & R_{ab}e_{i}^{a}e_{j}^{b}\\
T_{ij}= & T_{ab}e_{i}^{a}e_{j}^{b}\end{alignat}

\noindent with all space dependence in the frame vectors. Therefore
the Einstein equations can be rewritten:

\begin{align}
R_{00}= & kT_{00}-\frac{1}{2}kTg_{00}\\
R_{0i}= & kT_{0i}-\frac{1}{2}kTg_{0i}\\
R_{ab}= & k\left(T_{ab}-\frac{1}{2}T\gamma_{ab}\right)\end{align}

\noindent . If we have energy-momentum tensor\begin{align}
T_{\mu\nu}= & \left(p+\epsilon\right)u_{\mu}u_{\nu}-pg_{\mu\nu}\\
T= & \epsilon-3p\end{align}

\noindent then

\begin{align}
T_{00}= & \left(p+\epsilon\right)u_{0}u_{0}-pg_{00}\\
T_{0i}= & \left(p+\epsilon\right)u_{0}u_{i}-pg_{0i}\\
T_{ab}= & \left(p+\epsilon\right)u_{a}u_{b}-p\gamma_{ab}\end{align}

\noindent . If we then choose a synchronous Gaussian reference system,
as we always have freedom to do,

\begin{align}
g_{00}= & 1\\
g_{0i}= & 0\end{align}

\noindent so the Einstein equations read

\begin{align}
R_{00}= & k\left(p+\epsilon\right)u_{0}u_{i}k\left(p+\epsilon\right)u_{0}u_{0}-kpg_{00}-\frac{1}{2}k\left(\epsilon-3p\right)\\
R_{0i}= & k\left(p+\epsilon\right)u_{0}u_{i}\\
R_{ab}= & k\left(p+\epsilon\right)u_{a}u_{b}-kp\gamma_{ab}-\frac{1}{2}k\left(\epsilon-3p\right)\gamma_{ab}\end{align}

\noindent . If we then demand that our coordinate system be co-moving
with matter,

\begin{align}
u^{0}= & 1\\
u^{i}= & 0\label{eq:co-moving}\end{align}

\noindent then

\begin{align}
R_{00}= & \frac{1}{2}k\left(\epsilon+3p\right)\\
R_{0i}= & 0\\
R_{ab}= & \frac{1}{2}k\left(p-\epsilon\right)\gamma_{ab}\end{align}

\noindent . Let \begin{equation}
d_{ab}\equiv\frac{1}{2}\frac{\partial}{\partial t}g_{ij}e_{a}^{i}e_{b}^{j}=\frac{1}{2}\frac{d}{dt}\gamma_{ab}\end{equation}

\noindent and

\begin{equation}
d\equiv d_{ab}\gamma^{ab}\end{equation}

\noindent . The Christoffel symbols associated with our metric then
become \cite[ss. 97]{Landau-Lifschitz}

\begin{align}
\Gamma_{00}^{0}=\Gamma_{0i}^{0}=\Gamma_{00}^{i}= & 0\\
\Gamma_{ij}^{0}= & d_{ij}\\
\Gamma_{0j}^{i}= & d_{j}^{\, i}\\
\Gamma_{jk}^{i}= & \tilde{\Gamma}_{jk}^{i}\end{align}

\noindent where $\tilde{\Gamma}_{jk}^{i}$ are the Christoffel symbols
associated with the three-dimensional metric tensor $-g_{ij}$. The
Ricci tensor can then be written as\cite[ss. 97]{Landau-Lifschitz}:

\begin{align}
R_{00}= & -\dot{d}-d_{a}^{\, b}d_{b}^{\, a}\\
R_{0i}= & 0\\
R_{ab}= & \dot{d}_{ab}+dd_{ab}-2d_{ac}d_{b}^{\, c}-P_{ab}\end{align}

\noindent or explicitly\cite{GDI}

\begin{align}
\dot{d}+d_{a}^{\, b}d_{b}^{\, a}= & -\frac{1}{2}k\left(\epsilon+3p\right)\\
\dot{d}_{ab}+dd_{ab}-2d_{ac}d_{b}^{\, c}-P_{ab}= & \frac{1}{2}k\left(\epsilon-p\right)\gamma_{ab}\\
d_{a}^{\, b}C_{bc}^{\,\, a}= & 0\end{align}

\noindent where $P_{ij}$ is the three-dimensional Ricci tensor constructed
from $\tilde{\Gamma}_{jk}^{i}$.

\subsection{The curvature tensor for Bianchi IX}

Grishchuk explicitly gives the curvature tensors for all Bianchi types,
and a general method for easily deriving them, in \cite{Grishchuk curvature}.
These tensors can be stated in removable and non-removable parts,
the removable parts corresponding to time-dependent rotations of the
space. Let the symbol

\begin{equation}
\gamma_{abc}\equiv\gamma_{ad}C_{\, bc}^{a}\end{equation}
. Then where

\begin{equation}
\Gamma_{\, ab}^{c}\equiv\frac{1}{2}\gamma^{cd}\left(\gamma_{abd}+\gamma_{dab}-\gamma_{bda}\right)\end{equation}

\noindent (these are analogous to the Christoffel symbols of the full
space, but with different symmetry properties) the non-removable part
of the curvature tensor is given by

\begin{equation}
L_{ab}\equiv-2\Gamma_{\, a\left[b,c\right]}^{c}+2\Gamma_{\, d\left[b\right.}^{c}\Gamma_{\left.\,|a|c\right]}^{d}+2\Gamma_{\, ad}^{c}\Gamma_{\,\left[bc\right]}^{d}\end{equation}

\noindent where square brackets around the indices indicate the antisymmetric
part of the tensor; and the removable part is given by

\begin{equation}
b_{ab}\equiv\frac{1}{2}v_{c}C_{\, ba}^{c}+\frac{1}{2}\left(f_{a}v_{b}-f_{b}v_{a}\right)\end{equation}

\noindent and finally the curvature tensor

\begin{equation}
P_{ab}=L_{ab}-b_{bc}d_{a}^{\, c}-b_{ac}d_{b}^{\, c}-b_{ba}d\end{equation}

\noindent . In the co-moving case that $v_{a}=0$ we can simply state
$P_{ab}=H_{ab}$. For the particular case of Bianchi IX (the frame
vectors (\ref{eq:frame vectors})) and the curvature tensor when $v_{a}=0$
reads, for diagonal components:

\begin{equation}
P_{a}^{b}=\left[\frac{\left(\gamma_{fg}\eta^{fg}\right)^{2}}{2\left\Vert \gamma_{cd}\right\Vert }-\gamma^{fg}\eta_{fg}\right]\delta_{a}^{b}-\gamma^{bc}\eta_{ac}-\frac{\gamma_{af}\gamma_{gh}\eta^{fg}\eta^{bh}}{\left\Vert \gamma_{cd}\right\Vert }\end{equation}

\noindent and for non-diagonal components

\begin{equation}
P_{a}^{b}=-2\gamma^{cb}\eta_{ac}-\frac{1}{\left\Vert \gamma_{df}\right\Vert }\gamma_{ac}\gamma_{df}\eta^{bc}\eta^{df}\end{equation}

\noindent where $\left\Vert \gamma_{ab}\right\Vert $ is defined as
the determinant of $\gamma_{ab}$. The Einstein equations show that
when $v_{a}=0$ the non-diagonal components of $\gamma_{ab}$ must
be zero, so as a consequence of our Gaussian choice of coordinate
system therefore we can without loss of generality write the metric
for Bianchi IX\begin{equation}
\begin{array}{c}
\gamma_{11}=a_{F}^{2}e^{2\alpha}\\
\gamma_{22}=a_{F}^{2}e^{2\beta}\\
\gamma_{33}=a_{F}^{2}e^{2\gamma}\end{array}\label{eq:metric}\end{equation}

\noindent with all other space-space components zero, so explicitly
the the curvature tensor $P_{ab}$ for Bianchi IX reads

\begin{align}
P_{11}= & \frac{1}{2}e^{-2\delta}\left(-e^{4\alpha}+\left(e^{2\beta}-e^{2\gamma}\right)^{2}\right)e^{2\alpha}\\
P_{22}= & \frac{1}{2}e^{-2\delta}\left(-e^{4\beta}+\left(e^{2\gamma}-e^{2\alpha}\right)^{2}\right)e^{2\beta}\\
P_{33}= & \frac{1}{2}e^{-2\delta}\left(-e^{4\gamma}+\left(e^{2\alpha}-e^{2\beta}\right)^{2}\right)e^{2\gamma}\\
P_{ab}= & 0, & a\neq b\end{align}

\noindent and the contracted curvature scalar

\begin{equation}
P_{ab}\gamma^{ab}=2a_{F}^{-2}e^{-2\delta}\left[e^{4\alpha}+e^{4\beta}+e^{4\gamma}-2\left(e^{2\alpha+2\beta}+e^{2\beta+2\gamma}+e^{2\alpha+2\gamma}\right)\right]\end{equation}

\noindent .

\section{Einstein equations for Bianchi IX}

\subsection{Exact equations}

Let the symbol $\delta\equiv\alpha+\beta+\gamma$ for convenience
as in (\ref{eq:Delta definition}). For our chosen metric, we have
the auxiliary quantities

\begin{align}
d_{11}= & \left(a\dot{a}+a^{2}\dot{\alpha}\right)e^{2\alpha}\\
\dot{d}_{11}= & \left(\dot{a}^{2}+a\ddot{a}+4a\dot{a}\dot{\alpha}+a^{2}\ddot{\alpha}+2a^{2}\dot{\alpha}^{2}\right)e^{2\alpha}\\
d_{1}^{1}= & H+\dot{\alpha}\\
d= & 3H+\dot{\delta}\end{align}

\noindent and cyclic permutations in $\alpha,\beta,\gamma$ thereof
for 22- and 33-quantities. The full Einstein equations for Bianchi
IX read

\begin{align}
\left\{ \begin{array}{c}
\frac{3}{a_{F}^{2}}\left(\dot{a}_{F}^{2}+1\right)+\dot{\alpha}\dot{\beta}+\dot{\alpha}\dot{\gamma}+\dot{\beta}\dot{\gamma}+2\frac{\dot{a}_{F}}{a_{F}}\dot{\delta}+\\
+a_{F}^{-2}e^{-2\delta}\left[\begin{array}{c}
2\left(e^{2\alpha+2\beta}+e^{2\alpha+2\gamma}+e^{2\beta+2\gamma}\right)-\\
-e^{4\alpha}-e^{4\beta}-e^{4\gamma}-3e^{2\delta}\end{array}\right]\end{array}\right\}  & =k\epsilon\\
\left\{ \begin{array}{c}
\frac{\ddot{a}_{F}}{a_{F}}+2\frac{\dot{a}_{F}^{2}}{a_{F}^{2}}+\frac{2}{a_{F}^{2}}+\ddot{\alpha}+\frac{\dot{a}_{F}}{a_{F}}\left(3\dot{\alpha}+\dot{\delta}\right)+\dot{\alpha}\dot{\delta}+\\
+2a_{F}^{-2}e^{-2\delta}\left[e^{4\alpha}-\left(e^{2\beta}-e^{2\gamma}\right)^{2}-e^{2\delta}\right]\end{array}\right\}  & =\frac{1}{2}k\left(\epsilon-p^{\left(1\right)}\right)\\
\left\{ \begin{array}{c}
\frac{\ddot{a}_{F}}{a_{F}}+2\frac{\dot{a}_{F}^{2}}{a_{F}^{2}}+\frac{2}{a_{F}^{2}}+\ddot{\beta}+\frac{\dot{a}_{F}}{a_{F}}\left(3\dot{\beta}+\dot{\delta}\right)+\dot{\beta}\dot{\delta}+\\
+2a_{F}^{-2}e^{-2\delta}\left[e^{4\beta}-\left(e^{2\gamma}-e^{2\alpha}\right)^{2}-e^{2\delta}\right]\end{array}\right\}  & =\frac{1}{2}k\left(\epsilon-p^{\left(2\right)}\right)\\
\left\{ \begin{array}{c}
\frac{\ddot{a}_{F}}{a_{F}}+2\frac{\dot{a}_{F}^{2}}{a_{F}^{2}}+\frac{2}{a_{F}^{2}}+\ddot{\gamma}+\frac{\dot{a}_{F}}{a_{F}}\left(3\dot{\gamma}+\dot{\delta}\right)+\dot{\gamma}\dot{\delta}+\\
+2a_{F}^{-2}e^{-2\delta}\left[e^{4\gamma}-\left(e^{2\alpha}-e^{2\beta}\right)^{2}-e^{2\delta}\right]\end{array}\right\}  & =\frac{1}{2}k\left(\epsilon-p^{\left(3\right)}\right)\end{align}

\noindent . We can also define quantities as components of a gravitational
effective energy-momentum tensor:

\begin{align}
k\epsilon_{g}\equiv & -\left\{ \begin{array}{c}
\dot{\alpha}\dot{\beta}+\dot{\alpha}\dot{\gamma}+\dot{\beta}\dot{\gamma}+2\frac{\dot{a}_{F}}{a_{F}}\dot{\delta}+\\
+a_{F}^{-2}e^{-2\delta}\left[\begin{array}{c}
2\left(e^{2\alpha+2\beta}+e^{2\alpha+2\gamma}+e^{2\beta+2\gamma}\right)-\\
-e^{4\alpha}-e^{4\beta}-e^{4\gamma}-3e^{2\delta}\end{array}\right]\end{array}\right\} \label{eq:epsilong}\\
\frac{1}{2}k\left(\epsilon_{g}-p_{g}^{\left(1\right)}\right)\equiv & -\left\{ \begin{array}{c}
\ddot{\alpha}+\frac{\dot{a}_{F}}{a_{F}}\left(3\dot{\alpha}+\dot{\delta}\right)+\dot{\alpha}\dot{\delta}+\\
+2a_{F}^{-2}e^{-2\delta}\left[e^{4\alpha}-\left(e^{2\beta}-e^{2\gamma}\right)^{2}-e^{2\delta}\right]\end{array}\right\} \\
\frac{1}{2}k\left(\epsilon_{g}-p_{g}^{\left(2\right)}\right)\equiv & -\left\{ \begin{array}{c}
\ddot{\beta}+\frac{\dot{a}_{F}}{a_{F}}\left(3\dot{\beta}+\dot{\delta}\right)+\dot{\beta}\dot{\delta}+\\
+2a_{F}^{-2}e^{-2\delta}\left[e^{4\beta}-\left(e^{2\gamma}-e^{2\alpha}\right)^{2}-e^{2\delta}\right]\end{array}\right\} \\
\frac{1}{2}k\left(\epsilon_{g}-p_{g}^{\left(3\right)}\right)\equiv & -\left\{ \begin{array}{c}
\ddot{\gamma}+\frac{\dot{a}_{F}}{a_{F}}\left(3\dot{\gamma}+\dot{\delta}\right)+\dot{\gamma}\dot{\delta}+\\
+2a_{F}^{-2}e^{-2\delta}\left[e^{4\gamma}-\left(e^{2\alpha}-e^{2\beta}\right)^{2}-e^{2\delta}\right]\end{array}\right\} \\
kp_{g}^{\left(1\right)}\equiv & \left[\begin{array}{c}
2\ddot{\alpha}+6\frac{\dot{a}_{F}}{a_{F}}\dot{\alpha}+2\dot{\alpha}^{2}+\dot{\alpha}\dot{\beta}+\dot{\alpha}\dot{\gamma}-\dot{\beta}\dot{\gamma}+\\
+a_{F}^{-2}\left(\begin{array}{c}
5e^{2\left(\alpha-\beta-\gamma\right)}-3e^{2\left(\beta-\alpha-\gamma\right)}-3e^{2\left(\gamma-\alpha-\beta\right)}+\\
+6e^{-2\alpha}-2e^{-2\gamma}-2e^{-2\beta}-1\end{array}\right)\end{array}\right]\label{eq:pg1}\\
kp_{g}^{\left(2\right)}\equiv & \left[\begin{array}{c}
2\ddot{\beta}+6\frac{\dot{a}_{F}}{a_{F}}\dot{\beta}+2\dot{\beta}^{2}+\dot{\alpha}\dot{\beta}-\dot{\alpha}\dot{\gamma}+\dot{\beta}\dot{\gamma}+\\
+a_{F}^{-2}\left(\begin{array}{c}
5e^{2\left(\beta-\alpha-\gamma\right)}-3e^{2\left(\gamma-\beta-\alpha\right)}-3e^{2\left(\alpha-\beta-\gamma\right)}\\
+6e^{-2\beta}-2e^{-2\alpha}-2e^{-2\gamma}-1\end{array}\right)\end{array}\right]\label{eq:pg2}\\
kp_{g}^{\left(3\right)}\equiv & \left[\begin{array}{c}
2\ddot{\gamma}+6\frac{\dot{a}_{F}}{a_{F}}\dot{\gamma}+2\dot{\gamma}^{2}-\dot{\alpha}\dot{\beta}+\dot{\alpha}\dot{\gamma}+\dot{\beta}\dot{\gamma}+\\
+a_{F}^{-2}\left(\begin{array}{c}
5e^{2\left(\gamma-\beta-\alpha\right)}-3e^{2\left(\alpha-\gamma-\beta\right)}-3e^{2\left(\beta-\gamma-\alpha\right)}\\
+6e^{-2\gamma}-2e^{-2\beta}-2e^{-2\alpha}-1\end{array}\right)\end{array}\right]\label{eq:pg3}\end{align}

\noindent (all of which are zero when $\alpha=\beta=\gamma=0$). The
Bianchi identity $T_{\mu,\nu}^{\nu}$ demands $p_{g}^{\left(1\right)}=p_{g}^{\left(2\right)}=p_{g}^{\left(3\right)}$
so define the averaged gravitational pressure

\begin{align}
kp_{g}\equiv & \frac{1}{3}k\left(p_{g}^{\left(1\right)}+p_{g}^{\left(2\right)}+p_{g}^{\left(3\right)}\right)=\\
\equiv & \left[\begin{array}{c}
2\ddot{\delta}+6\frac{\dot{a}_{F}}{a_{F}}\dot{\delta}+2\left(\dot{\alpha}^{2}+\dot{\beta}^{2}+\dot{\gamma}^{2}\right)+\\
+2\left(\dot{\alpha}\dot{\beta}+\dot{\alpha}\dot{\gamma}+\dot{\beta}\dot{\gamma}\right)+\\
+a_{F}^{-2}\left(\begin{array}{c}
-e^{2\left(\alpha-\beta-\gamma\right)}-e^{2\left(\beta-\alpha-\gamma\right)}-e^{2\left(\gamma-\alpha-\beta\right)}\\
+2e^{-2\alpha}+2e^{-2\gamma}+2e^{-2\beta}-3\end{array}\right)\end{array}\right]\nonumber \end{align}

\noindent . Finally, 

\begin{equation}
k\left(\epsilon_{g}+3p_{g}\right)=2\ddot{\delta}+4\frac{\dot{a}_{F}}{a_{F}}\dot{\delta}+2\left(\dot{\alpha}^{2}+\dot{\beta}^{2}+\dot{\gamma}^{2}\right)\label{eq:e+3pg}\end{equation}

\noindent %
\footnote{Equation (\ref{eq:e+3pg}) corrects an error of sign in \cite[equation (27)]{GDI}.%
}. Define a quasi-conformal coordinate $\eta$ by $cdt\equiv a_{F}d\eta$;
note that this fixes the relationship between \emph{t}~and $\eta$
up to the level of the characteristic length $a_{i}$ and a constant
which can be set to zero. Given the impossibility of selecting a unique
and objective definition for the scale factor, we do \emph{not} define
the conformal time using such a function. Define a correction term
to the matter energy density \emph{q} such that

\begin{equation}
\epsilon=\epsilon_{F}\left(1+q\right)\end{equation}

\noindent . In $\eta$-time, the Einstein equations for Bianchi IX,
subtracting background terms on both sides, read:

\begin{align}
\left\{ \begin{array}{c}
\alpha^{\prime}\beta^{\prime}+\alpha^{\prime}\gamma^{\prime}+\beta^{\prime}\gamma^{\prime}+2\frac{a_{F}^{\prime}}{a_{F}}\delta^{\prime}+\\
+e^{-2\delta}\left[\begin{array}{c}
2\left(e^{2\alpha+2\beta}+e^{2\alpha+2\gamma}+e^{2\beta+2\gamma}\right)-\\
-e^{4\alpha}-e^{4\beta}-e^{4\gamma}-3e^{2\delta}\end{array}\right]\end{array}\right\}  & =a_{F}^{2}k\epsilon_{F}q\label{eq:Einstein EQ 0}\\
\left\{ \begin{array}{c}
\alpha^{\prime\prime}+\frac{a_{F}^{\prime}}{a_{F}}\left(2\alpha^{\prime}+\delta^{\prime}\right)+\alpha^{\prime}\delta^{\prime}+\\
+2e^{-2\delta}\left[e^{4\alpha}-\left(e^{2\beta}-e^{2\gamma}\right)^{2}-e^{2\delta}\right]\end{array}\right\}  & =\frac{1-w}{2}a_{F}^{2}k\epsilon_{F}q\label{eq:Einstein EQ 1}\\
\left\{ \begin{array}{c}
\beta^{\prime\prime}+\frac{a_{F}^{\prime}}{a_{F}}\left(2\beta^{\prime}+\delta^{\prime}\right)+\beta^{\prime}\delta^{\prime}+\\
+2e^{-2\delta}\left[e^{4\beta}-\left(e^{2\gamma}-e^{2\alpha}\right)^{2}-e^{2\delta}\right]\end{array}\right\}  & =\frac{1-w}{2}a_{F}^{2}k\epsilon_{F}q\label{eq:Einstein EQ 2}\\
\left\{ \begin{array}{c}
\gamma^{\prime\prime}+\frac{a_{F}^{\prime}}{a_{F}}\left(2\gamma^{\prime}+\delta^{\prime}\right)+\gamma^{\prime}\delta^{\prime}+\\
+2e^{-2\delta}\left[e^{4\gamma}-\left(e^{2\alpha}-e^{2\beta}\right)^{2}-e^{2\delta}\right]\end{array}\right\}  & =\frac{1-w}{2}a_{F}^{2}k\epsilon_{F}q\label{eq:Einstein EQ 3}\end{align}

\noindent . We also note the Einstein equations have an exact formal
solution \begin{equation}
k\epsilon=\left(Sa_{F}^{-3}e^{-\delta}\right)^{1+w}\label{eq:formal solution}\end{equation}

\noindent where \emph{S} is a constant of proportionality such that
$S^{1+w}$ has dimensionality of length to the $1+3w$ power. Finally,
if we define the gravitational equation of state $w_{g}\equiv p_{g}/\epsilon_{g}$
we see that necessarily $w_{g}=w$ and the Einstein equations can
be read as

\begin{equation}
kp_{g}^{\left(1\right)}+wa_{F}^{2}k\epsilon_{F}q=kp_{g}^{\left(2\right)}+wa_{F}^{2}k\epsilon_{F}q=kp_{g}^{\left(3\right)}+wa_{F}^{2}k\epsilon_{F}q=a_{F}^{2}k\epsilon_{F}q+k\epsilon_{g}=0\label{eq:backreaction}\end{equation}

\noindent . In other words, the effective energy-momentum tensor created
by cosmological gravitational waves equals minus the back-reaction
on matter energy density and pressure. Note that the quantity $k\epsilon_{g}/q$
is necessarily negative.

\subsection{Solutions to the Einstein equations at zero order}

For convenience, define the variable $x\equiv\frac{1+3w}{2}\eta$.
Then at zero order the Einstein equations for a Bianchi IX universe
have, for arbitrary constant equation of state, the following solution
and auxiliary quantities, which are identical to the solutions to
the Einstein equations in the unperturbed closed Friedmann cosmology:

\begin{align}
a_{F}= & a_{i}\left(\sin x\right)^{\frac{2}{1+3w}}\\
a_{F}^{\prime}= & a_{i}\left(\sin x\right)^{\frac{1-3w}{1+3w}}\cos x\\
a_{F}^{\prime\prime}= & \frac{1+3w}{2}a_{i}\left[\frac{1-3w}{1+3w}\left(\sin x\right)^{\frac{-6w}{1+3w}}\cos^{2}x-\left(\sin x\right)^{\frac{2}{1+3w}}\right]\\
a_{F}^{\prime}/a_{F}= & \cot x\\
H_{F}= & a_{i}^{-1}\cot x\csc x\\
Q_{F}= & \frac{1+3w}{2}\sec^{2}x\end{align}

\noindent . The quantity $a_{i}$ represents a characteristic scale
for the universe, and in the background case represents the radius
of curvature of the universe at the extent of its maximum expansion.
We treat $a_{i}$ as an arbitrary constant for the time being.

\subsection{Solutions at linear order}

To first order the Einstein equations take the form:

\begin{align}
2\frac{a_{F}^{\prime}}{a_{F}}\delta_{1}^{\prime}-2\delta_{1}= & a_{F}^{2}k\epsilon_{F}q_{1}\\
\alpha_{1}^{\prime\prime}+\frac{a_{F}^{\prime}}{a_{F}}\left(2\alpha_{1}^{\prime}+\delta_{1}^{\prime}\right)+8\alpha_{1}-4\delta_{1}= & \frac{1-w}{2}a_{F}^{2}k\epsilon_{F}q_{1}\\
\beta_{1}^{\prime\prime}+\frac{a_{F}^{\prime}}{a_{F}}\left(2\beta_{1}^{\prime}+\delta_{1}^{\prime}\right)+8\beta_{1}-4\delta_{1}= & \frac{1-w}{2}a_{F}^{2}k\epsilon_{F}q_{1}\\
\gamma_{1}^{\prime\prime}+\frac{a_{F}^{\prime}}{a_{F}}\left(2\gamma_{1}^{\prime}+\delta_{1}^{\prime}\right)+8\gamma_{1}-4\delta_{1}= & \frac{1-w}{2}a_{F}^{2}k\epsilon_{F}q_{1}\end{align}

\noindent where the subscript 1 denotes a first-order small quantity,
that is, a quantity small such that in the first approximation its
square is negligible. The formal solution (\ref{eq:formal solution})
gives us, to first order,

\begin{align}
a_{F}^{2}k\epsilon_{F}q_{1} & =-\left(1+w\right)S^{1+w}a_{F}^{-1-3w}\delta_{1}\end{align}

\noindent . Meanwhile, we can always choose to let \emph{S} take on
its Friedmannian value\cite{GDI}, so $S^{1+w}=3a_{i}^{1+3w}$. Therefore:

\begin{align}
2\frac{a_{F}^{\prime}}{a_{F}}\delta_{1}^{\prime}+\left[3\left(1+w\right)\csc^{2}x-2\right]\delta_{1}= & 0\\
\alpha_{1}^{\prime\prime}+2\frac{a_{F}^{\prime}}{a_{F}}\alpha_{1}^{\prime}+8\alpha_{1}+\frac{a_{F}^{\prime}}{a_{F}}\delta_{1}^{\prime}+\left(3\frac{1-w^{2}}{2}\csc^{2}x-4\right)\delta_{1}= & 0\\
\beta_{1}^{\prime\prime}+2\frac{a_{F}^{\prime}}{a_{F}}\beta_{1}^{\prime}+8\beta_{1}+\frac{a_{F}^{\prime}}{a_{F}}\delta_{1}^{\prime}+\left(3\frac{1-w^{2}}{2}\csc^{2}x-4\right)\delta_{1}= & 0\\
\gamma_{1}^{\prime\prime}+2\frac{a_{F}^{\prime}}{a_{F}}\gamma_{1}^{\prime}+8\gamma_{1}+\frac{a_{F}^{\prime}}{a_{F}}\delta_{1}^{\prime}+\left(3\frac{1-w^{2}}{2}\csc^{2}x-4\right)\delta_{1}= & 0\end{align}

\noindent which gives us the solution:

\begin{equation}
\delta_{1}=c_{1}\cos x\left(\csc x\right)^{\frac{3+3w}{1+3w}}\end{equation}

\noindent . The term governed by $c_{1}$ is a {}``removable'' perturbation,
that is, one not arising from a physical phenomenon but from small
changes in our selection of the scale factor. Grishchuk, Doroshkevich
\& Iudin argue\cite{GDI}, and Grishchuk later proves in the case
of high-frequency gravitational waves\cite{gw inflation}, that the
the removable perturbation arises in the transformation from \emph{t}-time
to conformal time, and represents a small change in the value of $\eta$
-- that is, from the constant that has been implicitly set to zero
in the relationship $dt=a_{F}d\eta$ so that $t=0$ coincides with
the metric singularity from $a_{F}\left(0\right)=0$. This coincides
with the argument made by Bardeen\cite{Bardeen} with regard to scalar
and vector perturbations with wavelengths longer than the Hubble radius;
Bardeen recommends a gauge choice minimizing shear. We always have
the freedom to set $c_{1}$ to zero but do not do so yet. In a radiation-dominated
universe, we have

\begin{equation}
\delta_{1}^{\mbox{radiation}}=c_{1}^{\mbox{radiation}}\cos\eta\csc^{2}\eta\end{equation}

\noindent and in a matter-dominated universe

\begin{equation}
\delta_{1}^{\mbox{matter}}=c_{1}^{\mbox{matter}}\cos\frac{\eta}{2}\csc^{3}\frac{\eta}{2}\end{equation}

\noindent . Therefore the full first-order functions can be written:

\begin{equation}
\alpha_{1}^{\prime\prime}+2\cot x\alpha_{1}^{\prime}+8\alpha_{1}=3c_{1}\left[\begin{array}{c}
1+\frac{1+w}{2}\left(\csc x\right)^{1+3w}-\\
-\frac{1-w^{2}}{2}\csc^{2}x\end{array}\right]\left(\csc x\right)^{\frac{3+3w}{1+3w}}\cos x\label{eq:Einstein 1st order}\end{equation}

\noindent \emph{etc.} Note that the right hand side contains \emph{no
}physical variables -- no characteristic length or energy density.
The Einstein equations at first order have the solutions (denoted
with a tilde for the $c_{1}=0$ case)

\begin{align}
\tilde{\alpha}_{1}^{\mbox{radiation}}= & \left(C_{\alpha1,1}\sin3\eta+C_{\alpha2,1}\cos3\eta\right)\csc\eta\nonumber \\
\tilde{\beta}_{1}^{\mbox{radiation}}= & \left(C_{\beta1,1}\sin3\eta+C_{\beta2,1}\cos3\eta\right)\csc\eta\\
\tilde{\gamma}_{1}^{\mbox{radiation}}= & \left(C_{\gamma1,1}\sin3\eta+C_{\gamma2,1}\cos3\eta\right)\csc\eta\nonumber \end{align}

\noindent and similarly for $\tilde{\beta},\tilde{\gamma}$ in a radiation-dominated
universe, and

\begin{equation}
\tilde{\alpha}_{1}^{\mbox{matter}}=\frac{C_{\alpha1,1}}{\sin\eta/2}\frac{d}{d\eta}\frac{\sin3\eta}{\sin\eta/2}+\frac{C_{\alpha2,1}}{\sin\eta/2}\frac{d}{d\eta}\frac{\cos3\eta}{\sin\eta/2}\label{eq:matter first order wave}\end{equation}

\noindent \emph{etc.} in a matter-dominated universe, in both cases
constrained by the condition $C_{\alpha1,1}+C_{\beta1,1}+C_{\gamma1,1}=C_{\alpha2,1}+C_{\beta2,1}+C_{\gamma2,1}=0$.
A general solution for any constant equation of state, in terms of
orthogonal polynomials in \emph{a}, exists but is far too cumbersome
to be of practical use in this work. We introduce the notation $C_{\alpha1,1}$
\emph{etc.} to be read in the following way: $C_{\alpha2,1}$ is an
arbitrary constant associated with the function $\alpha$, the first
index denoting the mode of the solution (1 for growing, 2 for decaying),
the second index denoting the order of the constant in an expansion
assuming $\alpha,\beta,\gamma\ll1$. For convenience, we will sometimes
write a generic solution to the differential equation (\ref{eq:Einstein 1st order})
as

\begin{equation}
\tilde{\alpha}_{1}=C_{\alpha1,1}y_{1}+C_{\alpha2,1}y_{2}\end{equation}

\noindent . These solutions can be written in a less symmetric but
easier-to-manipulate form:

\begin{align}
\tilde{\alpha}_{1}^{\mbox{radiation}}= & C_{\alpha1,1}\left(2\cos2\eta+1\right)+C_{\alpha2,1}\cos3\eta\csc\eta\\
\tilde{\alpha}_{1}^{\mbox{matter}}= & -\left[\begin{array}{c}
C_{\alpha1,1}\left(16\cos2\eta+10\cos\eta+9\right)+\\
+\frac{1}{4}C_{\alpha2,1}\csc^{3}\frac{\eta}{2}\left(5\cos\frac{7}{2}\eta-7\cos\frac{5}{2}\eta\right)\end{array}\right]\end{align}

\noindent \emph{etc}. When $\delta=0$ we recognize the homogeneous
first-order Einstein equations as describing weak gravitational waves
with wavenumber $n=3$ and a wave equation of the form

\begin{equation}
\nu^{\prime\prime}+2\cot\left(x\right)\nu^{\prime}+\left(n^{2}-1\right)\nu=0\end{equation}

\noindent , in line with \cite{GDI}'s description. In a radiation-dominated
universe we have explicitly for the full first-order solution:

\begin{equation}
\alpha_{1}^{\mbox{radiation}}=C_{\alpha1,1}\frac{\sin3\eta}{\sin\eta}+C_{\alpha2,1}\frac{\cos3\eta}{\sin\eta}+\frac{c_{1}}{3}\cos\eta\csc^{2}\eta\label{eq:alpha1 radiation}\end{equation}

\noindent \emph{etc.} and in a matter-dominated universe we have

\begin{equation}
\alpha_{1}^{\mbox{matter}}=\frac{C_{\alpha1,1}}{\sin\eta/2}\frac{d}{d\eta}\frac{\sin3\eta}{\sin\eta/2}+\frac{C_{\alpha2,1}}{\sin\eta/2}\frac{d}{d\eta}\frac{\cos3\eta}{\sin\eta/2}+\frac{c_{1}}{3}\cos\frac{\eta}{2}\csc^{3}\frac{\eta}{2}\label{eq:alpha1 matter}\end{equation}

\noindent . It is common to refer to the decaying {}``cos'' mode
of these gravitational waves as {}``singularity-destroying''\cite{GDI},
in that they diverge as $\eta\rightarrow0$, which could seem at first
to imply $\lim\underset{\eta\rightarrow0}{\gamma_{ab}}\rightarrow\infty$.
It is worth remembering that as the functions $\alpha,\beta,\gamma$
appear in the metric as exponents, that is, $\gamma_{11}=a_{F}^{2}e^{2\alpha}$
\emph{etc}., decaying functions are not necessarily {}``singularity-destroying''
for the following reasons:
\begin{itemize}
\item their divergence must overcome the convergence of the Friedmannian
term, which in the case of weak waves will occur when $w\leq2/3$
but not generally;
\item functions of the form $e^{-x^{-y}}$ for $x<0,y<0$ are non-analytic
near $x=0$, that is, they are not described by convergent Taylor
series in that region. 
\end{itemize}
As $C_{\alpha2,1}+C_{\beta2,1}+C_{\gamma2,1}=0$, either one or two
decaying terms preserve the $t=0$ singularity when the removable
perturbation is removed, in a manner analogous to that found in the
Kasner universe, in the case of weak gravitational waves (although
the price of this is a divergence later).

When discussing high-frequency, localized waves, it is easy to define
an amplitude of the waves by (for example) normalizing a root-mean-square
value over the wave's period. In the case of cosmological gravitational
waves however this procedure is not possible in an absolute sense
due to the diverging character of the decaying mode. Fortunately,
mathematical conditions on the relation of linear-order terms to quadratic-order
terms revealed at quadratic order (see \noun{section} \ref{sub:Solutions-at-quadratic})
cause the term {}``weak'' to give itself an objective meaning. If
we wish to normalize the growing modes, they have the following RMS
values:\begin{align}
y_{1}^{\mbox{RMS}}\equiv & \left[\frac{2}{\left(1+3w\right)\pi}\int_{0}^{\left(1+3w\right)\pi/2}y_{1}^{2}d\eta\right]^{1/2}\\
y_{1}^{\mbox{radiation,RMS}}= & \sqrt{3}\\
y_{1}^{\mbox{matter,RMS}}= & \sqrt{259}\approx16.1\end{align}

.

It is interesting to note that in matter, the decaying {}``cos''
mode of $\alpha_{1},\beta_{1},\gamma_{1}$ has the same $\eta$-dependence
as the removable perturbation; a cosmologist attempting to remove
what they assume, based on an incomplete picture of the sky, to be
a removable perturbation may inadvertently be suppressing evidence
of a gravitational wave!

Finally, the gravitational energy-momentum tensor's (entirely removable)
components read, to linear order:

\begin{align}
k\epsilon_{g\left(1\right)}= & 3\left(1+w\right)\frac{c_{1}}{a_{i}^{2}}\cos x\left(\csc x\right)^{\frac{9+9w}{1+3w}}\\
kp_{g\left(1\right)}^{\left(1\right)}=kp_{g\left(1\right)}^{\left(2\right)}=kp_{g\left(1\right)}^{\left(3\right)}= & 3w\left(1+w\right)\frac{c_{1}}{a_{i}^{2}}\cos x\left(\csc x\right)^{\frac{9+9w}{1+3w}}\end{align}

\noindent while the back-reaction of the gravitational waves at linear
order gives us matter EMT components which vary from background by:

\begin{equation}
q_{1}=-3\left(1+w\right)c_{1}\cos x\left(\csc x\right)^{\frac{5+9w}{1+3w}}\end{equation}

\noindent ; when removable perturbations have been removed, first-order
weak gravitational waves have no effect on the distribution of matter.

\subsection{Solutions at quadratic order\label{sub:Solutions-at-quadratic}}

The Einstein equations to quadratic order read:

\begin{align}
2\cot x\delta_{2}^{\prime}+\left[3\left(1+w\right)\csc^{2}x-2\right]\delta_{2}= & \left\{ \begin{array}{c}
\left[3\csc^{2}x\frac{\left(1+w\right)^{2}}{2}-2\right]\delta_{1}^{2}-\\
-\frac{1}{2}\left[\delta_{1}^{\prime2}-\left(\alpha_{1}^{\prime2}+\beta_{1}^{\prime2}+\gamma_{1}^{\prime2}\right)\right]+\\
+4\left(\alpha_{1}^{2}+\beta_{1}^{2}+\gamma_{1}^{2}\right)\end{array}\right\} \label{eq:T00-2}\\
\alpha_{2}^{\prime\prime}+\cot x\left(2\alpha_{2}^{\prime}+\delta_{2}^{\prime}\right)+8\alpha_{2}-4\delta_{2}= & \left[\begin{array}{c}
3\frac{1-w}{2}\csc^{2}x\left(-\left(1+w\right)\delta_{2}+\frac{\left(1+w\right)^{2}}{2}\delta_{1}^{2}\right)-\\
-\alpha_{1}^{\prime}\delta_{1}^{\prime}+8\left(\beta_{1}-\gamma_{1}\right)^{2}-\\
-16\alpha_{1}^{2}+16\alpha_{1}\delta_{1}-4\delta_{1}^{2}\end{array}\right]\label{eq:R11-2}\\
\beta_{2}^{\prime\prime}+\cot x\left(2\alpha_{2}^{\prime}+\delta_{2}^{\prime}\right)+8\beta_{2}-4\delta_{2}= & \left[\begin{array}{c}
3\frac{1-w}{2}\csc^{2}x\left(-\left(1+w\right)\delta_{2}+\frac{\left(1+w\right)^{2}}{2}\delta_{1}^{2}\right)-\\
-\beta_{1}^{\prime}\delta_{1}^{\prime}+8\left(\gamma_{1}-\alpha_{1}\right)^{2}-\\
-16\beta_{1}^{2}+16\beta_{1}\delta_{1}-4\delta_{1}^{2}\end{array}\right]\label{eq:R22-2}\\
\gamma_{2}^{\prime\prime}+\cot x\left(2\gamma_{2}^{\prime}+\delta_{2}^{\prime}\right)+8\gamma_{2}-4\delta_{2}= & \left[\begin{array}{c}
3\frac{1-w}{2}\csc^{2}x\left(-\left(1+w\right)\delta_{2}+\frac{\left(1+w\right)^{2}}{2}\delta_{1}^{2}\right)-\\
-\gamma_{1}^{\prime}\delta_{1}^{\prime}+8\left(\alpha_{1}-\beta_{1}\right)^{2}-\\
-16\gamma_{1}^{2}+16\gamma_{1}\delta_{1}-4\delta_{1}^{2}\end{array}\right]\label{eq:R33-2}\end{align}

\noindent . Taking the $\phantom{}^{\left(2\right)}T_{0}^{0}$ equation
(\ref{eq:T00-2}) first,

\begin{equation}
2\cot x\delta_{2}^{\prime}+\left[3\left(1+w\right)\csc^{2}x-2\right]\delta_{2}=\left\{ \begin{array}{c}
\left[3\csc^{2}x\frac{\left(1+w\right)^{2}}{2}-\frac{2}{3}\right]\delta_{1}^{2}-\frac{1}{3}\delta_{1}^{\prime2}+\\
+\frac{1}{2}\left(\tilde{\alpha}_{1}^{\prime2}+\tilde{\beta}_{1}^{\prime2}+\tilde{\gamma}_{1}^{\prime2}\right)+\\
+4\left(\tilde{\alpha}_{1}^{2}+\tilde{\beta}_{1}^{2}+\tilde{\gamma}_{1}^{2}\right)\end{array}\right\} \label{eq:delta equation}\end{equation}

\noindent . The homogeneous part $\tilde{\delta}_{2}$ of course has
the same form as at first order, representing a removable perturbation,
so

\begin{equation}
\tilde{\delta}_{2}=c_{2}\cos x\left(\csc x\right)^{\frac{3+3w}{1+3w}}\label{eq:-131}\end{equation}

\noindent . The complete solution in integral form is

\begin{align}
\delta_{2}= & \frac{1}{2}\cos x\left(\csc x\right)^{\frac{3+3w}{1+3w}}\times\\
 & \times\left\{ \int\left[\begin{array}{c}
\left[3\csc^{2}x\frac{\left(1+w\right)^{2}}{2}-\frac{2}{3}\right]\delta_{1}^{2}-\\
-\frac{1}{3}\delta_{1}^{\prime2}+\\
+\frac{1}{2}\left(\tilde{\alpha}_{1}^{\prime2}+\tilde{\beta}_{1}^{\prime2}+\tilde{\gamma}_{1}^{\prime2}\right)+\\
+4\left(\tilde{\alpha}_{1}^{2}+\tilde{\beta}_{1}^{2}+\tilde{\gamma}_{1}^{2}\right)\end{array}\right]\sec^{2}x\left(\sin x\right)^{\frac{4+6w}{1+3w}}d\eta+c_{2}\right\} \nonumber \end{align}

\noindent . Define the following pseudo-vectors and their Euclidean
dot products\begin{align}
\left(C_{\alpha1\left(1\right)},C_{\beta1\left(1\right)},C_{\gamma1\left(1\right)}\right) & \equiv\bm{\sigma}\\
\left(C_{\alpha2\left(1\right)},C_{\beta2\left(1\right)},C_{\gamma2\left(1\right)}\right) & \equiv\bm{\tau}\\
\left(C_{\alpha1\left(1\right)}^{2}+C_{\beta1\left(1\right)}^{2}+C_{\gamma1\left(1\right)}^{2}\right) & =\bm{\sigma\cdot\sigma} & \equiv\sigma^{2}\\
\left(C_{\alpha2,1}^{2}+C_{\beta2,1}^{2}+C_{\gamma2,1}^{2}\right) & =\bm{\tau\cdot\tau} & \equiv\tau^{2}\\
\left(C_{\alpha1,1}C_{\alpha2,1}+C_{\beta1,1}C_{\beta2,1}+C_{\gamma1,1}C_{\gamma2,1}\right) & =\bm{\sigma\cdot\tau}\end{align}

\noindent so

\begin{align}
\tilde{\alpha}_{1}^{2}+\tilde{\beta}_{1}^{2}+\tilde{\gamma}_{1}^{2}= & \sigma^{2}y_{1}^{2}+\tau^{2}y_{2}^{2}+2\bm{\sigma\cdot\tau}y_{1}y_{2}\\
\tilde{\alpha}_{1}^{\prime2}+\tilde{\beta}_{1}^{\prime2}+\tilde{\gamma}_{1}^{\prime2}= & \sigma^{2}y_{1}^{\prime2}+\tau^{2}y_{2}^{\prime2}+2\bm{\sigma\cdot\tau}y_{1}^{\prime}y_{2}^{\prime}\end{align}

\noindent . Note that the solution $\delta_{2}=0$ is excluded except
in the case of a vacuum; this further excludes the definition \begin{equation}
a^{2}\equiv\left(\gamma_{11}\gamma_{22}\gamma_{33}\right)^{1/3}\end{equation}

\noindent as a useful definition of the scale factor. More careful
examination reveals that the definition

\begin{equation}
a^{2}\equiv\gamma_{ab}\eta^{ab}\end{equation}

\noindent is incompatible with the second-order equation as well.
When all removable perturbations are set to zero,

\begin{align}
\delta_{2}^{\mbox{non-removable}}= & \cos x\left(\csc x\right)^{\frac{3+3w}{1+3w}}\times\nonumber \\
 & \times\left\{ \int\left[\begin{array}{c}
\sigma^{2}\left(2y_{1}^{2}+\frac{1}{4}y_{1}^{\prime2}\right)+\\
+\tau^{2}\left(2y_{2}^{2}+\frac{1}{4}y_{2}^{\prime2}\right)+\\
+\bm{\sigma\cdot\tau}\left(4y_{1}y_{2}+\frac{1}{2}y_{1}^{\prime}y_{2}^{\prime}\right)\end{array}\right]\tan^{2}x\left(\sin x\right)^{\frac{2}{1+3w}}d\eta\right\} \label{eq:delta2nr}\end{align}

\noindent .%
\footnote{The Einstein equations for weak gravitational waves in a Bianchi IX
universe have the elegant feature of being integrable in closed form,
always reducible to functions form $\sin\left(n\eta\right)\csc^{k}\left(\eta\right)$
and $\cos\left(n\eta\right)\csc^{k}\left(\eta\right)$. Theoreticians
working in regimes of higher-frequency gravitational waves in a flat
background may find it felicitous to approximate a Euclidean universe
as a closed one in order to avoid mathematical inconveniences associated
with the function $\mbox{sinc}\left(t\right)$!%
} We will discuss solutions to this equation term-by-term, noting that
these terms can be solved entirely from information we obtained at
first order.%
\footnote{Li and Schwarz\cite{Li back-reaction} obtain a similar result for
a flat universe, but apply their results to a different domain. The
averaging scheme they propose is not an applicable approach for cosmological
gravitational waves. The result is generally stated in \cite[ss. 96]{Landau-Lifschitz}.%
}

\subsubsection{Contributions from the removable perturbations}

Contributions from the removable perturbations at second order have
the explicit forms:

\paragraph{In a radiation-dominated universe:}

\begin{equation}
\delta_{2}^{\mbox{removable}}=-\frac{c_{1}^{2}}{12}\left(4\sin^{2}\frac{\eta}{2}+\tan^{2}\frac{\eta}{2}+\cot^{2}\frac{\eta}{2}-2\right)\csc^{2}\eta+c_{2}\cot\eta\csc\eta\end{equation}

Note that the terms deriving from the first-order removable perturbation
diverge as $\mathcal{O}\left(\eta^{-4}\right)$, while those from
the second-order removable perturbation diverge more slowly, as $\mathcal{O}\left(\eta^{-2}\right)$.

\paragraph{In a matter-dominated universe:}

\begin{equation}
\delta_{2}^{\mbox{removable}}=-\frac{c_{1}^{2}}{12}\left(3\csc^{4}\frac{\eta}{2}+8\csc^{2}\frac{\eta}{2}-10\right)\csc^{2}\frac{\eta}{2}+c_{2}\cot\frac{\eta}{2}\csc^{2}\frac{\eta}{2}\end{equation}

\noindent . Similarly, terms deriving from the first-order perturbation
diverge as $\mathcal{O}\left(\eta^{-6}\right)$ and so at small $\eta$
will dominate terms deriving from the second-order removable perturbation
which diverges as $\mathcal{O}\left(\eta^{-3}\right)$.

\subsubsection{Contributions from the growing mode}

Contributions from the growing mode have the following forms:

\begin{equation}
\delta_{2}^{\mbox{growing}}=\sigma^{2}\cot x\left(\csc x\right)^{\frac{2}{1+3w}}\int\tan^{2}x\left(\frac{1}{4}y_{1}^{\prime2}+2y_{1}^{2}\right)\left(\sin x\right)^{\frac{2}{1+3w}}d\eta\end{equation}

\noindent . We can already discern that the sign on $\delta_{2}^{\mbox{growing}}$
must be positive in a young universe.

\paragraph{In a radiation-dominated universe:}

\begin{equation}
\delta_{2}^{\mbox{growing,radiation}}=\sigma_{\mbox{radiation}}^{2}\cot\eta\csc\eta\left(-\frac{1}{3}\cos3\eta+\frac{1}{5}\cos5\eta+2\sec\eta\right)\end{equation}

\noindent ; note the diverging contribution of $\mathcal{O}\left(\eta^{-2}\right)$
from growing modes.

\paragraph{In a matter-dominated universe:}

\begin{equation}
\delta_{2}^{\mbox{growing,matter}}=\sigma_{\mbox{matter}}^{2}\cot\frac{\eta}{2}\csc^{2}\frac{\eta}{2}\left(\begin{array}{c}
-\frac{6063}{4}\eta+\frac{13001}{8}\sin\eta-\frac{3237}{8}\sin2\eta+\\
+\frac{933}{8}\sin3\eta-33\sin4\eta+\\
+\frac{32}{5}\sin5\eta+900\tan\frac{\eta}{2}\end{array}\right)\end{equation}

\noindent . In contrast to the radiation-dominated case, the growing
mode's contribution does not diverge in a matter-dominated universe
(the term in brackets equals $0+\mathcal{O}\left(\eta^{5}\right)$).
Approximating to lowest orders in $\eta$,

\begin{equation}
\delta_{2}^{\mbox{growing,matter}}\approx\sigma_{\mbox{matter}}^{2}\left(245\eta^{2}-\frac{21641}{84}\eta^{4}\right)\end{equation}

.

\subsubsection{Contributions from the decaying mode}

\paragraph{In a radiation-dominated universe}

In a radiation-dominated universe, the functions $y_{1}$ and $y_{2}$
have the property

\begin{equation}
y_{1}^{2}+y_{2}^{2}=\csc^{2}\eta\end{equation}

\noindent while the functions $y_{1}^{\prime}$ and $y_{2}^{\prime}$
are similarly related by

\begin{equation}
y_{1}^{\prime2}+y_{2}^{\prime2}=\left(8\sin^{2}\eta+1\right)\csc^{4}\eta\end{equation}

\noindent . This simplifies calculations as we can readily say

\begin{equation}
\delta_{2}^{\mbox{decaying}}=\tau^{2}\cot\eta\csc\eta\left(\frac{17}{4}\sec\eta+\frac{1}{4}\ln\tan\frac{\eta}{2}\right)-\frac{\tau^{2}}{\sigma^{2}}\delta_{2}^{\mbox{growing}}\label{eq:delta 2 decaying}\end{equation}

\noindent ; in a universe old enough that the diverging terms are
negligible, the decaying mode intrinsically decreases the scale factor
in the same way that the growing mode intrinsically increases it.

\paragraph{In a matter-dominated universe}

In a matter-dominated universe,

\begin{equation}
y_{1}^{2}+y_{2}^{2}=\csc^{4}\frac{\eta}{2}\left(9+\frac{1}{4}\cot^{2}\frac{\eta}{2}\right)\end{equation}

\noindent and

\begin{equation}
y_{1}^{\prime2}+y_{2}^{\prime2}=\frac{1}{16}\csc^{8}\frac{\eta}{2}\left(-608\cos\eta+140\cos2\eta+477\right)\end{equation}

\noindent so we can state

\begin{equation}
\delta_{2}^{\mbox{decaying,matter}}=\tau^{2}\cos\frac{\eta}{2}\csc^{3}\frac{\eta}{2}\left(\begin{array}{c}
18\eta+2450\tan\frac{\eta}{2}-\\
-\frac{10705}{48}\cot\frac{\eta}{2}-\\
-\frac{577}{96}\cot\frac{\eta}{2}\csc^{2}\frac{\eta}{2}\end{array}\right)-\frac{\tau^{2}}{\sigma^{2}}\delta_{2}^{\mbox{growing}}\end{equation}

\noindent . It is interesting to note that, due to the growing mode
contribution's much slower contribution to change in the scale factor,
the impact of the decaying mode on the dynamics of a young universe
can be many orders of magnitude greater than the impact of the growing
mode even when the decaying mode is several orders of magnitude weaker
than the growing mode. The ratio

\begin{equation}
\left|\frac{\delta_{2}^{\mbox{decaying,matter}}}{\delta_{2}^{\mbox{growing,matter}}}\right|\approx\frac{\tau_{\mbox{matter}}^{2}}{\sigma_{\mbox{matter}}^{2}}\eta^{-8}\end{equation}

\noindent so in a matter-dominated universe with $\eta\approx10^{-1}$
the decaying mode will have a greater impact on cosmic dynamics as
long as $\tau_{\mbox{matter}}^{2}>10^{-8}\sigma_{\mbox{matter}}^{2}$.

\subsubsection{Contributions from the $\bm{\sigma\cdot\tau}$ term}

The contributions are described by the equation

\begin{equation}
\delta_{2}^{\mbox{mixed}}=\bm{\sigma\cdot\tau}\cos x\left(\csc x\right)^{\frac{3+3w}{1+3w}}\int\left(4y_{1}y_{2}+\frac{1}{2}y_{1}^{\prime}y_{2}^{\prime}\right)\tan^{2}x\left(\sin x\right)^{\frac{2}{1+3w}}d\eta\end{equation}

and have the following explicit forms:

\paragraph{Radiation-dominated universe}

In a radiation-dominated universe,

\begin{equation}
\delta_{2}^{\mbox{mixed,radiation}}=\frac{16}{15}\bm{\sigma\cdot\tau}_{\mbox{radiation}}\sin\eta\cos\eta\left(3\cos2\eta+2\right)\end{equation}

\noindent .

\paragraph{Matter-dominated universe}

\noindent In a matter-dominated universe,

\begin{equation}
\delta_{2}^{\mbox{mixed}}=\bm{\sigma\cdot\tau}\cot\frac{\eta}{2}\csc^{2}\frac{\eta}{2}\left(\begin{array}{c}
-4\cos\eta-24\cos^{2}\eta-\cos3\eta-\\
-\frac{15}{2}\cos4\eta+5\cos5\eta\end{array}\right)\end{equation}

\noindent .

\subsubsection{Gravitational waves at second order}

Turning now to the $R_{a}^{b}$ equations (\ref{eq:R11-2}, \ref{eq:R22-2},
\ref{eq:R33-2}), to second order, the Einstein equations for $\epsilon-p^{\left(a\right)}$-terms
read:

\begin{equation}
\left\{ \begin{array}{c}
\alpha_{2}^{\prime\prime}+2\cot x\alpha_{2}^{\prime}+8\alpha_{2}\\
+\frac{1}{4}\left(\alpha_{1}^{\prime2}+\beta_{1}^{\prime2}+\gamma_{1}^{\prime2}\right)\\
-3\left[\frac{w}{2}\left(1+w\right)\csc^{2}x+1\right]\delta_{2}\end{array}\right\} =\left\{ \begin{array}{c}
\left[\begin{array}{c}
-3+\frac{1}{16}\left(1+3w\right)^{2}\tan^{2}x+\\
+\frac{3}{16}\left(1+w\right)\left(3w-1\right)+\\
+\left(1+w\right)^{2}\left(\frac{9}{16}-\frac{3}{4}w\right)\csc^{2}x\end{array}\right]\delta_{1}^{2}-\\
-\alpha_{1}^{\prime}\delta_{1}^{\prime}+\left(6\beta_{1}^{2}-16\beta_{1}\gamma_{1}+6\gamma_{1}^{2}\right)-\\
-18\alpha_{1}^{2}+16\alpha_{1}\delta_{1}\end{array}\right\} \end{equation}

\noindent \emph{etc.} If we suppress all removable terms, as we must
for any practical observation of second-order terms, and taking into
account (\ref{eq:delta2nr}), this further simplifies to

\begin{equation}
\alpha_{2}^{\prime\prime}+2\cot x\alpha_{2}^{\prime}+8\alpha_{2}-3\left[\frac{w}{2}\left(1+w\right)\csc^{2}x+1\right]\delta_{2}=\left[\begin{array}{c}
-26\alpha_{1}^{2}+\\
+14\beta_{1}^{2}+14\gamma_{1}^{2}-\\
-\frac{1}{4}\left(\alpha_{1}^{\prime2}+\beta_{1}^{\prime2}+\gamma_{1}^{\prime2}\right)\end{array}\right]\label{eq:Einstein 2nd order}\end{equation}

\noindent . Recalling the form of the gravitational waves including
the removable perturbation at first order, make the simple transformation
$\alpha_{2}\rightarrow\tilde{\alpha}_{2}+\frac{1}{3}\delta_{2}$ to
arrive at the equations:

\begin{equation}
\tilde{\alpha}_{2}^{\prime\prime}+2\cot x\tilde{\alpha}_{2}^{\prime}+8\tilde{\alpha}_{2}=40\left[\frac{1}{3}\left(\alpha_{1}^{2}+\beta_{1}^{2}+\gamma_{1}^{2}\right)-\alpha_{1}^{2}\right]\label{eq:2nd order wave}\end{equation}

\noindent \emph{etc}; we recognize that linear-order gravitational
waves act as a driving force on the waves at quadratic order. The
solution of this equation is straightforward but tedious and we arrive
at the following solutions:

\paragraph{In a radiation-dominated universe}

\begin{equation}
\alpha_{2}^{\mbox{radiation}}=\left[\begin{array}{c}
C_{\alpha1,2}\frac{\sin3\eta}{\sin\eta}+C_{\alpha2,2}\frac{\cos3\eta}{\sin\eta}+\\
+40\left(\frac{1}{3}\sigma^{2}-C_{\alpha1,1}^{2}\right)\left(\frac{1}{36}\frac{\sin3\eta}{\sin\eta}-\frac{1}{6}\eta\frac{\cos3\eta}{\sin\eta}\right)+\\
+40\left(\frac{1}{3}\tau^{2}-C_{\alpha2,1}^{2}\right)\left(\begin{array}{c}
\frac{1}{6}\eta\frac{\cos3\eta}{\sin\eta}+\frac{1}{36}\frac{\sin3\eta}{\sin\eta}+\frac{5}{24}+\\
+\frac{1}{16}\frac{\sin5\eta}{\sin\eta}-\frac{1}{6}\frac{\left(2\eta-\pi\right)\cos3\eta-2\sin3\eta\ln\left(2\sin\eta\right)}{\sin\eta}\end{array}\right)+\\
+40\left(\frac{2}{3}\bm{\sigma\cdot\tau}-2C_{\alpha1,1}C_{\alpha2,1}\right)\left(\begin{array}{c}
\frac{1}{6}\eta\frac{\sin3\eta}{\sin\eta}+\frac{1}{8}\cot\eta+\\
+\frac{1}{36}\frac{\cos3\eta}{\sin\eta}-\frac{1}{32}\frac{\cos5\eta}{\sin\eta}\end{array}\right)\end{array}\right]+\frac{1}{3}\delta_{2}\end{equation}

\noindent \emph{etc. }with the second-order constants $C_{\alpha1,2}$
\emph{etc. }constrained such that

\begin{equation}
C_{\alpha1,2}+C_{\beta1,2}+C_{\gamma1,2}=C_{\alpha1,2}+C_{\beta1,2}+C_{\gamma1,2}=0\end{equation}
. To lowest order in $\eta$ the solution for $\alpha_{2}$ reads

\begin{equation}
\alpha_{2}^{\mbox{radiation}}\approx\left[\begin{array}{c}
C_{\alpha1,2}\left(3-4\eta^{2}\right)+20\left(\frac{1}{3}\sigma^{2}-C_{\alpha1,1}^{2}\right)\left(-\frac{1}{6}+\frac{11}{9}\eta^{2}\right)\\
+C_{\alpha2,2}\eta^{-1}+\frac{20\pi}{3}\left(\frac{1}{3}\tau^{2}-C_{\alpha2,1}\right)\eta^{-1}+\\
+\frac{175}{36}\left(\frac{2}{3}\bm{\sigma\cdot\tau}-2C_{\alpha1,1}C_{\alpha2,1}\right)\eta^{-1}+\frac{1}{3}\delta_{2}^{\mbox{non-removable}}\end{array}\right]\label{eq:alpha2 radiation}\end{equation}

\noindent \emph{etc}. For the pure decaying mode, the contribution
from $\delta_{2}$ dominates, while for the pure growing mode and
the mixed term the contributions from the homogeneous parts of $\alpha_{2}$
dominate.

\paragraph{In a matter-dominated universe}

For a matter-dominated universe, the gravitational wave equation to
second order has the following solution%
\footnote{There is no {}``royal road'' to the explicit statement of this function,
which was derived by substitution and variation of parameters with
the assistance of a computer algebra system. With foreknowledge of
the form of the solution, the equation (\ref{eq:2nd order wave})
can be solved through the method of undetermined coefficients; this
requires solving a 21-dimensional linear system. (\ref{eq:2nd order wave})
may also admit a solution through the method of Fourier transforms,
but only under torture.%
}:

\begin{equation}
\alpha_{2}=\left(\begin{array}{c}
C_{\alpha1,2}\csc\frac{\eta}{2}\frac{d}{d\eta}\frac{\sin3\eta}{\sin\eta/2}+C_{\alpha2,2}\csc\frac{\eta}{2}\frac{d}{d\eta}\frac{\cos3\eta}{\sin\eta/2}+\\
+\alpha_{2}^{\mbox{growing}}+\alpha_{2}^{\mbox{decaying}}+\alpha_{2}^{\mbox{mixed}}+\frac{1}{3}\delta_{2}^{\mbox{non-removable}}\end{array}\right)\label{eq:alpha2 matter}\end{equation}

\begin{equation}
\alpha_{2}^{\mbox{growing}}\equiv5\left(\frac{1}{3}\sigma^{2}-C_{\alpha1,1}^{2}\right)\left\{ \begin{array}{c}
\frac{1}{70}\sum_{n=0}^{10}g_{n}\cos n\eta+\\
+\frac{1}{56}\csc^{3}\frac{\eta}{2}\left[\begin{array}{c}
\eta\left(\begin{array}{c}
-1128960\cos\frac{5\eta}{2}+\\
+806400\cos\frac{7\eta}{2}\end{array}\right)+\\
+\sum_{n=0}^{11}h_{n}\sin\left(\frac{2n+1}{2}\eta\right)\end{array}\right]\end{array}\right\} \end{equation}

\begin{multline*}
g_{0}=32900,g_{1}=443310,g_{2}=90230,g_{3}=354221,g_{4}=20195,g_{5}=248918,\\
g_{6}=-57025,g_{7}=68911,g_{8}=-37880,g_{9}=15440,g_{10}=-22400\\
h_{0}=1166543,h_{1}=-1664285,h_{2}=888216,h_{3}=990580,h_{4}=-1262310,h_{5}=677390,\\
h_{6}=-363895,h_{7}=197841,h_{8}=-116900,h_{9}=66864,h_{10}=-34304,h_{11}=8960\end{multline*}

\begin{equation}
\alpha_{2}^{\mbox{growing}}\approx\left(\frac{1}{3}\sigma^{2}-C_{\alpha1,1}^{2}\right)\left(82630-\frac{4513087}{7}\eta^{2}\right)\label{eq:-159}\end{equation}

\begin{equation}
\alpha_{2}^{\mbox{decaying}}\equiv\frac{1}{245}\left(\frac{1}{3}\tau^{2}-C_{\alpha2,1}^{2}\right)\csc^{4}\frac{\eta}{2}\left(\begin{array}{c}
-\frac{\eta}{2}\tan\frac{\eta}{2}\sum_{n=0}^{4}j_{n}\cos^{n}\eta+\\
+\sum_{n=0}^{6}k_{n}\cos^{n}\eta+\\
+\ln\left(-2\sin^{2}\frac{\eta}{2}\right)\sum_{n=0}^{4}l_{n}\cos^{n}\eta\end{array}\right)\label{eq:a2 decaying}\end{equation}

\begin{multline*}
j_{0}=-34020,j_{1}=-17010,j_{2}=153090,j_{3}=22680,j_{4}=-113400\\
k_{0}=58329,k_{1}=-514422,k_{2}=368937,k_{3}=675396,\\
k_{4}=-678540,k_{5}=31500,k_{6}=61250\\
l_{0}=-5670,l_{1}=102060,l_{2}=-73710,l_{3}=-136080,l_{4}=113400\end{multline*}

\begin{equation}
\alpha_{2}^{\mbox{mixed}}\equiv\frac{4}{105}\left(\frac{1}{3}\bm{\sigma\cdot\tau}-C_{\alpha1,1}C_{\alpha2,1}\right)\csc^{2}\frac{\eta}{2}\left(\begin{array}{c}
\frac{\eta}{2}\sum_{n=0}^{3}m_{n}\cos\eta-\\
-\cot\frac{\eta}{2}\sum_{n=0}^{5}n_{n}\cos^{n}\eta\end{array}\right)\end{equation}

\begin{multline*}
m_{0}=2310,m_{1}=-39270,m_{2}=-9240,m_{3}=46200\\
n_{0}=-936,n_{1}=15693,n_{2}=30204,n_{3}=-58700,n_{4}=-25200,n_{5}=42000\end{multline*}

\[
\alpha_{2}^{\mbox{mixed}}\approx-\frac{32}{105}\left(\frac{1}{3}\bm{\sigma\cdot\tau}-C_{\alpha1,1}C_{\alpha2,1}\right)\eta^{-3}\sum_{n=0}^{5}n_{n}\]

\noindent \emph{etc}. The statement of the solutions to the gravitational
wave equations to quadratic order in the matter-dominated universe
are original to this work; the radiation-dominated quadratic order
wave equations were presented in \cite{GDI}. Note that $\sum_{n}l_{n}=\sum_{n}m_{n}=0$.

Most interesting is the presence of ln-terms in (\ref{eq:delta 2 decaying})
and (\ref{eq:a2 decaying}), which on the one hand indicate the appearance
of the power-law behavior of metric coefficients which typify the
Kasner universe and the BKL universe in its quasi-isotropic phase;
but which on the other hand show the breakdown of our approximation
scheme and the limit of regular perturbation theory in solving the
problem to hand; the Taylor expansion of the growing mode of $\alpha_{2}$
indicates further that waves must be very weak ($\left\Vert \sigma\right\Vert =\mathcal{O}\left(10^{-4}\right)$)
for the approximation scheme to be rigorously valid. In any case,
indications are that the growing mode of hypothetical cosmological
gravitational waves should be very much stronger than the decaying
mode (see \noun{section} \ref{sub:Acceleration-in-the}); we will
not need to make use of the second-order solutions for the decaying
mode and from here on will treat the decaying mode as being linear-order
weak, that is, $C_{\alpha2,1}^{2}\approx C_{\beta2,1}^{2}\approx C_{\gamma2,1}^{2}\approx C_{\alpha2,2}\approx C_{\beta2,2}\approx C_{\gamma2,2}\approx\tau^{2}\approx0$.

\subsection{Strong growing waves in the quasi-isotropic regime\label{sub:Strong-growing-waves}}

\cite[part 3]{GDI} begins the development of equations for a radiation-dominated
universe describing strong gravitational waves in Bianchi IX. Similar
equations in a matter-dominated universe are useful in considering
observed acceleration, as $\Delta Q\approx-1$.

Consider the equations (\ref{eq:Einstein EQ 0}-\ref{eq:Einstein EQ 3}).
Assume a solution of the form

\begin{align}
\alpha= & \sum_{n=0}^{\infty}c_{2n}^{\alpha}\eta^{2n}\nonumber \\
\beta= & \sum_{n=0}^{\infty}c_{2n}^{\beta}\eta^{2n}\label{eq:series}\\
\gamma= & \sum_{n=0}^{\infty}c_{2n}^{\gamma}\eta^{2n}\nonumber \end{align}

\noindent with the terms $c_{n}^{\xi}$ constants. It is convenient
to define $e^{2c_{0}^{\alpha}}\equiv A,e^{2c_{0}^{\beta}}\equiv B,e^{2c_{0}^{\gamma}}\equiv G$.
In a matter-dominated universe, to lowest two orders the solutions
read

\begin{align}
\alpha\approx & c_{0}^{\alpha}+\frac{1}{20}\left[1-\frac{1}{ABG}\left(5A^{2}-3B^{2}-3G^{2}+6BG-2AB-2AG\right)\right]\eta^{2}\nonumber \\
\beta\approx & c_{0}^{\beta}+\frac{1}{20}\left[1-\frac{1}{ABG}\left(5B^{2}-3G^{2}-3A^{2}+6AG-2BG-2AB\right)\right]\eta^{2}\label{eq:quasi-isotropic}\\
\gamma\approx & c_{0}^{\gamma}+\frac{1}{20}\left[1-\frac{1}{ABG}\left(5G^{2}-3A^{2}-3B^{2}+6AB-2AG-2BG\right)\right]\eta^{2}\nonumber \end{align}

\noindent where $c_{0}^{\alpha},c_{0}^{\beta},c_{0}^{\gamma}$ are
arbitrary; if we want to preserve the Friedmannian value of \emph{S}
then we need \begin{equation}
c_{0}^{\alpha}+c_{0}^{\beta}+c_{0}^{\gamma}=0\label{eq:strong normalization}\end{equation}
\cite{GDI}. We always have the freedom to set one of these to zero
by a simple scaling of the metric; this preserves the two degrees
of freedom for the gravitational wave.

If we apply the condition (\ref{eq:strong normalization}) and set
the parameter $c_{0}^{\gamma}=0$ by scaling, then the strong growing-mode
waves are described by

\begin{align}
c_{0}^{\alpha}\in & \mathbb{R}\label{eq:normalized 1}\\
c_{0}^{\beta}= & -c_{0}^{\alpha}\\
c_{0}^{\gamma}= & 0\\
c_{2}^{\alpha}= & \frac{1}{20}\left(-5A^{2}+2A+6-6A^{-1}+3A^{-2}\right)\\
c_{2}^{\beta}= & \frac{1}{20}\left(3A^{2}-6A+6+2A^{-1}-5A^{-2}\right)\\
c_{2}^{\gamma}= & \frac{1}{20}\left(3A^{2}+2A-10+2A^{-1}+3A^{-2}\right)\label{eq:normalized 6}\end{align}

with the single parameter $c_{0}^{\alpha}$ determining the whole
system. Note that setting $c_{0}^{\gamma}=0$ does not imply $\gamma^{\prime}=0$.
We can also qualitatively say that for any value of \emph{A}, two
of functions $\alpha,\beta,\gamma$ will be positive, as will $\delta$,
unless $A=1$ (the background case), in the regime that $A\eta$ is
sufficiently small that $A^{3}\eta^{3}$ is negligible.

The functions (\ref{eq:series}) are linearly independent with $y_{2}^{\mbox{matter}}$
to lowest order in $\eta$ and therefore can be used together to describe
a matter-dominated universe with arbitrarily strong growing gravitational
waves and weak decaying gravitational waves up to order $\eta^{2}$,
as long as the series (\ref{eq:series}) converge.

\subsection{Dynamics\label{sub:Dynamics}}

As in the Kasner universe (see \noun{section} \ref{sub:Scale-factor}),
it is useful to generalize quantities pertaining to the expansion
of space which are spherically symmetric in Friedmannian cosmology.

In terms of our statement of the metric (\ref{eq:metric}), the generalized
dynamical quantities for our space are

\begin{equation}
a_{ab}=a_{F}\left(\begin{array}{ccc}
e^{\alpha} & 0 & 0\\
0 & e^{\beta} & 0\\
0 & 0 & e^{\gamma}\end{array}\right)\end{equation}

\begin{equation}
\bar{a}=\frac{1}{3}a_{F}\left(e^{\alpha}+e^{\beta}+e^{\gamma}\right)\end{equation}

\begin{equation}
H_{ab}=\left(\begin{array}{ccc}
\dot{a}_{F}/a_{F}+\dot{\alpha} & 0 & 0\\
0 & \dot{a}_{F}/a_{F}+\dot{\beta} & 0\\
0 & 0 & \dot{a}_{F}/a_{F}+\dot{\gamma}\end{array}\right)\label{eq:generalized H}\end{equation}

\begin{equation}
\bar{H}=\frac{\dot{a}_{F}}{a_{F}}+\frac{1}{3}\dot{\delta}\end{equation}

\begin{equation}
Q_{1}^{1}\equiv\frac{d}{dt}H^{1c}\eta_{1c}-\delta_{1}^{1}=-\frac{\left(\begin{array}{c}
\ddot{a}_{F}/a_{F}+2H_{F}\dot{\alpha}+\\
+\ddot{\alpha}+\dot{\alpha}^{2}\end{array}\right)}{\left(H_{F}+\dot{\alpha}\right)^{2}}\label{eq:generalized Q}\end{equation}

\emph{etc.}

\begin{equation}
\bar{Q}=-\frac{1}{3}\left(\begin{array}{c}
\frac{\ddot{a}_{F}/a_{F}+2H_{F}\dot{\alpha}+\ddot{\alpha}+\dot{\alpha}^{2}}{\left(H_{F}+\dot{\alpha}\right)^{2}}+\\
+\frac{\ddot{a}_{F}/a_{F}+2H_{F}\dot{\beta}+\ddot{\beta}+\dot{\beta}^{2}}{\left(H_{F}+\dot{\beta}\right)^{2}}+\\
+\frac{\ddot{a}_{F}/a_{F}+2H_{F}\dot{\gamma}+\ddot{\gamma}+\dot{\gamma}^{2}}{\left(H_{F}+\dot{\gamma}\right)^{2}}\end{array}\right)\end{equation}

\noindent . Our goal in undertaking the arduous task of solving the
Einstein equations has been to derive the impact of long-wavelength
gravitational waves on cosmic dynamics, particularly acceleration.
We are now in a position to begin to discuss this impact.

Let each quantity in section (\ref{sub:Dynamics}) be expanded out
into a background term plus corrections, such that for example

\begin{equation}
a_{ab}\approx a_{ab}^{\left(0\right)}+a_{ab}^{\left(1\right)}+a_{ab}^{\left(2\right)}\end{equation}

\noindent . Then the zero-order, background terms are simply\begin{align}
a_{ab}^{\left(0\right)}= & a_{F}\eta_{ab}\\
H_{ab}^{\left(0\right)}= & H_{F}\eta_{ab}\\
\phantom{}^{\left(0\right)}Q_{a}^{b}= & Q_{F}\delta_{a}^{b}\end{align}

\noindent . While the gravitational energy-momentum tensor vanishes
at first order with the removal of removable perturbations, the presence
of weak gravitational waves can affect observed dynamic quantities.
At first order:

\begin{equation}
a_{ab}^{\left(1\right)}=\left(\begin{array}{ccc}
\alpha_{1} & 0 & 0\\
0 & \beta_{1} & 0\\
0 & 0 & \gamma_{1}\end{array}\right)\end{equation}

\begin{equation}
\bar{a}_{\left(1\right)}=\frac{1}{3}\delta_{1}\end{equation}

\begin{equation}
H_{ab}^{\left(1\right)}=\left(\begin{array}{ccc}
\dot{\alpha}_{1} & 0 & 0\\
0 & \dot{\beta}_{1} & 0\\
0 & 0 & \dot{\gamma}_{1}\end{array}\right)\end{equation}

\begin{equation}
\bar{H}_{\left(1\right)}=\frac{1}{3}\dot{\delta}_{1}\end{equation}

\begin{equation}
\phantom{}^{\left(1\right)}Q_{1}^{1}=-H_{F}^{-1}\left[2\left(Q_{F}+1\right)\dot{\alpha}_{1}+H_{F}^{-1}\ddot{\alpha}_{1}\right]\end{equation}

\emph{etc.}, 

\begin{equation}
\bar{Q}_{\left(1\right)}=-\frac{1}{3}H_{F}^{-1}\left[2\left(Q_{F}+1\right)\dot{\delta}_{1}+H_{F}^{-1}\ddot{\delta}_{1}\right]\end{equation}

\noindent Thus we illustrate the need for truly representative sky
coverage in considering the problem of acceleration: gravitational
waves can contribute to anisotropic acceleration even when they do
not affect the distribution of matter. In domains when the first derivatives
of a wave is small (that is, near peaks and troughs of the wave),
the accelerative effect will not be accompanied by a large change
in the Hubble flow. As before, a failure to completely suppress the
removable perturbation may lead to incorrect evaluation of the strength
of decaying modes. To first order, non-zero contribution to the average
over the whole sky of the perturbations is removable; first-order
weak gravitational waves in Bianchi IX do not produce isotropic acceleration.

To quadratic order, the dynamic quantities have the forms

\begin{equation}
a_{ab}^{\left(2\right)}=a_{F}\left(\begin{array}{ccc}
\alpha_{2}+\alpha_{1}^{2}/2 & 0 & 0\\
0 & \beta_{2}+\beta_{1}^{2}/2 & 0\\
0 & 0 & \gamma_{2}+\gamma_{1}^{2}/2\end{array}\right)\end{equation}

\begin{equation}
\bar{a}_{2}=\frac{1}{3}a_{F}\left[\delta_{2}+\frac{1}{2}\left(\alpha_{1}^{2}+\beta_{1}^{2}+\gamma_{1}^{2}\right)\right]\end{equation}

\begin{equation}
H_{ab}^{\left(2\right)}=\left(\begin{array}{ccc}
\dot{\alpha}_{2} & 0 & 0\\
0 & \dot{\beta}_{2} & 0\\
0 & 0 & \dot{\gamma}\end{array}\right)\end{equation}

\begin{equation}
\bar{H}_{\left(2\right)}=\frac{1}{3}\dot{\delta}_{2}\end{equation}

\begin{equation}
\phantom{}^{\left(2\right)}Q_{1}^{1}=-H_{F}^{-1}\left[\begin{array}{c}
2\left(Q_{F}+1\right)\dot{\alpha}_{2}+H_{F}^{-1}\ddot{\alpha}_{2}-\\
-3H_{F}^{-1}\left(Q_{F}+1\right)\dot{\alpha}_{1}^{2}-2H_{F}^{-2}\dot{\alpha}_{1}\ddot{\alpha}_{1}\end{array}\right]\end{equation}

\noindent \emph{etc.,}

\begin{equation}
\bar{Q}_{\left(2\right)}=-\frac{1}{3}H_{F}^{-1}\left[\begin{array}{c}
2\left(Q_{F}+1\right)\dot{\delta}_{2}+H_{F}^{-1}\ddot{\delta}_{2}-\\
-3H_{F}^{-1}\left(Q_{F}+1\right)\left(\dot{\alpha}_{1}^{2}+\dot{\beta}_{1}^{2}+\dot{\gamma}_{1}^{2}\right)-\\
-2H_{F}^{-2}\left(\dot{\alpha}_{1}\ddot{\alpha}_{1}+\dot{\beta}_{1}\ddot{\beta}_{1}+\dot{\gamma}_{1}\ddot{\gamma}_{1}\right)\end{array}\right]\end{equation}

\noindent . At second order we begin to see a consequence of the non-linearity
of the Bianchi IX Einstein equations which is potentially very important
in the study of cosmic dynamics: isotropic changes to the Hubble parameter
and to acceleration from anisotropic metric terms. With our knowledge
of the Einstein equations at first and second order (\ref{eq:Einstein 1st order},\ref{eq:delta equation},\ref{eq:Einstein 2nd order})
we can show this explicitly:

\begin{equation}
\phantom{}^{\left(2\right)}Q_{1}^{1}=-\tan x\left\{ \begin{array}{c}
\left(3w+\left(1+3w\right)\tan^{2}x\right)\alpha_{2}^{\prime}-8\tan x\alpha_{2}-\\
-40\tan x\alpha_{1}^{2}+\frac{3}{2}\left(1-3w-\left(1+3w\right)\tan^{2}x\right)\tan x\alpha_{1}^{\prime2}+\\
+16\tan^{2}x\alpha_{1}^{\prime}\alpha_{1}+\\
+\tan x\left[\begin{array}{c}
3\left[\frac{w}{2}\left(1+w\right)\csc^{2}x+1\right]\delta_{2}+\\
+14\left(\alpha_{1}^{2}+\beta_{1}^{2}+\gamma_{1}^{2}\right)-\frac{1}{4}\left(\alpha_{1}^{\prime2}+\beta_{1}^{\prime2}+\gamma_{1}^{\prime2}\right)\end{array}\right]\end{array}\right\} \end{equation}

\noindent \emph{etc. }and

\begin{equation}
\bar{Q}_{\left(2\right)}=\frac{1}{3}\tan^{2}x\left\{ \begin{array}{c}
\frac{1}{2}\left(1+3w\right)^{2}\sec^{2}x\delta_{2}-\\
-2\left(1+3w\right)\sec^{2}x\left(\alpha_{1}^{2}+\beta_{1}^{2}+\gamma_{1}^{2}\right)+\\
+\frac{1}{4}\left[1+15w+5\left(1+3w\right)\tan^{2}x\right]\left(\alpha_{1}^{\prime2}+\beta_{1}^{\prime2}+\gamma_{1}^{\prime2}\right)-\\
-16\tan x\left(\alpha_{1}^{\prime}\alpha_{1}+\beta_{1}^{\prime}\beta_{1}+\gamma_{1}^{\prime}\gamma_{1}\right)\end{array}\right\} \end{equation}

\noindent . Isotropic acceleration with quadratic-order strength arises
from the non-linear interaction of linear-order gravitational waves;
but in the regime of $\left|\alpha\right|,\left|\beta\right|,\left|\gamma\right|\ll1$
the gravitational waves at linear order will dominate measurement
of cosmological parameters. 

In a matter-dominated universe with $\eta$ small the deceleration
terms become, defining

\begin{equation}
\Delta Q_{b}^{a}\equiv Q_{b}^{a}-Q_{F}\delta_{b}^{a}\end{equation}

\begin{equation}
\Delta\bar{Q}\equiv\bar{Q}-Q_{F}\end{equation}

\begin{equation}
\Delta Q_{1,\mbox{matter}}^{1}\approx-\frac{\eta^{2}}{4}\left\{ \begin{array}{c}
\tan\frac{\eta}{2}\left(\alpha_{1}^{\prime}+\alpha_{2}^{\prime}\right)-8\left(\alpha_{1}+\alpha_{2}\right)-\\
-40\alpha_{1}^{2}+\frac{3}{2}\alpha_{1}^{\prime2}+16\tan\frac{\eta}{2}\alpha_{1}^{\prime}\alpha_{1}+\\
+3\delta_{2}+14\left(\alpha_{1}^{2}+\beta_{1}^{2}+\gamma_{1}^{2}\right)-\frac{1}{4}\left(\alpha_{1}^{\prime2}+\beta_{1}^{\prime2}+\gamma_{1}^{\prime2}\right)\end{array}\right\} \end{equation}

\begin{equation}
\Delta\bar{Q}^{\mbox{matter}}\approx\frac{1}{48}\eta^{2}\left[\begin{array}{c}
2\delta_{2}-8\left(\alpha_{1}^{2}+\beta_{1}^{2}+\gamma_{1}^{2}\right)+\\
+\left(\alpha_{1}^{\prime2}+\beta_{1}^{\prime2}+\gamma_{1}^{\prime2}\right)-\\
-16\tan\frac{\eta}{2}\left(\alpha_{1}^{\prime}\alpha_{1}+\beta_{1}^{\prime}\beta_{1}+\gamma_{1}^{\prime}\gamma_{1}\right)\end{array}\right]\end{equation}

\noindent \emph{etc.} Explicitly, these will have the lowest-order
forms:

\begin{equation}
\Delta Q_{1,\mbox{matter}}^{1}\approx-\frac{\eta^{2}}{4}\left(\begin{array}{c}
C_{\alpha1,1}\left(280-259\eta^{2}\right)+\\
+C_{\alpha2,1}\left(16\eta^{-3}-72\eta^{-1}+\frac{2251}{5}\eta\right)+\\
+C_{\alpha1,1}^{2}\left(710040-4687753\eta^{2}\right)+\\
+\sigma^{2}\left(-609590/3+\frac{4402405}{3}\eta^{2}\right)\end{array}\right)\end{equation}

\begin{equation}
\Delta\bar{Q}^{\mbox{matter}}\approx\frac{\sigma^{2}}{24}\eta^{2}\left(-4900+2983\eta^{2}\right)\end{equation}

\noindent . These results are encouraging as if we choose $\left\Vert \sigma\right\Vert \sim10^{-4}$
(in order to make the gravitational waves weak) and $\eta\sim10^{-2}$to
match (\ref{sub:Cosmological-parameters}), we obtain $\Delta Q_{1,\mbox{matter}}^{1}\sim-10^{-6}$,
which has the right sign as well as all the contributions at both
first and second orders going in the {}``right'' direction, toward
acceleration. It is particularly encouraging that both growing and
decaying modes contribute to acceleration to their lowest orders in
$\eta$.

\subsection{Back-reaction}

Of interest in discussing the problem of acceleration is the effective
equation of state of the gravitational waves' contribution to the
energy density. Empirically, the equation of state of dark energy
seems to be close to $w_{X}=-1$ (see \noun{section} \ref{sub:Analysis}),
where the quantity $w_{x}$ is related to the source of the energy
such that the source evolves with regard to the scale factor at a
rate of $a^{-3\left(1+w_{X}\right)}$. As noted in (\noun{section}
\ref{sub:Scale-factor}) there is no unique way to define the scale
factor; but a condition of quasi-isotropy is that expansion in every
direction in the current epoch is proportional, that is, that they
evolve as the same power of time. If the decaying mode of the cosmological
gravitational wave is weak, then this evolution will be proportional
to the Friedmannian scale factor.

To quadratic order, (\ref{eq:epsilong}) reads

\begin{equation}
k\epsilon_{g}^{\left(2\right)}=3\left(1+w\right)a_{F}^{-2}\csc^{2}x\delta_{2}\end{equation}

\noindent and so by (\ref{eq:backreaction})

\begin{equation}
q_{\left(2\right)}=-\left(1+w\right)a_{F}^{-2}\delta_{2}\end{equation}

\noindent . When the growing mode is dominant, $\delta_{2}$ is always
positive in a matter-dominated universe, and therefore $q_{\left(2\right)}$
is negative and thus the back-reaction appears to have negative energy
density; a significant {}``mixed'' $\bm{\sigma\cdot\tau}$ term,
however, can easily introduce intervals where $q_{\left(2\right)}>0$.

In a matter-dominated universe and when the growing mode is dominant,
$q_{\left(2\right)}\propto\eta^{-2}$ which, if the universe is evolving
with a scale as $a_{F}\propto\eta^{2}$, implies an equation of state
for the back-reaction of $w_{X}=-1/3$ (as compared to an equation
of state for a cosmological constant of $w_{X}=-1$). While no investigation
of the equation of state of dark energy includes this value within
its highest confidence interval, measurements of $w_{X}$ remain tentative,
with large errors and high sensitivity both to single data points
and to the algorithm for curve-fitting models to the data (see \noun{section}
\ref{sub:Analysis}).

The dominant term in (\ref{eq:epsilong}) is the $a_{F}^{\prime}/a_{F}$-term.
This stands in stark contrast to the commonly-considered case of gravitational
waves in a background so slowly moving compared to the period of the
waves that $\dot{a}_{F}\approx0$, in which instance the quadratic
combination of first-derivative terms dominates.

In regimes of stronger growing-mode gravitational waves, though, the
scale factor as defined in (\ref{eq:generalized scale factor}) will
be more dominated by terms of higher, even order and so $a_{ab}\propto\eta^{4}$
or higher. As the growing mode increases in strength, the equation
of state decreases asymptotically toward a limit of $w_{X}=-1$; if
the scale factor grows as $\eta^{2s}$, the equation of state for
the back-reaction is given by $w_{X}=\left(1/3s\right)-1$. As acceleration
is empirically $Q_{0}=-0.6$, this implies that in real life the gravitational
wave strength is of order unity and therefore the effective equation
of state is close to $-1$. Thus, the quasi-isotropic Bianchi IX model
with strong growing-mode gravitational waves and weak or zero decaying-mode
waves is compatible with the observed data on the equation of state
of dark energy, without the invocation of a cosmological constant;
the theory would be invalidated by definitive measurements of $w_{X}<-1$.

In any case, the fact of $w_{X}<0$ allows us to draw a conclusion
regarding cosmic evolution. \cite{MTW} notes Kasner-like cosmologies
go through two stages of evolution:
\begin{enumerate}
\item A {}``vacuum'' stage, where matter's influence is, due to its evolution
as $a^{-4}$, weak compared to the influence of the anisotropic expansion
and contraction, influence which, in light of (\noun{section} \ref{sub:Gravitational-wave-nature}),
we now understand to be the result of gravitational waves in the BKL
universe;
\item a {}``matter'' stage, where expansion isotropizes\cite{Tsagas}
and is driven by, first relativistic ($w=1/3$), then cold, non-relativistic
($w=0$) matter. Formally, the contribution of curvature to cosmic
evolution becomes important in this era ($w_{K}=-1/3$), but as the
influence of curvature will be isotropic in Bianchi IX and the radius
of curvature is very large compared to the Hubble radius (see \noun{section}
\ref{sub:Cosmological-parameters}), curvature will not have a practical
influence on observations in and of itself.%
\footnote{Formally we can also say that, due to the action of proton decay and
positron annihilation, after sufficient time the $w=0$ phase will
return to a $w=1/3$ phase where the universe is filled with neutrinos
and photons. Following this period there will be another return to
$w=0$ as these free particles are absorbed by black holes. As these
black holes evaporate by the process of Hawking radiation, there will
then be a final return to $w=1/3$. \cite{Five Ages} gives a popular-science
presentation of the universe in these phases, but as it was written
only shortly after the discovery of acceleration its treatment of
dark energy is highly speculative.%
} To this second stage we can add a third stage:
\item A {}``dark energy'' stage, in which growing modes of the cosmological
gravitational waves which drove the initial isotropy return as the
dominant influence on cosmic evolution.
\end{enumerate}

\subsection{Amplification of gravitational waves\label{sub:Amplification-of-gravitational}}

Grishchuk observed\cite{Grishchuk 1974} that when the background
of a cosmology containing gravitational waves varies rapidly, weak
gravitational waves can be amplified where they would otherwise, in
a slowly-moving background, decay rapidly\cite{Lifshitz 46}. With
regard to the Bianchi IX cosmology this is significant as when the
growing mode of a cosmological gravitational wave dominates, the leading
term in the gravitational energy density is of the form $\left(a_{F}^{\prime}/a_{F}\right)\delta_{2}^{\prime}=\mathcal{O}\left(\mbox{constant}\right)$.
Cosmological observations (see \noun{section} \ref{sub:Cosmological-parameters})
indicate the universe has $\eta<\mathcal{O}\left(10^{-1}\right)$.
In this regime, the term $a_{F}^{\prime}/a_{F}=\cot\left(\eta/2\right)\approx2/\eta$,
which is dependent on the rate of change of the background, is arbitrarily
large; therefore, weak waves may have an effect orders of magnitude
greater than their amplitude. Similarly, the decaying mode of gravitational
waves can have prominent or even dominant power in a sufficiently
young universe even when the amplitude of the decaying mode is smaller
than that of the growing mode.

\section{Conclusions}

Solutions have been presented for the gravitational wave equation
for a Bianchi IX universe perturbed to quadratic order from the closed
Friedmann case. Quadratic order is the limit of perturbation theory's
applicability to explore nearly-Friedmannian Bianchi IX when decaying
modes are sufficiently strong that they are not negligible.

At quadratic order, the non-linear interaction of the gravitational
waves produces isotropic changes to dynamic quantities. While this
isotropic change is likely to be dominated in any particular direction
by linear-order contributions from the gravitational waves, in the
regime of strong gravitational waves they will become more important
and potentially even dominant. Where \cite{Hobill Guo} discussed
the possibility of acceleration in a non-vacuum Bianchi IX universe
only qualitatively, we have shown it explicitly as well as illustrating
a clear link between acceleration and the gravitational waves which
are intrinsic to Bianchi IX in its full generality.

It is curious to note that the order-$\eta^{2}$ approximation we
have made in (\noun{section} \ref{sub:Strong-growing-waves}), $\alpha$
and $\delta$ in the normalization we have chosen take the form of
Alexander polynomials\cite{Alexander,Kawauchi}, although not Alexander
polynomials for any knot of fewer than 11 crossings. Whether this
mathematical observation is significant or coincidental is a subject
for further debate, but as gravitational waves in Bianchi IX are moving
equatorially around our background 3-sphere\cite{GDI}, and as a sub-class
of knots (the {}``torus knots'') are constructed by wrapping one
2-torus around another it is conceivable there could be a connection.

Back-reaction from growing modes of the gravitational waves appears
to have negative energy density and an equation of state compatible
with that observed for dark energy, especially in the regime of strong
gravitational waves and quasi-isotropic expansion; when gravitational
waves are strong, they become the dominant contributor to the evolution
of the cosmos in an era following the era of matter domination.

Therefore, from the perspective of cosmic dynamics, cosmological gravitational
waves in a quasi-isotropic Bianchi IX universe are a viable candidate
for dark energy, without the invocation of a cosmological constant
and without requiring any modification of the theory of relativity.
An analysis of the impact of these gravitational waves on the cosmic
microwave background is necessary in order to determine whether constraints
from the CMB are compatible with the observed data on acceleration.

\part{The Cosmic Microwave Background of a Bianchi IX universe\label{sec:Impact-on-the}}

While long-wavelength gravitational waves can cause acceleration in
a Bianchi IX universe, the effect of such waves must be compatible
with the observed cosmic microwave background in order to represent
a practical model for explaining observed acceleration.

Sachs \& Wolfe initiated\cite{Sachs-Wolfe} the systematic study of
the effect of perturbations on the CMB, following a formalism developed
by Kristian \& Sachs\cite{Kristian & Sachs}. Sachs \& Wolfe's work
developed the theory of scalar, vector and tensor perturbations on
the CMB in a flat almost-isotropic universe to first order.

Sachs \& Wolfe's work was generalized by Anile \& Motta\cite{Anile Motta S-W}
to the almost-isotropic closed and open Friedmann cosmologies, again
at first order. While Anile \& Motta begin to consider the impact
of long-wavelength gravitational waves on the CMB, they choose to
explore the impact of waves with scales much smaller than the Hubble
radius. Anile \& Motta subsequently\cite{Anile Motta waves} ruled
out the existence of these waves at significant strengths in the observable
universe.

Doroshkevitch, Lukash \& Novikov considered the impact of an anisotropic
universe on the CMB in the case of the Bianchi VII, VIII and IX models\cite{DLN anisotropy},
and concluded that a Bianchi IX model was potentially {}``compatible
with observations, only if there was some secondary heating of the
intergalactic gas''. Doroshkevitch \emph{et al}'s most important
calculations are carried out on the assumption, then widespread, of
$\Omega_{M}\approx1$ and as such are of limited applicability; interestingly,
though, in their conclusions they note that if $\Omega_{M}<1$, {}``$\Delta T/T$
will be close to the maximum value only in a small 'spot' with an
angular size $\theta\approx4\Omega$'' (where by {}``small'' they
give the example of $\Omega_{M}\approx0.1\implies\theta\approx23^{\circ}$). 

Sung \& Coles analytically and computationally explore the impact
of various unperturbed Bianchi models, including Bianchi IX, on the
CMB\cite{Sung & Coles}. They report the useful theorem that {}``a
gravitational field alone is not able to generate polarization'',
but do not consider the general case of Bianchi IX, only the isotropic
case equivalent to the closed Friedmann universe.

\section{Geodesic equations}

The effect of the metric on the CMB is determined by examining the
change in geodesics of light rays relative to an isotropic, background
case. Let the subscript \emph{E }denote a function evaluated at the
time of the emission of a photon, and the subscript \emph{R} denote
that function evaluated at the time of the photon's reception. Then
the change in the temperature of the background radiation \emph{T}
is given by

\begin{equation}
T_{R}/T_{E}=\frac{1}{z+1}\end{equation}

\noindent . Consider the path of a light ray; let this be a four-vector
denoted by $k^{\mu}$ such that $k^{\mu}k_{\mu}=0$, with the light
ray received in the direction $k_{R}^{i}=e^{i}$. The geodesic equation
for the time part of $k^{\mu}$ in a Bianchi cosmology reads

\begin{equation}
\frac{dk^{0}}{d\lambda}+\Gamma_{\, ij}^{0}k^{i}k^{j}=0\label{eq:full time geodesic}\end{equation}

\noindent and the equations for the space part of the vector read

\begin{equation}
\frac{dk^{a}}{d\lambda}+\Gamma_{\,00}^{a}+\Gamma_{\,0i}^{a}k^{i}+\Gamma_{\, i0}^{a}k^{i}+\Gamma_{\, bc}^{a}k^{b}k^{c}=0\end{equation}
Recalling (\ref{eq:co-moving}) and (\ref{eq:metric}) the Christoffel
symbols

\begin{equation}
\Gamma_{\, ij}^{0}=\frac{1}{2}\gamma_{ab,0}e_{i}^{a}e_{j}^{b}\end{equation}
, $\Gamma_{\,00}^{a}=\Gamma_{\,0i}^{a}=\Gamma_{\, i0}^{a}=0$ and
the Ricci rotation coefficients read%
\footnote{The symbol $\varepsilon_{abc}$ represents the Levi-Civita symbol
defined such that $\varepsilon_{123}=1$%
}:

\begin{align}
\Gamma_{\, bc}^{a}= & \frac{1}{2}\left(\delta_{f}^{a}\epsilon_{bcd}+\gamma^{ag}\gamma_{cd}\epsilon_{gbf}-\gamma^{ag}\gamma_{db}\epsilon_{cgf}\right)\eta^{df}\nonumber \\
\Gamma_{\,23}^{1}= & \frac{1}{2}\left(\gamma^{11}\left(\gamma_{33}-\gamma_{22}\right)+1\right)= & \frac{1}{2}\left(e^{2\gamma-2\alpha}-e^{2\beta-2\alpha}+1\right)\nonumber \\
\Gamma_{\,32}^{1}= & \frac{1}{2}\left(\gamma^{11}\left(\gamma_{33}-\gamma_{22}\right)-1\right)= & \frac{1}{2}\left(e^{2\gamma-2\alpha}-e^{2\beta-2\alpha}-1\right)\nonumber \\
\Gamma_{\,31}^{2}= & \frac{1}{2}\left(\gamma^{22}\left(\gamma_{11}-\gamma_{33}\right)+1\right)= & \frac{1}{2}\left(e^{2\alpha-2\beta}-e^{2\gamma-2\beta}+1\right)\\
\Gamma_{\,13}^{2}= & \frac{1}{2}\left(\gamma^{22}\left(\gamma_{11}-\gamma_{33}\right)-1\right)= & \frac{1}{2}\left(e^{2\alpha-2\beta}-e^{2\gamma-2\beta}-1\right)\nonumber \\
\Gamma_{\,12}^{3}= & \frac{1}{2}\left(\gamma^{33}\left(\gamma_{22}-\gamma_{11}\right)+1\right) & \frac{1}{2}\left(e^{2\beta-2\gamma}-e^{2\alpha-2\gamma}+1\right)\nonumber \\
\Gamma_{\,21}^{3}= & \frac{1}{2}\left(\gamma^{33}\left(\gamma_{22}-\gamma_{11}\right)-1\right)= & \frac{1}{2}\left(e^{2\beta-2\gamma}-e^{2\alpha-2\gamma}-1\right)\nonumber \end{align}

\noindent with all others zero; note that the form of the rotation
coefficients guarantees that only anisotropic parts of the metric
tensor will have an effect on $k^{i}$ (and therefore $\delta$-terms,
whether removable or non-removable always vanish in the geodesic equations;
recall \noun{section} \ref{sub:Scale-factor}). Using the same method
of conformally-related objects as described in (\cite[part IIe]{Sachs-Wolfe}),
define the vector $\bar{k}^{\mu}:a_{F}^{2}\bar{k}^{\mu}=k^{\mu}$
and the tensor $\bar{\gamma}_{ab}:a_{F}^{2}\bar{\gamma}_{ab}=\gamma_{ab}$;
recall that $k_{R}^{0}=-k_{R}^{i}k_{i}^{R}=1$. This gives us geodesic
equations:

\begin{align}
\frac{d\bar{k}^{0}}{d\lambda}+\frac{1}{2}\bar{\gamma}_{ab,0}\bar{k}^{a}\bar{k}^{b}= & 0\label{eq:time geodesic}\\
\frac{d\bar{k}^{1}}{d\lambda}+\left(e^{2\gamma-2\alpha}-e^{2\beta-2\alpha}\right)\bar{k}^{2}\bar{k}^{3}= & 0\label{eq:k1}\\
\frac{d\bar{k}^{2}}{d\lambda}+\left(e^{2\alpha-2\beta}-e^{2\gamma-2\beta}\right)\bar{k}^{1}\bar{k}^{3}= & 0\label{eq:k2}\\
\frac{d\bar{k}^{3}}{d\lambda}+\left(e^{2\beta-2\gamma}-e^{2\alpha-2\gamma}\right)\bar{k}^{1}\bar{k}^{2}= & 0\label{eq:k3}\end{align}

\noindent . Despite the symmetry of these equations, their nonlinearity
has inhibited the discovery of exact solutions and research into their
properties is ongoing; see for example \cite{Nilsson}. However, with
solution up to quadratic order for the metric in hand (\ref{eq:alpha1 radiation},\ref{eq:alpha1 matter},\ref{eq:alpha2 radiation},\ref{eq:alpha2 matter}),
we can explicitly solve the equations in the case of weak waves. Let
$\bar{k}^{a}=\bar{k}_{R}^{a}+\Delta\bar{k}^{a}\left(\lambda\right)$.
Expanding out the geodesic equations to second order in the metric:

\begin{align}
\frac{d\Delta\bar{k}_{1}^{0}}{d\lambda}+\frac{1}{2}\left[\alpha_{1}^{\prime}\left(\bar{k}_{R}^{1}\right)^{2}+\beta_{1}^{\prime}\left(\bar{k}_{R}^{2}\right)^{2}+\gamma_{1}^{\prime}\left(\bar{k}_{R}^{3}\right)^{2}\right]= & 0\\
\frac{d\Delta\bar{k}_{1}^{1}}{d\lambda}+2\left(\gamma_{1}-\beta_{1}\right)\bar{k}_{R}^{2}\bar{k}_{R}^{3}= & 0\\
\frac{d\Delta\bar{k}_{2}^{1}}{d\lambda}+2\left(\alpha_{1}-\gamma_{1}\right)\bar{k}_{R}^{1}\bar{k}_{R}^{3}= & 0\\
\frac{d\Delta\bar{k}_{3}^{1}}{d\lambda}+2\left(\beta_{1}-\alpha_{1}\right)\bar{k}_{R}^{1}\bar{k}_{R}^{2}= & 0\end{align}

\begin{align}
\frac{d\Delta\bar{k}_{2}^{0}}{d\lambda}+\frac{1}{2}\left[\begin{array}{c}
\left(\alpha_{2}^{\prime}+2\alpha_{1}^{\prime}\alpha_{1}\right)\left(\bar{k}_{R}^{1}\right)^{2}+2\bar{k}_{R}^{1}\alpha_{1}^{\prime}\Delta\bar{k}_{1}^{1}+\\
+\left(\beta_{2}^{\prime}+2\beta_{1}^{\prime}\beta_{1}\right)\left(\bar{k}_{R}^{2}\right)^{2}+2\bar{k}_{R}^{2}\beta_{1}^{\prime}\Delta\bar{k}_{1}^{2}+\\
+\left(\gamma_{2}^{\prime}+2\gamma_{1}^{\prime}\gamma_{1}\right)\left(\bar{k}_{R}^{3}\right)^{2}+2\bar{k}_{R}^{3}\gamma_{1}^{\prime}\Delta\bar{k}_{1}^{3}\end{array}\right]= & 0\\
\frac{d\Delta\bar{k}_{2}^{1}}{d\lambda}+2\left[\begin{array}{c}
\left(\gamma_{1}-\beta_{1}\right)\left(\bar{k}_{R}^{2}\Delta\bar{k}_{1}^{3}+\bar{k}_{R}^{3}\Delta\bar{k}_{1}^{2}\right)+\\
+\left(\gamma_{2}-\beta_{2}+3\gamma_{1}^{2}-3\beta_{1}^{2}\right)\bar{k}_{R}^{2}\bar{k}_{R}^{3}\end{array}\right]= & 0\\
\frac{d\Delta\bar{k}_{2}^{2}}{d\lambda}+2\left[\begin{array}{c}
\left(\alpha_{1}-\gamma_{1}\right)\left(\bar{k}_{R}^{3}\Delta\bar{k}_{1}^{1}+\bar{k}_{R}^{1}\Delta\bar{k}_{1}^{3}\right)+\\
+\left(\alpha_{2}-\gamma_{2}+3\alpha_{1}^{2}-3\gamma_{1}^{2}\right)\bar{k}_{R}^{1}\bar{k}_{R}^{3}\end{array}\right]= & 0\\
\frac{d\Delta\bar{k}_{2}^{3}}{d\lambda}+2\left[\begin{array}{c}
\left(\beta_{1}-\alpha_{1}\right)\left(\bar{k}_{R}^{1}\Delta\bar{k}_{1}^{2}+\bar{k}_{R}^{2}\Delta\bar{k}_{1}^{1}\right)+\\
+\left(\beta_{2}-\alpha_{2}+3\beta_{1}^{2}-3\alpha_{1}^{2}\right)\bar{k}_{R}^{1}\bar{k}_{R}^{2}\end{array}\right]= & 0\end{align}

\noindent . To first order, the equations are trivially solved by
choosing $\lambda=\eta$ as the affine parameter; the problem of determining
$d\lambda/d\eta$ is overcome by our choice of reference system, the
lack of vector perturbations and the homogeneity of space: 

\begin{align}
\Delta\bar{k}_{1}^{0}= & -\frac{1}{2}\left[\alpha_{1}\left(\bar{k}_{R}^{1}\right)^{2}+\beta_{1}\left(\bar{k}_{R}^{2}\right)^{2}+\gamma_{1}\left(\bar{k}_{R}^{3}\right)^{2}\right]_{\eta=\eta_{E}}^{\eta=\eta_{R}}\label{eq:Delta k^0 first order}\\
= & -\frac{1}{2}\left[\tilde{\alpha}_{1}\left(\bar{k}_{R}^{1}\right)^{2}+\tilde{\beta}_{1}\left(\bar{k}_{R}^{2}\right)^{2}+\tilde{\gamma}_{1}\left(\bar{k}_{R}^{3}\right)^{2}+\frac{1}{3}\delta_{1}\right]_{\eta=\eta_{E}}^{\eta=\eta_{R}}\nonumber \\
\Delta\bar{k}_{1}^{1}= & 2\bar{k}_{R}^{2}\bar{k}_{R}^{3}\int_{\eta_{E}}^{\eta_{R}}\left(\beta_{1}-\gamma_{1}\right)d\eta\\
\Delta\bar{k}_{1}^{2}= & 2\bar{k}_{R}^{3}\bar{k}_{R}^{1}\int_{\eta_{E}}^{\eta_{R}}\left(\gamma_{1}-\alpha_{1}\right)d\eta\\
\Delta\bar{k}_{1}^{3}= & 2\bar{k}_{R}^{1}\bar{k}_{R}^{2}\int_{\eta_{E}}^{\eta_{R}}\left(\alpha_{1}-\beta_{1}\right)d\eta\end{align}

\noindent . The relationship (\ref{eq:Delta k^0 first order}) explicitly
shows the quadrupolar nature of changes to the CMB alluded to in \cite{DLN anisotropy}.
An unremoved removable perturbation changes the temperature of the
whole sky isotropically; this confirms the effect noted by Hwang \&
Noh\cite{Hwang Sachs-Wolfe}.

The equations for quadratic-order corrections read

\begin{equation}
\frac{d\Delta\bar{k}_{2}^{0}}{d\lambda}+\frac{1}{2}\left[\begin{array}{c}
\left(\alpha_{2}^{\prime}+2\alpha_{1}^{\prime}\alpha_{1}\right)\left(\bar{k}_{R}^{1}\right)^{2}+2\bar{k}_{R}^{1}\alpha_{1}^{\prime}\Delta\bar{k}_{1}^{1}+\\
+\left(\beta_{2}^{\prime}+2\beta_{1}^{\prime}\beta_{1}\right)\left(\bar{k}_{R}^{2}\right)^{2}+2\bar{k}_{R}^{2}\beta_{1}^{\prime}\Delta\bar{k}_{1}^{2}+\\
+\left(\gamma_{2}^{\prime}+2\gamma_{1}^{\prime}\gamma_{1}\right)\left(\bar{k}_{R}^{3}\right)^{2}+2\bar{k}_{R}^{3}\gamma_{1}^{\prime}\Delta\bar{k}_{1}^{3}\end{array}\right]=0\end{equation}

\noindent which due to the cancellation of the terms in the right
column integrates trivially to

\begin{equation}
\Delta\bar{k}_{2}^{0}=-\frac{1}{2}\left[\left(\alpha_{2}+\alpha_{1}^{2}\right)\left(\bar{k}_{R}^{1}\right)^{2}+\left(\beta_{2}+\beta_{1}^{2}\right)\left(\bar{k}_{R}^{2}\right)^{2}+\left(\gamma_{2}+\gamma_{1}^{2}\right)\left(\bar{k}_{R}^{3}\right)^{2}\right]_{\eta=\eta_{E}}^{\eta=\eta_{R}}\end{equation}

\noindent (reiterating the quadrupolar character of the change to
the CMB, but generalizing it to anisotropic expansion); meanwhile
for the space part of the vector

\begin{align}
\frac{d\Delta\bar{k}_{2}^{1}}{d\lambda}+2\left[\begin{array}{c}
\left(\gamma_{1}-\beta_{1}\right)\left(\bar{k}_{R}^{2}\Delta\bar{k}_{1}^{3}+\bar{k}_{R}^{3}\Delta\bar{k}_{1}^{2}\right)+\\
+\left(\gamma_{2}-\beta_{2}+3\gamma_{1}^{2}-3\beta_{1}^{2}+2\delta_{1}\left(\beta_{1}-\gamma_{1}\right)\right)\bar{k}_{R}^{2}\bar{k}_{R}^{3}\end{array}\right]= & 0\\
\frac{d\Delta\bar{k}_{2}^{2}}{d\lambda}+2\left[\begin{array}{c}
\left(\alpha_{1}-\gamma_{1}\right)\left(\bar{k}_{R}^{3}\Delta\bar{k}_{1}^{1}+\bar{k}_{R}^{1}\Delta\bar{k}_{1}^{3}\right)+\\
+\left(\alpha_{2}-\gamma_{2}+3\alpha_{1}^{2}-3\gamma_{1}^{2}+2\delta_{1}\left(\gamma_{1}-\alpha_{1}\right)\right)\bar{k}_{E}^{3}\bar{k}_{E}^{1}\end{array}\right]= & 0\\
\frac{d\Delta\bar{k}_{2}^{3}}{d\lambda}+2\left[\begin{array}{c}
\left(\beta_{1}-\alpha_{1}\right)\left(\bar{k}_{R}^{1}\Delta\bar{k}_{1}^{2}+\bar{k}_{R}^{2}\Delta\bar{k}_{1}^{1}\right)+\\
+\left(\beta_{2}-\alpha_{2}+3\beta_{1}^{2}-3\alpha_{1}^{2}+2\delta_{1}\left(\alpha_{1}-\beta_{1}\right)\right)\bar{k}_{R}^{1}\bar{k}_{R}^{2}\end{array}\right]= & 0\end{align}

\noindent which has solutions

\begin{align}
\Delta\bar{k}_{2}^{1}= & -2\left\{ \begin{array}{c}
2\bar{k}_{R}^{1}\int_{\eta_{E}}^{\eta_{R}}\left[\left(\tilde{\gamma}_{1}-\tilde{\beta}_{1}\right)\left(\begin{array}{c}
\left(\bar{k}_{R}^{2}\right)^{2}\int^{\eta}\left(\tilde{\alpha}_{1}-\tilde{\beta}_{1}\right)d\bar{\eta}+\\
+\left(\bar{k}_{R}^{3}\right)^{2}\int^{\eta}\left(\tilde{\gamma}_{1}-\tilde{\alpha}_{1}\right)d\bar{\eta}\end{array}\right)d\eta\right]+\\
+\bar{k}_{R}^{2}\bar{k}_{R}^{3}\int_{\eta_{E}}^{\eta_{R}}\left(\gamma_{2}-\beta_{2}+\tilde{\gamma}_{1}^{2}-\tilde{\beta}_{1}^{2}+2\tilde{\alpha}_{1}\left(\tilde{\beta}_{1}-\tilde{\gamma}_{1}\right)\right)d\eta\end{array}\right\} \\
\Delta\bar{k}_{2}^{2}= & -2\left\{ \begin{array}{c}
2\bar{k}_{R}^{2}\int_{\eta_{E}}^{\eta_{R}}\left[\left(\tilde{\alpha}_{1}-\tilde{\gamma}_{1}\right)\left(\begin{array}{c}
\left(\bar{k}_{R}^{3}\right)^{2}\int^{\eta}\left(\tilde{\beta}_{1}-\tilde{\gamma}_{1}\right)d\bar{\eta}+\\
+\left(\bar{k}_{R}^{1}\right)^{2}\int^{\eta}\left(\tilde{\alpha}_{1}-\tilde{\beta}_{1}\right)d\bar{\eta}\end{array}\right)d\eta\right]+\\
+\bar{k}_{R}^{3}\bar{k}_{R}^{1}\int_{\eta_{E}}^{\eta_{R}}\left(\alpha_{2}-\gamma_{2}+\tilde{\alpha}_{1}^{2}-\tilde{\gamma}_{1}^{2}+2\tilde{\beta}_{1}\left(\tilde{\gamma}_{1}-\tilde{\alpha}_{1}\right)\right)d\eta\end{array}\right\} \\
\Delta\bar{k}_{2}^{3}= & -2\left\{ \begin{array}{c}
2\bar{k}_{R}^{3}\int_{\eta_{E}}^{\eta_{R}}\left[\left(\tilde{\beta}_{1}-\tilde{\alpha}_{1}\right)\left(\begin{array}{c}
\left(\bar{k}_{R}^{1}\right)^{2}\int^{\eta}\left(\tilde{\gamma}_{1}-\tilde{\alpha}_{1}\right)d\bar{\eta}+\\
+\left(\bar{k}_{R}^{2}\right)^{2}\int^{\eta}\left(\tilde{\beta}_{1}-\tilde{\gamma}_{1}\right)d\bar{\eta}\end{array}\right)d\eta\right]+\\
+\bar{k}_{R}^{1}\bar{k}_{R}^{2}\int_{\eta_{E}}^{\eta_{R}}\left(\beta_{2}-\alpha_{2}+\tilde{\beta}_{1}^{2}-\tilde{\alpha}_{1}^{2}+2\tilde{\gamma}_{1}\left(\tilde{\alpha}_{1}-\tilde{\beta}_{1}\right)\right)d\eta\end{array}\right\} \end{align}

.

\section{Redshift and CMB variations}

The geodesic of a light ray is related to its observed redshift by

\begin{equation}
z+1=\frac{\left(k^{\mu}u_{\mu}\right)_{R}}{\left(k^{\mu}u_{\mu}\right)_{E}}\end{equation}

\noindent \cite{Sachs-Wolfe}. Having determined $u_{0}=1$ and $u_{i}=0$
this simplifies to

\begin{equation}
z+1=\frac{a_{F}\left(\eta_{R}\right)}{a_{F}\left(\eta_{E}\right)}\bar{k}_{R}^{0}\end{equation}

\noindent so, to quadratic order,

\begin{equation}
z+1\approx\frac{a_{F}\left(\eta_{R}\right)}{a_{F}\left(\eta_{E}\right)}\left\{ 1-\frac{1}{2}\left[\begin{array}{c}
\left(\alpha_{1}+\alpha_{2}+\alpha_{1}^{2}\right)\left(\bar{k}_{R}^{1}\right)^{2}+\\
+\left(\beta_{1}+\beta_{2}+\beta_{1}^{2}\right)\left(\bar{k}_{R}^{2}\right)^{2}+\\
+\left(\gamma_{1}+\gamma_{2}+\gamma_{1}^{2}\right)\left(\bar{k}_{R}^{3}\right)^{2}\end{array}\right]_{\eta=\eta_{E}}^{\eta=\eta_{R}}\right\} \end{equation}

\noindent . Meanwhile, the temperature field

\begin{equation}
\frac{T_{R}}{T_{E}}=\frac{1}{z+1}\approx\frac{a_{F}\left(\eta_{E}\right)}{a_{F}\left(\eta_{R}\right)}\left\{ \begin{array}{c}
1+\frac{1}{2}\left[\alpha_{1}\left(\bar{k}_{R}^{1}\right)^{2}+\beta_{1}\left(\bar{k}_{R}^{2}\right)^{2}+\gamma_{1}\left(\bar{k}_{R}^{3}\right)^{2}\right]+\\
+\frac{1}{4}\left[\alpha_{1}\left(\bar{k}_{R}^{1}\right)^{2}+\beta_{1}\left(\bar{k}_{R}^{2}\right)^{2}+\gamma_{1}\left(\bar{k}_{R}^{3}\right)^{2}\right]^{2}+\\
+\frac{1}{2}\left[\begin{array}{c}
\left(\alpha_{2}+\alpha_{1}^{2}\right)\left(\bar{k}_{R}^{1}\right)^{2}+\\
+\left(\beta_{2}+\beta_{1}^{2}\right)\left(\bar{k}_{R}^{2}\right)^{2}+\\
+\left(\gamma_{2}+\gamma_{1}^{2}\right)\left(\bar{k}_{R}^{3}\right)^{2}\end{array}\right]\end{array}\right\} _{\eta=\eta_{E}}^{\eta=\eta_{R}}\end{equation}

\noindent so

\begin{equation}
\frac{\Delta T}{T_{R}}\approx\frac{a_{F}\left(\eta_{E}\right)}{a_{F}\left(\eta_{R}\right)}\left\{ \begin{array}{c}
\frac{1}{2}\left[\alpha_{1}\left(\bar{k}_{R}^{1}\right)^{2}+\beta_{1}\left(\bar{k}_{R}^{2}\right)^{2}+\gamma_{1}\left(\bar{k}_{R}^{3}\right)^{2}\right]+\\
+\frac{1}{4}\left[\alpha_{1}\left(\bar{k}_{R}^{1}\right)^{2}+\beta_{1}\left(\bar{k}_{R}^{2}\right)^{2}+\gamma_{1}\left(\bar{k}_{R}^{3}\right)^{2}\right]^{2}+\\
+\frac{1}{2}\left[\begin{array}{c}
\left(\alpha_{2}+\alpha_{1}^{2}\right)\left(\bar{k}_{R}^{1}\right)^{2}+\\
+\left(\beta_{2}+\beta_{1}^{2}\right)\left(\bar{k}_{R}^{2}\right)^{2}+\\
+\left(\gamma_{2}+\gamma_{1}^{2}\right)\left(\bar{k}_{R}^{3}\right)^{2}\end{array}\right]\end{array}\right\} _{\eta=\eta_{E}}^{\eta=\eta_{R}}\label{eq:deltaT/T}\end{equation}

\noindent .

\section{Comparison with the observed CMB\label{sub:Comparison-with-the}}

Five-year and seven-year results\cite{WMAP 7-year,WMAP 5-year} from
WMAP\cite{WMAP what is it} give the best picture to date of the CMB.
The WMAP observations reconfirm the constraint of the quantity $\Delta T/T<10^{-4}$\cite{Wright CMB};
any change to the CMB from acceleration must be equal to or smaller
than this value in order to be compatible with observations, placing
an additional constraint on cosmological models. This implies that
in the current epoch, and in the absensce of further special alignment,
$\left|\alpha\right|,\left|\beta\right|,\left|\gamma\right|\lesssim10^{-5}$.
In a matter dominated universe, under ordinary circumstances, this
implies (since $\eta\lesssim10^{-1}$; see \noun{section} \ref{sub:Cosmological-parameters})\begin{align}
\left|C_{\alpha1,1}y_{1}^{\mbox{matter}}\right|\lesssim10^{-5}\implies & \left|C_{\alpha1,1}\right|\lesssim10^{-6}\\
\left|C_{\alpha2,1}y_{2}^{\mbox{matter}}\right|\lesssim10^{-5}\implies & \left|C_{\alpha2,1}\right|\lesssim10^{-8}\end{align}

\noindent ; meanwhile in a radiation-dominated universe,

\begin{align}
\left|C_{\alpha1,1}y_{1}^{\mbox{radiation}}\right|\lesssim10^{-5}\implies & \left|C_{\alpha1,1}\right|\lesssim10^{-5}\\
\left|C_{\alpha2,1}y_{2}^{\mbox{radiation}}\right|\lesssim10^{-5}\implies & \left|C_{\alpha2,1}\right|\lesssim10^{-6}\end{align}

\noindent . The coefficients associated with the decaying mode are
constrained to be smaller than those associated with the growing mode
without further theoretical considerations.

\subsection{CMB anomalies}

Since the publication of the latest generation of CMB maps\cite{Tegmark CMB map},
numerous claims have been made (for example, \cite{Tegmark CMB map,Axis of Evil,Cold Spot II,Eriksen dipole})
of anomalous structure in the CMB. While the WMAP team argue\cite{WMAP anomalies}
that these phenomena are not of statistical significance, if a quasi-isotropic
Bianchi IX universe could produce any of the perceived patterns it
would point the way toward further observational studies of the CMB
to determine cosmological parameters, and establish the quasi-isotropic
Bianchi IX universe as a viable model for cosmology.

In all cases, we emphasize that the most likely explanation for any
perceived pattern in the CMB which is not shown to be statistically
significant is the null hypothesis: that is, the human perceptive
phenomenon of pareidolia, the same phenomenon responsible for observing
familiar shapes in clouds or the {}``Man in the Moon''.

\subsubsection{Cold spots, {}``fingers'' and the {}``Axis of Evil''}

Two compact, supposedly anomalous areas of low temperature have been
noted in the CMB, the so called {}``cold spots''.

The first of these (called Cold Spot I in \cite{WMAP anomalies})
is a region\cite{WMAP first-year results} covering approximately
15000 square degrees in the direction of the galactic center, much
of which is 194 microkelvin\cite{Tegmark CMB map} colder than the
CMB mean temperature ($\Delta T/T_{R}=-7.12\times10^{-5}$).

Particularly noteworthy regarding Cold Spot I is its membership in
one of four {}``fingers'' spaced at roughly 90-degree angles around
the galactic equator, intersticed by four areas of higher ($\Delta T/T_{R}=7.12\times10^{-5}$)
temperature%
\footnote{The CMB dipole is defined as such a way as to be traceless, so $\int\Delta T_{\mbox{quadrupole}}/T_{R}dS=0$.%
}. Qualitatively, such a pattern is roughly consistent with the expected
pattern if two of the functions $\alpha,\beta,\gamma>0$ and if two
of the the principle axes of the metric tensor lie on the axes of
the cold and hot zones (implying the third axis points along the {}``Axis
of Evil'', see below). The so-called {}``Cold Spot II'' reported
by Vielva et al.\cite{Cold Spot II,Vielva} also forms part of these
{}``finger'' structures\cite{WMAP anomalies}.

Cold Spot I also has the angular size \cite{DLN anisotropy} predicts
for the observed value of $\Omega_{M}\approx.3$. 

Due to the coincidence of the cold spot with the direction of the
galactic center, there are no optical observations in its direction
(see \noun{figure} \ref{Flo:Fig1}), and therefore there is no data
on cosmic acceleration in the direction of Cold Spot I.

(\noun{Equation} \ref{eq:deltaT/T}) implies that any cold spot resulting
from anisotropy in the metric should be accompanied by an identical
cold spot at a point antipodal to the original spot. Tegmark's examination\cite{Tegmark CMB map}
of the one-year WMAP data on the CMB low-order multipoles revealed
an alignment between the CMB quadrupole and octupole in the direction
of $\left(l,b\right)\approx\left(-110^{\circ},60^{\circ}\right)$
along which the quadrupole is nearly zero, an axis which Land \& Maguiejo
found\cite{Axis of Evil} extended to the 16-pole and 32-pole as well;
the alignment has been dubbed the {}``Axis of Evil''. While examination
of the three-year WMAP data\cite{Axis of Evil 2} found the Axis of
Evil to be of lower significance than initially thought (94\%-98\%),
it still persists; the WMAP team's discussion of the alignment\cite[pt. 7]{WMAP anomalies}
admits the {}``remarkability'' of this alignment and, while assigning
its existence to chance, does not attempt to explain the {}``Axis
of Evil'' in full.

The Axis of Evil, which in equatorial coordinates\cite[p. 43]{Galactic to equatorial}
lies close to RA 10:44 Dec $+7.6^{\circ}$, falls within the zone
in which redshift data has been collected for measurement of the cosmic
deceleration parameter. To simplest linear approximation with a pure
growing mode, (that is, that the functions $\alpha$ and $\alpha^{\prime}$
are both small such that $\alpha^{2}\approx0$) this alignment rules
out a CMB arising from cosmological gravitational waves as a source
of cosmic acceleration. However, the fact of the alignment of the
quadrupole, octopole, 16-pole and 32-pole indicates that non-linear
contributions of gravitational waves to acceleration are not ruled
out.

The question of the overall magnitude of the quadrupole, which is
only 14\% of the expected value\cite{Tegmark CMB map,COBE quadrupole},
has also been raised. The WMAP team\cite[ pt. 4]{WMAP anomalies}
agree with Tegmark that the depressed quadrupole falls within the
95\% confidence interval for simulations of the CMB, but do not attempt
an explanation for the unusually strong octopole term. Long-wavelength
gravitational waves can easily explain both through judicious choice
the arbitrary constants $C_{\alpha1,1}$ etc. in a manner compatible
with the CMB. Efstathiou\cite{Low quadrupole} supposes that the depressed
quadrupole could be an indication of a closed universe; however the
relationships he proposes generate zero contributions to the CMB power
spectrum from the genuinely cosmological, intrinsic $n=3$ waves found
in Bianchi IX, and any observational test using his framework must
rely on correct evaluation of gauge terms whose effective wavelengths
must be far longer than the cosmic horizon. Furthermore, Efstathiou's
conclusion that a closed universe would automatically require a scrapping
of current inflationary models is contradicted by others; for example
Guth argues that a universe that is closed but with a very large radius
of curvature is not ruled out\cite{Guth}.

The quasi-isotropic Bianchi IX model cannot provide an explanation
for hemispherical dipole asymmetry claimed by Ericksen \emph{et. al.}\cite{Eriksen dipole}.

\section{Conclusions}

The long-wavelength gravitational waves intrinsic to a quasi-isotropic
Bianchi IX will cause a change in the cosmic microwave background
with a distinctive quadrupolar signature. A radially-symmetric pattern
of light deflections in the CMB resulting from shear may also be observed.

The almost-isotropic Bianchi IX model can be compatible with the CMB
as observed, and can provide an explanation for perceived anomalies
observed in the CMB by COBE and WMAP. However, the existence of these
anomalies beyond the level of statistical noise is not certain; a
possible route of cross-disciplinary research is open in the form
of examination of the phenomenon of pareidolia as applied to the CMB.

Models of quasi-isotropic Bianchi IX relying on pure growing modes
or pure decaying modes of the gravitational waves cannot simultaneously
explain observed cosmic acceleration and the observed cosmic microwave
background. Research into the non-linear regime of the Bianchi IX
cosmology may elucidate the existence of a model of an accelerating
universe in which isotropy is almost preserved.

\part{An accelerating Bianchi IX universe preserving an almost-isotropic
CMB}

In order for a Bianchi IX universe to both appear nearly isotropic
in the cosmic microwave background and to accelerate through the existence
of long-wavelength gravitational waves, it must fulfill two conditions.
The first is that the function $k^{0}\left(\eta_{R}\right)$ must
have absolute value less than the limit imposed by observations of
the cosmic microwave background, $\Delta T/T_{R}$. The second is
that at least one of the functions $Q_{a}^{b}<0$. It is possible
for both these conditions to be simultaneously filled while remaining
compatible with other observational constraints on cosmological parameters.

The idea of long-wavelength gravitational waves causing anisotropy
in the CMB has been proposed, but not applied to the Bianchi IX universe.
Grishchuk \& Zel'dovich consider the possibility of long-wavelength
gravitational waves existing in a Friedmann universe without violating
the limits imposed by the CMB\cite{Long waves CMB}, but do not apply
their work to the gravitational waves of cosmological character which
appear in some homogeneous cosmologies. Campanelli \emph{et. al.}
suggest that such a universe could exist and propose a Taub-type Bianchi
I universe which also includes anisotropic dark energy as an initial
explanation for the observed CMB, complementing Rodrigues\cite{Rodrigues}.
Critically, they do not consider gravitational waves as a generator
of the anisotropy and treat the parameters of the Taub universe as
if dark energy were simply established by fiat. Similarly, Kovisto
and Mota\cite{Kovisto} do not look beyond the Bianchi I model and
instead fall back on exotic theories to explain dark energy.

\section{Cosmological parameters\label{sub:Cosmological-parameters}}

WMAP\cite{WMAP 5-year,WMAP 7-year} has produced an all-sky survey
of the CMB which, if the universe is almost Friedmannian, can be used
to constrain cosmological parameters.

Let the radius of curvature $a_{0}$ and conformal time $\eta$ of
the background Friedmann cosmology be treated as a free parameters;
assume a closed universe. The WMAP seven-year data gives

\begin{align}
H_{0}= & 70.4_{-1.4}^{+1.3}\mbox{km/s/Mpc}\\
\Omega_{K}= & -.0025\pm0.0109\end{align}

\noindent (WMAP's analysis includes the value of $\Omega_{K}$ measured
by baryon acoustic oscillations reported in \cite{BAO peak}). The
radius of curvature, Hubble parameter and curvature energy density
are related by

\begin{equation}
a_{0}=H_{0}^{-1}\sqrt{-\Omega_{K}^{-1}}\end{equation}

\noindent while the Hubble parameter, radius of curvature and $\eta$-time
are related by

\begin{equation}
H_{0}a_{0}=\cot\left(\eta_{0}/2\right)\end{equation}

\noindent . Therefore we have limiting values (as defined by the 95\%
confidence boundary of the WMAP observations)

\begin{align}
a_{0}\geq & 1.12\times10^{29}\mbox{cm}\\
\eta_{0}\leq & 0.0266\end{align}

\noindent and highest-confidence values

\begin{align}
a_{0}= & 2.68\times10^{29}\mbox{cm}\\
\eta_{0}= & 0.00499\end{align}

\noindent . Meanwhile, the ratio of Hubble radius to radius of curvature
is at least

\begin{equation}
H_{0}a_{0}\geq8.67\end{equation}

\noindent with a best-fit value of

\begin{equation}
H_{0}a_{0}=20.0\end{equation}

\noindent . In other words, if the universe is closed, then the cosmological
gravitational waves of the Bianchi IX cosmology are of much, much
longer wavelength than the observable universe.

Finally, from the value of the redshift of decoupling, $z_{\mbox{last scattering}}=1090$,
we can say by  (\ref{eq:Redshift definition}) that

\begin{equation}
\eta_{R}/\eta_{E}\approx33.0\end{equation}

\noindent . As the available data, including that from supernovae
(see \noun{table} \ref{tab:Summary-of-results}), does not exclude
a flat universe, we are always free, in developing the theory of Bianchi
IX and acceleration, to set the parameter $\eta$ as close to zero
as necessary; doing so will not, in and of itself, violate observations,
but will instead be constrained by the impact of the decaying mode
of the gravitational waves on the CMB.

\section{Compatibility with the redshift}

Of all the observed cosmological parameters observed by WMAP and other
probes of the CMB, the ones that are directly observed are $\Delta T/T_{R}$
and $z_{\mbox{last scattering}}$. From these we can say that in the
current epoch the universe appears isotropic and that its expansion
since last scattering has, on average to the present time, been isotropic;
neither of these facts necessarily imply that the overall expansion
was isotropic at any time before the present. Instead, the condition
of quasi-isotropy simply implies that

\begin{equation}
\frac{dk^{0}}{d\eta}+\frac{1}{2}\gamma_{ab,0}k^{a}k^{b}\approx0\end{equation}

\noindent . This implies that shear is small, so\begin{align}
k^{a}\approx & k_{0}^{a}\\
z+1\approx & a_{F}\left(\eta_{R}\right)/a_{F}\left(\eta_{E}\right)\label{eq:approx redshift}\end{align}

\noindent as in the background Friedmann case.

We can obtain a near-zero value to the wave functions in the present
epoch by admitting the presence of both growing and decaying modes
in the gravitational waves. We want the condition (assuming $\Delta T/T_{R}$
is positive; in the case that it is negative the inequalities must
be reversed)

\noindent \begin{equation}
0\leq\frac{a_{F}\left(\eta_{E}\right)}{a_{F}\left(\eta_{R}\right)}e^{\alpha\left(\eta_{E}\right)-\alpha\left(\eta_{R}\right)}\leq\left|\Delta T/T_{R}\right|\end{equation}
 \& similarly for $\beta,\gamma$. In its full form this equation
is transcendental even when discussing weak waves, but expanding (\ref{eq:matter first order wave})
to lowest surviving order in $\eta$, we obtain

\begin{equation}
\left|37C_{\alpha1,1}\left(\eta_{R}^{2}-\eta_{E}^{2}\right)+4C_{\alpha2,1}\left(\eta_{R}^{-3}-\eta_{E}^{-3}\right)\right|\leq\left|\Delta T/T_{R}\right|\end{equation}

\noindent . In a young universe, the times of emission and reception
of a light ray are related by $\eta_{E}\approx\eta_{R}\left(z+1\right)^{-1/2}$
so

\begin{equation}
\left|37C_{\alpha1,1}\left(1-\left(z+1\right)^{-1}\right)\eta_{R}^{2}+4C_{\alpha2,1}\left(1-\left(z+1\right)^{3/2}\right)\eta_{R}^{-3}\right|\leq\left|\Delta T/T_{R}\right|\end{equation}

\noindent . Let:
\begin{itemize}
\item $10^{-g}$ be the amplitude of the growing mode $C_{\alpha1,1}$,
so $C_{\alpha1,1}=\mbox{sgn}\left(C_{\alpha1,1}\right)10^{-g}$;
\item $10^{-d}$ be the amplitude of the decaying mode $C_{\alpha2,1}$,
so $C_{\alpha2,1}=\mbox{sgn}\left(C_{\alpha2,1}\right)10^{-d}$;
\item $10^{-b}$ be the value of $\eta_{R}$;
\item $10^{-T}$ be the value of $\left|\Delta T/T_{R}\right|$
\end{itemize}
so noting that $z\sim1000=10^{3}$ our condition becomes approximately

\begin{equation}
\left|\mbox{sgn}\left(C_{\alpha1,1}\right)10^{-2b-g+3/2}-\mbox{sgn}\left(C_{\alpha2,1}\right)10^{3b-d+5}\right|\lesssim\left|10^{-T}\right|\end{equation}

\noindent . When the amplitude of the growing mode term dominates,
this approximate inequality is satisfied by

\begin{equation}
-2b-g+3/2\lesssim-T\label{eq:growing CMB constraint}\end{equation}

\noindent ; when the decaying mode dominates, the inequality is satisfied
by

\begin{equation}
3b-d+5\lesssim-T\label{eq:decaying CMB constraint}\end{equation}

\noindent . WMAP constrains $T\approx4$ (the difference between lowest
and highest temperatures is $2\Delta T_{R}/T=1.4\times10^{-4}$) and
$b\gtrsim1$. This constrains the growing and decaying modes, when
they act on their own, to:

\begin{align}
g\gtrsim & 7/2\\
d\gtrsim & 12\end{align}

\noindent . There exists a third possibility, in which the growing
and decaying contributions are, in the current epoch, of equal size
and opposite sign. For this to be the case, we need

\begin{equation}
-2b-g+3/2\approx3b-d+5\end{equation}

\noindent ; this approach relies on the observation of amplification
of weak gravitational waves in rapidly-changing backgrounds (see \noun{section}
\ref{sub:Amplification-of-gravitational}). Since \emph{b} is a free
parameter this approximate equation can always be satisfied, but we
still need to satisfy the constraints of the CMB.

\section{Acceleration in the Bianchi IX universe\label{sub:Acceleration-in-the}}

\subsection{Order of magnitude estimates for gravitational wave amplitudes\label{sub:Order-of-magnitude}}

Meanwhile, consider the tensorial deceleration parameter:

\begin{equation}
Q_{1}^{1}\equiv-\frac{\ddot{a}_{11}a_{11}}{\left(\dot{a}_{11}\right)^{2}}=\left[Q_{0}-2\frac{a_{F}}{\dot{a}_{F}}\dot{\alpha}-\frac{a_{F}^{2}}{\dot{a}_{F}^{2}}\left(\ddot{\alpha}+\dot{\alpha}^{2}\right)\right]\left(1+2\frac{a_{F}}{\dot{a}_{F}}\dot{\alpha}+\frac{a_{F}^{2}}{\dot{a}_{F}^{2}}\dot{\alpha}^{2}\right)^{-1}\label{eq:exact acceleration}\end{equation}

\noindent \& similarly for $Q_{2}^{2},Q_{3}^{3}$; this relationship
is exact. Evaluating (\ref{eq:generalized Q}) gives to lowest surviving
order in $\eta$

\begin{equation}
\Delta Q_{11,\mbox{growing}}^{\left(1\right)}\approx-70C_{\alpha1,1}\eta^{2}\end{equation}

\begin{equation}
\Delta Q_{11,\mbox{decaying}}^{\left(1\right)}\approx\frac{19}{2}C_{\alpha2,1}\eta^{-1}\end{equation}

\noindent . With an observed $\Delta Q_{11}\approx-1$ we can write:

\begin{equation}
10^{0}\approx\mbox{sgn}\left(C_{\alpha1,1}\right)10^{9/5-2b-g}-\mbox{sgn}\left(C_{\alpha2,1}\right)10^{1-d+b}\label{eq:acceleration equation}\end{equation}

\noindent . In the case of the growing mode dominating we need $\mbox{sgn}\left(C_{\alpha1,1}\right)=+1$
and $9/5+2b-g\approx0$. This forms a system of equations with (\ref{eq:growing CMB constraint})
so we have, at the limit of the allowed CMB perturbation,

\begin{equation}
\begin{cases}
9/5-2b-g\approx0\\
-2b-g+11/2\approx0\end{cases}\implies\mbox{no solution}\end{equation}

\noindent ; the growing mode cannot, on its own, cause the observed
acceleration and be compatible with the CMB. For the decaying mode,
we need $\mbox{sgn}\left(C_{\alpha2,1}\right)=-1$ and have

\begin{equation}
\begin{cases}
1-d+b\approx0\\
3b-d+5\approx-4\end{cases}\implies\begin{cases}
b\approx & -5\\
d\approx & -6\end{cases}\implies\begin{cases}
\eta\sim1\times10^{5}\\
C_{\alpha2,1}\sim1\times10^{6}\end{cases}\end{equation}

\noindent which is a nonsense result. Therefore neither the growing
or decaying modes, on their own, can both cause observed acceleration
and preserve the CMB. In the cases of the two modes having comparable
effect on the metric and opposite sign, though, we can solve (\ref{eq:acceleration equation})
with $\mbox{sgn}\left(C_{\alpha1,1}\right)=+1$, $\mbox{sgn}\left(C_{\alpha2,1}\right)=-1$
and 

\begin{equation}
10^{0}\approx-\left(C_{\alpha1,1}\right)10^{9/5-2b-g}+\left(C_{\alpha2,1}\right)10^{1-d+b}\end{equation}

\begin{equation}
\begin{cases}
g-d\approx-5b-7/2\\
9/5-2b-g\approx0\end{cases}\implies d\approx\frac{17}{10}+7b,g\approx2b-\frac{52}{10}\end{equation}

\noindent when the growing mode dominates the change in acceleration;
this sets estimated limits on the parameters (since $b\gtrsim2$):

\begin{align}
C_{\alpha1,1}\gtrsim & 2\times10^{1}\\
C_{\alpha2,1}\lesssim & 2\times10^{-16}\end{align}

\noindent . When the decaying mode dominates the change in acceleration,

\begin{equation}
\begin{cases}
g-d\approx-5b-7/2\\
1-d+b\approx0\end{cases}\implies g\approx-4b-5/2,d\approx b+1\end{equation}

\noindent which constrains the parameters

\begin{align}
C_{\alpha1,1}\gtrsim & 3\times10^{10}\\
C_{\alpha2,1}\lesssim & 1\times10^{-3}\end{align}

\noindent . While the values for the growing mode are far greater
than those for what could be called {}``weak'' waves (recalling
the constraints of \noun{section} \ref{sub:Solutions-at-quadratic}),
our educated estimate for $C_{\alpha1,1}$ in the growing-mode dominated
regime aligns nicely with the necessary strong-wave growing-mode value
for $\Delta Q_{1}^{1}$ disregarding the CMB. Therefore we can turn
to an analysis in the quasi-isotropic regime.

\subsection{Quasi-isotropic, strong growing mode acceleration}

We apply the same reasoning as in the previous section, but we are
aware of constraints (from \cite{Tegmark CMB map}) not just on the
CMB in the direction of the observed acceleration (which we continue
to assign as the {}``$\alpha$'' or $e_{i}^{1}$ direction) but
on the CMB in the other two (the {}``beta'' and {}``gamma'' directions):

\begin{align}
\Delta T_{\alpha}/T_{R}+\frac{1}{2}\left(\Delta T_{\beta}/T_{R}+\Delta T_{\gamma}/T_{R}\right)= & 7.1\times10^{-5}\label{eq:t1}\\
2\Delta T_{\beta}/T_{R}= & 1.4\times10^{-4}\label{eq:t2}\\
2\Delta T_{\gamma}/T_{R}= & 1.4\times10^{-4}\label{eq:t3}\\
Q_{1}^{1}= & -0.6\label{eq:a1}\\
\eta_{R}\lesssim & 3\times10^{-2}\end{align}

\noindent .

In this and all regimes to follow we can also approximate $Q_{F}\approx Q_{F}^{\mbox{flat}}=1/2$
to the limit of precision given the constraints on $\eta$; $Q_{F}$
will be 1\% stronger than $Q_{F}^{\mbox{flat}}$ only when $\eta\approx0.51$.
Between the constraints (\ref{eq:t1}-\ref{eq:a1}) and the average
over the sky of $\Delta T/T_{R}=0$, we have four equations with seven
unknowns ($\eta$, $c_{0}^{\alpha},c_{0}^{\beta},c_{0}^{\gamma}$,
$C_{\alpha2},C_{\beta2},C_{\gamma2}$). These equations are, explicitly
(see equations \ref{eq:alpha1 matter}, \ref{eq:generalized Q}, \ref{eq:deltaT/T}):

\begin{align}
7.1\times10^{-5}\gtrsim & \left(\eta_{E}/\eta_{R}\right)^{2}\left(e^{\alpha\left(\eta_{E}\right)-\alpha\left(\eta_{R}\right)}-1\right)\label{eq:alpha practical}\\
1.4\times10^{-4}\gtrsim & \left(\eta_{E}/\eta_{R}\right)^{2}\left(e^{\beta\left(\eta_{E}\right)-\beta\left(\eta_{R}\right)}-1\right)\\
1.4\times10^{-4}\gtrsim & \left(\eta_{E}/\eta_{R}\right)^{2}\left(e^{\gamma\left(\eta_{E}\right)-\gamma\left(\eta_{R}\right)}-1\right)\\
Q_{1}^{1}= & \frac{Q_{F}-\tan\left(\eta_{R}/2\right)\alpha_{R}^{\prime}-\tan^{2}\left(\eta_{R}/2\right)\alpha_{R}^{\prime\prime}-\tan^{2}\left(\eta_{R}/2\right)\alpha_{R}^{\prime2}}{1+2\tan\left(\eta_{R}/2\right)\alpha_{R}^{\prime}+\tan^{2}\left(\eta_{R}/2\right)\alpha_{R}^{\prime2}}\label{eq:acceleration practical}\end{align}

\noindent .

Trivially, we can see that in the limit of $\alpha,\beta,\gamma\rightarrow\infty$,
we must have $Q_{1}^{1}\approx Q_{2}^{2}\approx Q_{3}^{3}\rightarrow-1$;
if acceleration is driven by growing modes of long-wavelength gravitational
waves then in the long run the universe asymptotically approaches
de Sitter expansion as if driven by a cosmological constant, indicating
a solution in the regime of quasi-isotropy.

Consider the quasi-isotropic solution to the growing mode of the Einstein
equations, normalized as in \noun{equations} (\ref{eq:normalized 1}-\ref{eq:normalized 6}).
In the regime where $c_{0}^{\alpha}$ is sufficiently large that $A\gg1$,
we can approximate

\begin{align}
c_{2}^{\alpha}\approx & -\frac{1}{4}A^{2}\\
c_{2}^{\beta}\approx c_{2}^{\gamma}\approx & \frac{3}{20}A^{2}\end{align}

\noindent (an identical argument, with the functions $\alpha$ and
$\beta$ transposing their roles, applies for the case where $c_{0}^{\alpha}<0$).
From these terms we can also approximate the next order terms in the
series:

\begin{align}
c_{4}^{\alpha}\approx & \frac{521}{5600}A^{4}\\
c_{4}^{\beta}\approx c_{4}^{\gamma}\approx & -\frac{15}{224}A^{4}\end{align}

\noindent . Approximating equation (\ref{eq:acceleration practical})
to order $A^{4}\eta^{4}$ we obtain the relationships

\begin{align*}
Q_{1}^{1}\left(\eta_{R}\right)= & \frac{Q_{F}+\frac{3}{8}A^{2}\eta_{R}^{2}-\left(\frac{521}{1120}+\frac{1}{16}\right)A^{4}\eta_{R}^{4}}{1-\frac{1}{2}A^{2}\eta_{R}^{2}+\left(\frac{1}{16}+\frac{521}{1400}\right)A^{4}\eta_{R}^{4}}+\mathcal{O}\left(\left(\frac{1}{2}A\eta_{R}\right)^{6}\right)\\
Q_{2}^{2}\left(\eta_{R}\right)\approx Q_{3}^{3}\left(\eta_{R}\right)= & \frac{Q_{F}-\frac{9}{40}A^{2}\eta_{R}^{2}+\left(\frac{75}{224}-\frac{9}{400}\right)A^{4}\eta_{R}^{4}}{1+\frac{3}{10}A^{2}\eta_{R}^{2}+\left(\frac{9}{400}-\frac{15}{56}\right)A^{4}\eta_{R}^{4}}+\mathcal{O}\left(\left(\frac{1}{2}A\eta_{R}\right)^{6}\right)\end{align*}

\noindent . When $Q_{1}^{1}\left(\eta_{R}\right)=-0.6$ then $A\eta_{R}\approx1.5\pm0.2$
($c_{0}^{\alpha}\gtrsim1.9$), within the limit of applicability of
the expansion and also in the regime where the infinite series (\ref{eq:series})
converge. Thus, we have shown analytically that long-wavelength gravitational
waves can explain cosmic acceleration if that acceleration is anisotropic.

We can also make the following qualitative assessments about acceleration.
Firstly, its time-evolution is non-monotonic. In the $\alpha$ direction,
the universe will at first exhibit slightly increased deceleration,
before starting to accelerate. In the $\beta$ and $\gamma$ directions,
deceleration will asymptotically increase toward infinity but then
acceleration will decrease from infinity, quickly converging on the
strong-field value of $Q_{2}^{2}=Q_{3}^{3}=-1$. Acceleration in the
$\alpha$ direction begins at $A\eta\approx1.2$ and the universe
accelerates in every direction after $A\eta\approx1.6$; thus the
supposition that acceleration is a recent phenomenon is supported.

A universe that is accelerating in every direction is within the region
allowed by the model. \noun{Figure} (\ref{fig:Along-one-direction,})
illustrates the evolution of the deceleration parameters as a function
of time. The constraints placed on the decaying mode in (\noun{section}
\ref{sub:Cosmological-parameters}) and the upper limit on $\eta_{R}$
show that the decaying mode of long-wavelength gravitational waves
has not played a significant role in cosmic acceleration; in the epoch
of last scattering, the deceleration parameter was almost isotropic
and had a close to Friedmannian value.

\begin{figure}
\begin{centering}
\includegraphics{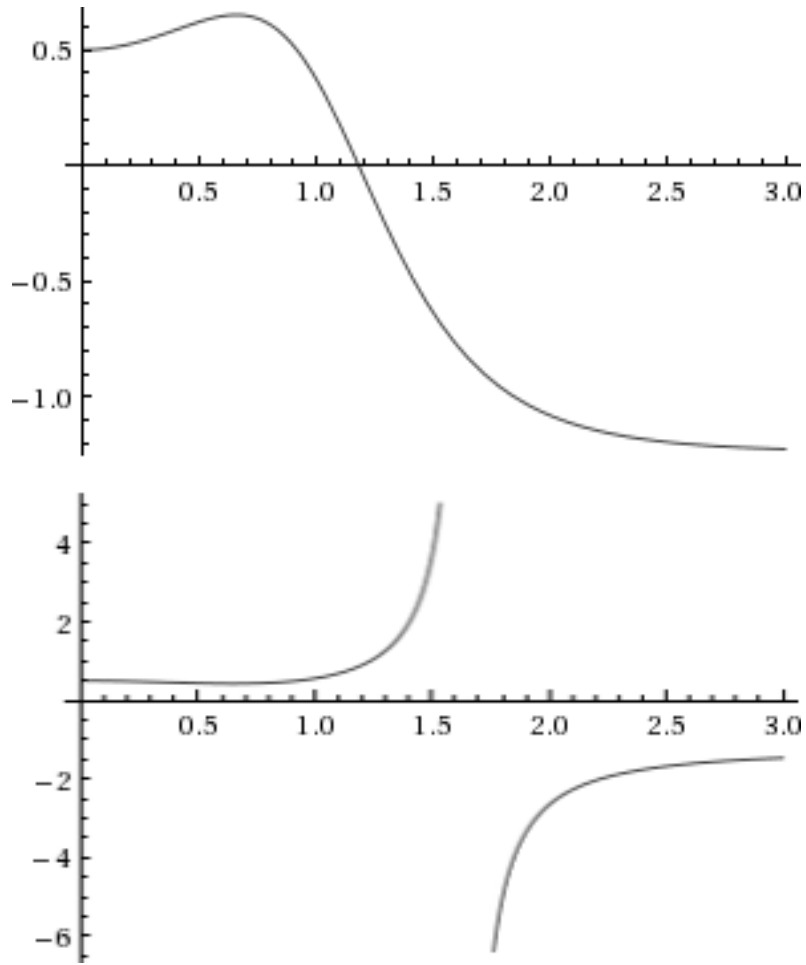}
\par\end{centering}

\caption{Deceleration parameter versus time}
\label{fig:Along-one-direction,}Along one direction, the universe
at first decelerates, then quickly begins accelerating. Along the
other two directions, the deceleration parameter goes to infinity
before converging from negative infinity to the value $-1$. The vertical
axis of each graph gives $Q_{a}^{b}$; the horizontal axis is in units
of $A\eta$.
\end{figure}

We now turn our attention to the preservation of the CMB. We have
three equations in three unknowns, taking the lowest term in the decaying
mode and the lowest two terms in the growing mode:\begin{align}
7.1\times10^{-5}\gtrsim & 4\left(\eta_{E}/\eta_{R}\right)^{2}\left[\begin{array}{c}
c_{2}^{\alpha}\left(\eta_{E}^{2}-\eta_{R}^{2}\right)+\frac{1}{2}\left(c_{2}^{\alpha}\right)^{2}\left(\eta_{E}^{2}-\eta_{R}^{2}\right)^{2}+\\
+c_{4}^{\alpha}\left(\eta_{E}^{4}-\eta_{R}^{4}\right)+C_{\alpha2,1}\left(\eta_{E}^{-3}-\eta_{R}^{-3}\right)\end{array}\right]\\
1.4\times10^{-4}\gtrsim & 4\left(\eta_{E}/\eta_{R}\right)^{2}\left[\begin{array}{c}
c_{2}^{\beta}\left(\eta_{E}^{2}-\eta_{R}^{2}\right)+\frac{1}{2}\left(c_{2}^{\beta}\right)^{2}\left(\eta_{E}^{2}-\eta_{R}^{2}\right)^{2}+\\
+c_{4}^{\beta}\left(\eta_{E}^{4}-\eta_{R}^{4}\right)+C_{\beta2,1}\left(\eta_{E}^{-3}-\eta_{R}^{-3}\right)\end{array}\right]\\
1.4\times10^{-4}\gtrsim & 4\left(\eta_{E}/\eta_{R}\right)^{2}\left[\begin{array}{c}
c_{2}^{\gamma}\left(\eta_{E}^{2}-\eta_{R}^{2}\right)+\frac{1}{2}\left(c_{2}^{\gamma}\right)^{2}\left(\eta_{E}^{2}-\eta_{R}^{2}\right)^{2}+\\
+c_{4}^{\gamma}\left(\eta_{E}^{4}-\eta_{R}^{4}\right)+C_{\gamma2,1}\left(\eta_{E}^{-3}-\eta_{R}^{-3}\right)\end{array}\right]\end{align}

\noindent . As $30\eta_{E}\approx\eta_{R}$ we can further approximate

\begin{align}
7.1\times10^{-5}\gtrsim & \frac{1}{225}\left[\frac{1}{4}A^{2}\eta_{R}^{2}+\left(\frac{1}{32}-\frac{521}{5600}\right)A^{4}\eta_{R}^{4}+27000C_{\alpha2,1}\eta_{R}^{-3}\right]\\
1.4\times10^{-4}\gtrsim & \frac{1}{225}\left[-\frac{3}{20}A^{2}\eta_{R}^{2}+\left(\frac{9}{800}+\frac{15}{224}\right)A^{4}\eta_{R}^{4}+27000C_{\beta2,1}\eta_{R}^{-3}\right]\\
1.4\times10^{-4}\gtrsim & \frac{1}{225}\left[-\frac{3}{20}A^{2}\eta_{R}^{2}+\left(\frac{9}{800}+\frac{15}{224}\right)A^{4}\eta_{R}^{4}+27000C_{\gamma2,1}\eta_{R}^{-3}\right]\end{align}

. If we take the inequalities as approximate equivalences and use
$A\eta_{R}\approx1.5$ then this system has solutions

\begin{align*}
C_{\beta2,1}\approx & C_{\gamma2,1}\approx & -3\times10^{-7}\eta_{R}^{3}\\
C_{\alpha2,1}\approx &  & -9\times10^{-6}\eta_{R}^{3}\end{align*}

which is compatible with the estimates of (\noun{section} \ref{sub:Order-of-magnitude}).
That $C_{\alpha2,1}+C_{\beta2,1}+C_{\gamma2,1}\neq0$ is a consequence
of the impossibility of \emph{a priori} choosing an {}``unperturbed''
temperature against which to compare anisotropic CMB fluctuations;
the significance of non-linear terms means we cannot at the same time
have the average over the whole sky of $\Delta T/T_{R}=0$ and have
$\delta_{1}=0$, recalling \noun{(equation} \ref{eq:Delta k^0 first order}). 

We exhaust almost all the freedom in the system (\ref{eq:alpha practical}-\ref{eq:acceleration practical})
in choosing to explain the {}``Axis of Evil'' at the same time as
acceleration; if this requirement is dropped and we treat CMB variations
as insignificant then a broad range of solutions opens up. In particular,
if the ratio of growing mode to decaying mode is approximately equal
for all three of $\alpha,\beta,\gamma$ we always have sufficient
freedom to choose a $\eta$ that reduces CMB variation to below the
level of detectability, at the expense of {}``tuning'' the universe
to place us as observers in the era when the CMB is nearly isotropic.

\subsubsection{Compatibility with an almost-isotropic Hubble flow}

The objection could be raised that the necessity of the universe contracting
along two axes demands that a large region of the sky be blue-shifted,
which would surely have been observed. This problem can be made to
vanish into statistical noise by the choice of a sufficiently small
$\eta$ as (\ref{eq:generalized H}) implies $a_{F}H_{11}=a_{F}^{\prime}/a_{F}+\alpha^{\prime}\approx2\left(\eta^{-1}+c_{2}^{\alpha}\eta\right)$
\emph{etc}.

\section{Conclusions}

It is possible for a Bianchi IX universe with initial conditions $c_{0}^{\alpha},c_{0}^{\beta},c_{0}^{\gamma}\sim1$
to display the acceleration observed in our universe while not only
remaining compatible with the observed CMB but providing and explanation
for potentially meaningful patterns in the CMB, specifically the so-called
{}``Axis of Evil'' and its associated phenomena such as cold spots.
These conditions can be attained without additional constraints on
the cosmological parameter of $\Omega_{K}$, a parameter which is
subject to further scrutiny and potentially tightening toward the
flat universe case of $\Omega_{K}=0$.

The method of combining strong growing modes with linear-order weak
decaying modes of cosmological gravitational waves is borne out by
observational data, which imply a difference of at least 17 orders
of magnitude in amplitude between the growing and decaying modes.
In the current epoch, decaying modes of cosmological gravitational
waves can be neglected entirely. However, in the time close to last
scattering, these modes may have participated at a strength comparable
to the growing modes. Furthermore, the action of growing or decaying
modes on their own is ruled out as an explanation for acceleration
as neither on its own can preserve the CMB.

The question of how the ratio of growing mode to decaying mode is
equal along all three principle axes of the metric tensor is answered
easily if we postulate that cosmological gravitational waves present
at the earliest moment in time were all in phase (the easiest way
to do this is to postulate that they consisted of pure growing modes).
As the functions $\alpha$, $\beta$ and $\gamma$ would have all
crossed the boundary from a $w=1/3$ medium to a $w=0$ medium at
the same time, they would thus have remained in phase after last scattering,
implying equal growing-to-decaying ratios for all three functions.
As this transition happened in the very young universe ($\eta_{E}\lesssim2\times10^{-3}$),
the decaying mode that exists after last scattering would be very
small.

The nonlinearity of Bianchi IX causes growing modes with initial values
of order unity to develop exponentially and cause very powerful effects.
The structure of the equations also indicates that multiple sets of
initial conditions can lead to the same set of cosmological parameters.
In light of the requirement of this model that both strong growing
modes and weak but non-zero decaying modes of the gravitational waves
exist, the possibility that these long-wavelength gravitational waves
constituted the {}``pump field'' of inflation\cite{gw inflation}
in the early universe should be explored.

The model proposed can be tested and falsified by observation of acceleration
in areas of the sky $90^{\circ}$ from the highly-observed field;
in areas of the sky away from the currently-observed acceleration,
we will see either a very large deceleration parameter or a negative
one. From the analysis of acceleration data in (\noun{Part} \ref{sec:Observations-of-acceleration})
it is easy to see that, in the current state of observations, there
are several possible areas of the sky where evidence of a gravitational-wave
nature of cosmic acceleration could be sitting undetected.

\part{Conclusions}

\section{Directions for future research}

The possibility of explaining cosmic acceleration through a Bianchi
IX cosmological model opens up numerous possibilities for future research,
both theoretical and observational.

While the difficulties with carrying out a full-sky optical survey
of supernovae are understandable, experimental verification or falsification
of a Bianchi IX model for acceleration requires nearly full sky coverage
at high \emph{z} to discover or rule out regions of anisotropy in
the acceleration field. Infrared astronomy with wide sky coverage,
for example WFIRST\cite{WFIRST}, presents the best possibility for
these new observations through traditional astronomy. The Einstein
telescope provides the tantalizing possibility of independent verification
of the properties of dark energy through the examination of gravitational
radiation.\cite{Zhao}

Meanwhile, the available supernova data can be re-examined for signs
of acceleration, although given the comparatively small datasets in
any particular area other than the highly-observed field and the equatorial
bias in the distribution of the data this re-examination is less likely
to produce definitive results. Célérier is justified in her criticisms\cite{really cosmological constant?}
of the assumptions being made in proposed models of cosmological acceleration;
it is curious that the authors if \cite{local Hubble Bubble} reasoned,
with 44 low-\emph{z }sources, that {}``poor coverage at low and moderate
Galactic latitudes {[}...{]} makes it practically impossible to distinguish
between a peculiar monopole and a quadrupole'' but that \cite{Riess 1998},
which shares two authors with \cite{local Hubble Bubble}, does not
even mention the possibility of cosmographic bias in its smaller sample
of high-\emph{z} sources.

Consideration should be given to the question of why cosmographic
bias exists, and whether it points to an unexpected privileging of
the observer: namely, the fact that modern observatories are hosted
only in regions of the Earth that can afford to host them.

Perturbative methods for solving the Einstein equations for weak gravitational
waves in Bianchi IX can be considered exhausted, having reached the
limit of practical utility at quadratic order. Further analytic explorations
should concentrate on the quasi-isotropic approach. The fact of Bianchi
IX's easy reduction to a system of non-linear second-order ordinary
differential equations combined with the divergence of Taylor series
describing strong gravitational waves point toward either a Fourier-series
approach or numerical methods for further analysis; the likelihood
of chaotic behavior\cite{Mixmaster} in Bianchi IX, though, merits
caution in the selection of initial conditions for any simulation.

Numerical examination of the quasi-isotropic regime should also be
pursued for a fuller exploration of the space allowing for anisotropic
acceleration while preserving an almost-isotropic cosmic microwave
background. The next generation of microwave anisitropy probe should
settle the question of whether the {}``Axis of Evil'' and similar
phenomena are genuine artifacts or statistical noise; in the meantime,
the question of pareidolia in relation to the CMB has not been explored
and deserves formal examination in order to raise awareness within
the scientific community of the issue.

Overall, any theory is only as good as its ability to predict future
results. Cosmic acceleration needs to be more closely examined, not
only for time dependence, but for spatial dependence, before any theory
can emerge as preferred.

\section{Implications of the Bianchi IX cosmological model}

Since the discovery of cosmic acceleration, a wide range of scalar
theories, ranging from the mundane to the exotic, have been put forward
to explain the phenomenon. While the fact of acceleration, the discovery
of which was the logical culmination of the hunt for the {}``missing
mass'' of the universe above and beyond that provided by dark matter,
necessarily implies the slaughter of at least one sacred cow, the
community of physicists has no consensus over which should be sacrificed
the most readily.

Attempts to surrender homogeneity are physically the best-grounded
but philosophically the most rash. Certainly the idea of a purely
homogeneous cosmology is an approximation; but a universe which is
not on average homogeneous, that is, where the homogeneous regions
are rare exceptions, is one in which cosmology as a science ceases
to be possible. The {}``Swiss cheese'' universe has the advantage
of making use of a known, exact solution to the Einstein equations
and at least avoids the exceptionalism of the {}``Hubble bubble''
proposal, but defeats itself on the grounds of testability.

Meanwhile, postulation of exotic states of matter has been done too
enthusiastically for the evidence available. The simple fact of noting
that the available data on acceleration was anisotropic exposes as
irrational exuberance the rush to explain the phenomenon through the
medium of a substance which has never been seen or even indicated
in the laboratory, and whose theoretical justification is far beyond
testability. The willingness of many to see acceleration as a falsification
of the theory of general relativity looks all the more bizarre when
counterposed with the unwillingness to explore gravitational-wave
solutions to the problem.

The objection could be raised that asserting acceleration to potentially
be anisotropic, in the weak sense of the word {}``isotropy'', violates
the cosmological principle by saying that our telescopes are privileged
observers, in that our observational field happens to align with an
axis of acceleration. This is no more so true than the {}``privilege''
hypothesized by, for example, Riess \emph{et. al.} when they assert,
from a few dozen data points, that acceleration is a recent phenomenon,
and that implicitly we are privileged observers in time for taking
up cosmology just as the universe has begun to exhibit this behavior.
While a cosmological constant is the simplest explanation for $w_{X}=-1$
on mathematical grounds, the lack of physical justification for a
non-zero cosmological constant puts it in the same class as scalar-field
theories. The simple fact is, $w_{X}=-1$ is, in the long run, the
natural equation of state for any function which grows faster than
the matter-driven terms in the background cosmology. The idea of the
{}``Big Rip''\cite{Big Rip}, while intellectually (and emotionally)
intriguing, makes the same mistake in the other direction, privileging
observers to be alive just as the universe is beginning to tear itself
apart. In this sense, a $w_{X}=-1$ field is the best preserver of
the cosmological principle, and when the cosmological constant has
been excluded the simplest explanation for acceleration comes from
a tensorial field.

Similarly, when cosmic flatness is called into question -- and it
has never been and can never be definitively proven, it can only be
disproven -- the next-simplest model is the closed model. Recall that
the Bianchi models are distinguished by their symmetries, and of all
the Bianchi models with Friedmann universes as special cases, Bianchi
IX has the most symmetric symmetries, obeying a {}``handedness''
rule students learn before their first year of university. The fact
of this {}``handedness'' -- parity -- may even provide a neat explanation
of the CP violation in particle physics\cite{Handedness}, as Grishchuk
alluded to\cite{GDI}.

The least speculative fact revealed by the assessment of available
acceleration data is that more data is needed, from broader areas
of the sky. The anticipated launch of WFIRST is likely to prove more
momentous for cosmology than the flight of WMAP; WMAP largely reconfirmed
what we already believed we knew, but WFIRST and SNAP will clearly
illustrate how much we do not know. We also need techniques to see
deeper into the sky and measure the distance-redshift relationship
further into the past; the standard ladder of baryon acoustic oscillations\cite{BAO peak}
combined with better redshift data from WiggleZ may provide the necessary
window.

That Bianchi IX could in principle contain accelerating regimes was
never really in doubt. Numerical and qualitative analysis has indicated
this ever since \cite{GDI} noted that the vacuum equations contained
a regular minimum, implying a positive first derivative for the Hubble
parameter. The character of the acceleration has now been more properly
investigated, bringing with it the possibility of a purely gravitational
explanation for inflation, especially in light of the divergence of
$\delta$ constructed only from growing modes in the radiation-dominated
universe. An exploration of the differences between Bianchi I and
Bianchi IX in a universe filled with ultra-relativistic matter could
make Bianchi IX into a panacea for all the major problems of large-scale
cosmology. 

The unwillingness of the perturbed Bianchi IX cosmology to support
decaying-mode gravitational waves stronger than linear order is puzzling,
especially as the BKL universe always has a divergent term. The BKL
universe, though, never reaches a singularity, and so the divergence
of the a decaying mode never has time to take effect. Furthermore,
the power law contraction along one axis could always be explained
by a {}``growing'' (non-diverging) function with negative coefficients,
due to the exponential term in the metric.

The impact of strong waves on the CMB, meanwhile, also requires deeper
explanation. Preservation of the CMB's apparent anisotropy at first
glance appears to require some {}``tuning'', a particular growing-decaying
ratio which merits deeper questioning; there is also the outstanding
matter of why we happen to live in one of the few periods of time
when the CMB appears nearly isotropic. Clever examination of the symmetries
of Bianchi IX may reveal a more satisfying answer, although the ability
of Bianchi IX to explain CMB anomalies is one of its most satisfying
features.

Most fundamentally, the biggest impact of the Bianchi IX theory of
cosmic acceleration is the expansion of the cosmologist's parameter
space. While in scalar models the only parameter truly open for discussion
is the function describing the equation of state of dark energy, the
gravitational waves of the Bianchi IX universe have four degrees of
freedom; while the strength a non-zero cosmological constant has some
theoretical justification in fundamental physics independent of large-scale
cosmology, there is no immediately apparent reason why the gravitational
waves in Bianchi IX should have any particular amplitude. As always
in cosmology, we need more information than we have.

\section*{Acknowledgments}

I thank Leonid Grishchuk and Peter Coles for continuing advice, support
and supervision throughout this research. Dr Rockhee Sung of University
of Cape Town provided helpful conversations.

\part*{\pagebreak{}Appendix\label{par:Appendix:-Supernovae-used}}

\begin{longtable}{|l||l||l|}
\caption{\label{tab:Locations-of-observations}Supernova observations used
in analysis of acceleration}
\endfirsthead
\cline{2-3} 
\multicolumn{1}{l|}{} & Right ascension, J2000 & Declination, J2000\tabularnewline
\hline 
\multicolumn{3}{|l|}{\emph{Riess 1998 supernovae:\cite{Riess evidence}}}\tabularnewline
\hline 
SN1994U & 13:04:56 & \textminus{}6:3:39\tabularnewline
\hline 
SN1997bp & 12:46:54 & \textminus{}10:21:27\tabularnewline
\hline 
SN1996V & 11:21:31 & 2:48:40\tabularnewline
\hline 
SN1994C & 07:56:40 & 44\textdegree{} 52' 19\textquotedblright{}\tabularnewline
\hline 
SN1995M & 09:38:42 & \textminus{}11:39:52\tabularnewline
\hline 
SN1995ae & 23:16:56 & \textminus{}1:55:24\tabularnewline
\hline 
SN1994B & 08:20:41 & 15:43:49\tabularnewline
\hline 
SN1995ao & 02:57:31 & \textminus{}0:18:40\tabularnewline
\hline 
SN1995ap & 03:12:28 & 0:41:43\tabularnewline
\hline 
SN1996R & 11:16:10 & 0:11:39\tabularnewline
\hline 
SN1996T & 10:05:28 & \textminus{}6:32:36\tabularnewline
\hline 
SN1997I & 04:59:37 & \textminus{}2:50:58\tabularnewline
\hline 
SN1997ap & 13:47:10 & 2:23:57\tabularnewline
\hline 
\multicolumn{3}{|l|}{\emph{SDSS-II SNIa observations:\cite{SDSS-II technical}}}\tabularnewline
\hline 
(Corner 1) & 20:00:00 & 1:15:00\tabularnewline
\hline 
(Corner 2) & 20:00:00 & \textminus{}1:15:00\tabularnewline
\hline 
(Corner 3) & 04:00:00 & 1:15:00\tabularnewline
\hline 
(Corner 4) & 04:00:00 & \textminus{}1:15:00\tabularnewline
\hline 
\multicolumn{3}{|l|}{\emph{ESSENCE windows:\cite{ESSENCE}}}\tabularnewline
\hline 
waa1 & 23:29:52.92 & \textminus{}08:38:59.7\tabularnewline
\hline 
waa2 & 23:27:27.02 & \textminus{}08:38:59.7\tabularnewline
\hline 
waa3 & 23:25:01.12 & \textminus{}08:38:59.7\tabularnewline
\hline 
waa5 & 23:27:27.02 & \textminus{}09:14:59.7\tabularnewline
\hline 
waa6 & 23:25:01.12 & \textminus{}09:14:59.7\tabularnewline
\hline 
waa7 & 23:30:01.20 & \textminus{}09:44:55.9\tabularnewline
\hline 
waa8 & 23:27:27.02 & \textminus{}09:50:59.7\tabularnewline
\hline 
waa9 & 23:25:01.12 & \textminus{}09:50:59.7\tabularnewline
\hline 
wbb1 & 01:14:24.46 & 00:51:42.9\tabularnewline
\hline 
wbb3 & 01:09:36.40 & 00:46:43.3\tabularnewline
\hline 
wbb4 & 01:14:24.46 & 00:15:42.9\tabularnewline
\hline 
wbb5 & 01:12:00.46 & 00:15:42.9\tabularnewline
\hline 
wbb6 & 01:09:00.16 & 00:10:43.3\tabularnewline
\hline 
wbb7 & 01:14:24.46 & \textminus{}00:20:17.1\tabularnewline
\hline 
wbb8 & 01:12:00.46 & \textminus{}00:20:17.1\tabularnewline
\hline 
wbb9 & 01:09:36.40 & \textminus{}00:25:16.7\tabularnewline
\hline 
wcc1 & 02:10:00.90 & \textminus{}03:45:00.0\tabularnewline
\hline 
wcc2 & 02:07:40.60 & \textminus{}03:45:00.0\tabularnewline
\hline 
wcc3 & 02:05:20.30 & \textminus{}03:45:00.0\tabularnewline
\hline 
wcc4 & 02:10:01.20 & \textminus{}04:20:00.0\tabularnewline
\hline 
wcc5 & 02:07:40.80 & \textminus{}04:20:00.0\tabularnewline
\hline 
wcc7 & 02:10:01.55 & \textminus{}04:55:00.0\tabularnewline
\hline 
wcc8 & 02:07:41.03 & \textminus{}04:55:00.0\tabularnewline
\hline 
wcc9 & 02:05:20.52 & \textminus{}04:55:00.0\tabularnewline
\hline 
wdd2 & 02:31:00.25 & \textminus{}07:48:17.3\tabularnewline
\hline 
wdd3 & 02:28:36.25 & \textminus{}07:48:17.3\tabularnewline
\hline 
wdd4 & 02:34:30.35 & \textminus{}08:19:18.2\tabularnewline
\hline 
wdd5 & 02:31:00.25 & \textminus{}08:24:17.3\tabularnewline
\hline 
wdd6 & 02:28:36.25 & \textminus{}08:24:17.3\tabularnewline
\hline 
wdd7 & 02:33:24.25 & \textminus{}08:55:18.2\tabularnewline
\hline 
wdd8 & 02:31:00.25 & \textminus{}09:00:17.3\tabularnewline
\hline 
wdd9 & 02:28:36.25 & \textminus{}09:00:17.3\tabularnewline
\hline 
\multicolumn{3}{|l|}{\emph{HST supernovae:\cite{HST}}}\tabularnewline
\hline 
SCP05D0 & 02:21:42.066 & \textminus{}03:21:53.12\tabularnewline
\hline 
SCP06H5  & 14:34:30.140 & 34:26:57.30\tabularnewline
\hline 
SCP06K0  & 14:38:08.366 & 34:14:18.08\tabularnewline
\hline 
SCP06K18  & 14:38:10.665 & 34:12:47.19\tabularnewline
\hline 
SCP06R12  & 02:23:00.083 & \textminus{}04:36:03.05\tabularnewline
\hline 
SCP06U4 & 23:45:29.430 & \textminus{}36:32:45.75\tabularnewline
\hline 
SCP06C1 & 12:29:33.013 & 01:51:36.67\tabularnewline
\hline 
SCP06F12  & 14:32:28.749 & 33:32:10.05\tabularnewline
\hline 
SCP05D6  & 02:21:46.484 & \textminus{}03:22:56.18\tabularnewline
\hline 
SCP06G4 & 14:29:18.744 & 34:38:37.39\tabularnewline
\hline 
SCP06A4  & 22:16:01.078 & \textminus{}17:37:22.10\tabularnewline
\hline 
SCP06C0  & 12:29:25.655 & 01:50:56.59\tabularnewline
\hline 
SCP06G3  & 14:29:28.430 & 34:37:23.15\tabularnewline
\hline 
SCP06H3 & 14:34:28.879 & 34:27:26.62\tabularnewline
\hline 
SCP06N33  & 02:20:57.699 & \textminus{}03:33:23.98\tabularnewline
\hline 
SCP05P1  & 03:37:50.352 & \textminus{}28:43:02.67\tabularnewline
\hline 
SCP05P9 & 03:37:44.513 & \textminus{}28:43:54.58\tabularnewline
\hline 
SCP06X26  & 09:10:37.888 & 54:22:29.06\tabularnewline
\hline 
SCP06Z5 & 22:35:24.967 & \textminus{}25:57:09.61\tabularnewline
\hline 
\multicolumn{3}{|l|}{\emph{Riess {}``gold'' dataset:\cite{Riess far supernovae,Riess Gold updated}}}\tabularnewline
\hline 
Window 1 & 03:32:30 & \textminus{}27:46:50:00\tabularnewline
\hline 
Window 2 & 12:37:00 & 62:10:00\tabularnewline
\hline
\end{longtable}

\end{document}